\documentclass[aps,
notitlepage,twocolumn,showpacs,superscriptaddress,nofootinbib,preprintnumbers]{revtex4-1}  % for review and submission
\usepackage{graphicx}  % needed for figures
\usepackage{dcolumn}   % needed for some tables
\usepackage{bm}        % for math
\usepackage{amssymb}   % for math
\usepackage{amsmath}
\usepackage{slashed}
\usepackage{mathtools}
\usepackage[colorlinks=true, linkcolor=.]{hyperref}
\usepackage{cleveref}
\usepackage{tikz}
\usetikzlibrary{patterns}

\DeclarePairedDelimiter{\floor}{\lfloor}{\rfloor}

\allowdisplaybreaks

\newcommand{\cE}{\mathcal{E}}
\newcommand{\cO}{\mathcal{O}}
\newcommand{\cM}{\mathcal{M}}

\newcommand{\Vtot}{ V_{\rm tot}}
\newcommand{\Qtot}{ Q_{\rm tot}}
\newcommand{\Fa}{F^{(\mathfrak{a}_{1})}}

\begin{document}

% the following line is for submission, including submission to the arXiv!!
%\hspace{5.2in} \mbox{Fermilab-Pub-04/xxx-E}

\title{Second post-Minkowskian scattering to all orders in spin }
\title{Classical gravitational scattering at $\cO(G^{2}S_{1}^{\infty}S_{2}^{\infty})$}
%\input author_list.tex       % D0 authors (remove the first 3 lines
                             % of this file prior to submission, they
			     % contain a time stamp for the authorlist)
%\author{Andreas Helset}                             % (includes institutions and visitors)

%\email[]{ahelset@caltech.edu}

%\affiliation{Walter Burke Institute for Theoretical Physics, California Institute of Technology, Pasadena, CA 91125}

\author{Rafael Aoude}
\email[]{rafael.aoude@uclouvain.be}
\affiliation{Centre for Cosmology, Particle Physics and Phenomenology (CP3),\\
Universit\'{e} catholique de Louvain, 1348 Louvain-la-Neuve, Belgium}
\author{Kays Haddad}
\email[]{kays.haddad@physics.uu.se}
\affiliation{Department of Physics and Astronomy, Uppsala University, \\
Box 516, 75120 Uppsala, Sweden}
\affiliation{Nordita, Stockholm University and KTH Royal Institute of Technology, \\
Hannes Alfv\'{e}ns v\"{a}g 12, 10691 Stockholm, Sweden}
\author{Andreas Helset}
\email[]{ahelset@caltech.edu}
\affiliation{Walter Burke Institute for Theoretical Physics,
California Institute of Technology,\\ Pasadena, CA 91125, USA}

\date{\today}

\begin{abstract}
	We calculate the scattering of two rotating objects with the linear-in-curvature spin-induced multipoles of Kerr black holes at $\cO(G^2)$ and all orders in the spins of both objects.
    This is done including the complete set of contact terms potentially relevant to Kerr-black-hole scattering at $\cO(G^{2})$.
    As such, Kerr black holes 
    %are 
    should be described by this scattering amplitude for a specific choice of values for the contact-term coefficients.
    %We provide the data needed to search for structure in the amplitude and thereby identify the interactions of Kerr black holes.
    The inclusion of all potential contact terms means this amplitude allows for a comprehensive search for structures emerging for certain values of the coefficients, and hence special properties that might be exhibited by Kerr-black-hole scattering.
    Our result can also act as a template for comparison for future computations of classical gravitational high-spin scattering.
\end{abstract}


\preprint{ CP3-23-19}
\preprint{ UUITP-10/23}
\preprint{ CALT-TH-2023-010}

\maketitle

%%%%%%%%%%%%%%%%%%%%%%%%%%%%%%%%%%%%%%%%%%%%%%%%%%%%%%%
%%%%%%% Begin Introduction %%%%%%%%%%%%%%%%%%%%%%%%%%%%
%%%%%%%%%%%%%%%%%%%%%%%%%%%%%%%%%%%%%%%%%%%%%%%%%%%%%%%

\section{Introduction}

Black holes are unique objects. 
By the no-hair theorem \cite{PhysRev.164.1776,PhysRevLett.26.331}, they are described solely by their charge, mass, and angular momentum, where only the latter two are necessary for describing astrophysical (Kerr) black holes \cite{Hansen:1974zz}.
Despite this special property, stellar-mass black holes are ubiquitous in nature, as they are the endpoint of the life cycle of sufficiently-massive stars. 
Moreover, experimental advances have made these previously-elusive objects accessible for direct study.
While the shadow of a supermassive black hole has been captured in images \cite{EventHorizonTelescope:2019dse}, stellar-mass black holes are most readily observed through gravitational waves emitted during compact binary coalescence \cite{LIGOScientific:2016aoc}.

The no-hair theorem paints an image of black holes which is not dissimilar to quantum fields.
The latter can also be characterized by three quantities: charges under gauge groups, mass, and spin quantum number.
Such parallels have not gone unnoticed, with the theoretical study of classical black holes surging in recent years simultaneously with observational breakthroughs
\cite{Cachazo:2017jef,Vines:2017hyw,Guevara:2018wpp,Chung:2018kqs,Kosower:2018adc,Vines:2018gqi,KoemansCollado:2019ggb,Maybee:2019jus,Guevara:2019fsj,Siemonsen:2019dsu,Arkani-Hamed:2019ymq,Damgaard:2019lfh,Aoude:2020onz,Chung:2020rrz,Guevara:2020xjx,Kosmopoulos:2021zoq,Bautista:2021wfy,Haddad:2021znf,Herrmann:2021lqe,Herrmann:2021tct,Chiodaroli:2021eug,Brandhuber:2021eyq,Alessio:2022kwv,Aoude:2022trd,Aoude:2022thd,FebresCordero:2022jts,Bellazzini:2022wzv,Bautista:2022wjf,Cangemi:2022bew,Bjerrum-Bohr:2023jau,Brandhuber:2023hhy,Herderschee:2023fxh,Elkhidir:2023dco,Georgoudis:2023lgf}.
In particular, techniques from scattering amplitudes have been used with great success to analytically calculate the interaction Hamiltonian for two black holes \cite{Cheung:2018wkq,Bern:2019nnu,Cristofoli:2019neg,Bern:2019crd,Bern:2020buy,Chen:2021kxt,Bern:2021yeh,Bern:2022kto,Bautista:2023szu,Barack:2023oqp}, relevant for describing the gravitational emission from black-hole mergers.

In the process of employing amplitudes techniques for high-precision calculations, special qualities of amplitudes relevant to Kerr black holes have been uncovered. 
By matching to the classical computation of ref.~\cite{Vines:2017hyw}, a particular scattering amplitude has been shown to describe the dynamics of a Kerr black hole at leading order in Newton's constant, including all spin effects \cite{Arkani-Hamed:2017jhn}.
This amplitude has an intricate structure where the full dependence on the spin is captured by an exponential factor \cite{Guevara:2018wpp,Guevara:2019fsj,Arkani-Hamed:2019ymq,Aoude:2020onz}. Intriguingly, the amplitude which turned out to describe Kerr black holes is singled out by a notion of maximal simplicity \cite{Arkani-Hamed:2017jhn}.

Progress beyond leading order in Newton's constant has also been made for Kerr black holes, which depends on higher-point amplitudes.
When high orders in the spin-multipole expansion are of interest, the necessary amplitudes are most conveniently constructed using recursive techniques \cite{Britto:2004ap,Britto:2005fq}. 
This, however, introduces unphysical poles that must be removed from the amplitude \cite{Arkani-Hamed:2017jhn}, which were encountered in the specific context of Kerr black holes in refs.~\cite{Guevara:2018wpp,Chung:2018kqs}.
The unphysical poles in the opposite-helicity Compton amplitude have been removed to all orders in the spin for black holes \cite{Aoude:2022trd} (see also refs.~\cite{Chung:2018kqs,Falkowski:2020aso,Chiodaroli:2021eug,Cangemi:2022bew,Bautista:2022wjf}) and neutron stars \cite{Haddad:2023ylx}.
When constructing the same-helicity Compton amplitude recursively, unphysical poles do not develop for Kerr-black-hole spin-induced multipoles \cite{Johansson:2019dnu}.
In order to capture all dynamics of the Compton amplitude, allowing for a complete set of contact terms with a priori unfixed coefficients comes hand-in-hand with removing unphysical poles.

As the amplitudes approach to general relativity is an effective one, the contact-term coefficients of the Compton amplitude must be unambiguously fixed through a matching calculation to a setup where the identity of the black hole is well established.
For example, the Compton amplitude could be matched to classical amplitudes derived from solutions of the Teukolsky equation \cite{Teukolsky:1973ha}. This program was initiated in refs.~\cite{Dolan:2008kf,Bautista:2021wfy,Bautista:2022wjf}.

Alternatively, one can seek an amplitudes-based definition of Kerr black holes.
Building off of the simplicity of the Kerr amplitude at leading order in the coupling and the uniqueness of Kerr black holes in the scheme of astrophysical objects, it is not unreasonable to anticipate that Kerr amplitudes at higher order in the coupling exhibit some identifying properties.
Therefore, a jumping-off point for an amplitudes-based understanding of Kerr black holes is to ask which values of the contact terms endow the amplitudes with special features.

Searching for structure in scattering amplitudes is a data-driven endeavor.
The aim of this paper is to provide the data to be mined, and hence build the superstructure which we expect to contain an analytic description of Kerr-black-hole scattering.
We do this by calculating the scattering amplitude of two rotating objects to $\cO(G^2)$ and all orders in both the relative velocity and their spins.
To zero in on the case of Kerr-black-hole scattering, we build this amplitude from the three-point amplitude known to describe Kerr black holes, and allow only the subset of all contact terms in the Compton amplitude which can contribute to Kerr-black-hole scattering at $\cO(G^{2})$.
This extends a previous all-order-in-spin result \cite{Aoude:2022thd} which considered one rotating and one nonrotating object.
Moreover, the inclusion of general contact terms relaxes a key assumption in that paper.

The layout of the paper is as follows. We start by discussing the on-shell formalism for heavy particles. Then we set up the calculation of the $\cO(G^2)$ scattering amplitude from unitarity cuts. We present the result for the all-order-in-spin scattering amplitude, and expand it in certain limits and kinematic regimes for illustration. Finally, we compare to the existing literature before concluding. The notation used throughout the paper is defined in the appendix.

%%%%%%%%%%%%%%%%%%%%%%%%%%%%%%%%%%%%%%%%%%%%%%%%%%%%%%%
%%%%%%% End Introduction %%%%%%%%%%%%%%%%%%%%%%%%%%%%%%
%%%%%%%%%%%%%%%%%%%%%%%%%%%%%%%%%%%%%%%%%%%%%%%%%%%%%%%

%%%%%%%%%%%%%%%%%%%%%%%%%%%%%%%%%%%%%%%%%%%%%%%%%%%%%%%
%%%%%%% Begin Heavy Particles %%%%%%%%%%%%%%%%%%%%%%%%%
%%%%%%%%%%%%%%%%%%%%%%%%%%%%%%%%%%%%%%%%%%%%%%%%%%%%%%%

\section{Heavy Particles}\label{sec:heavy}

Here we summarize the on-shell formalism for heavy particles. For more details, see refs.~\cite{Aoude:2020onz,Aoude:2022trd}.\footnote{Effective field theories with heavy particles applied to classical systems have recently been referred to as "Heavy-mass Effective Field Theories" or "HEFT" (see, e.g., refs.~\cite{Brandhuber:2021eyq,Bjerrum-Bohr:2023jau,Brandhuber:2023hhy,Herderschee:2023fxh}). We instead use the name Heavy Particle Effective Theory (HPET), which is more faithful to the long-standing literature regarding effective field theories with heavy particles \cite{Georgi:1990um,Luke:1992cs,Manohar:2000dt,Heinonen:2012km}. This name also avoids overlapping with the Higgs Effective Field Theory (HEFT) acronym appearing in beyond-the-Standard-Model literature \cite{Brivio:2017vri}.}
%%%%%%%%
The discussion of a heavy particle starts with decomposing the momentum of the heavy particle into two parts, one large and one small:
\begin{align}
	p_{\mu} = m v_{\mu} + k_{\mu} .
\end{align}
The velocity $v^{\mu}$ satisfies $v^2 = 1$.
The mass $m$ is assumed to be much larger than the typical interaction scale in the problem, while the residual momentum $k_{\mu}$ is on the order of this scale.
An expansion in $|k|/m$ thus naturally follows.

This is then folded in with the on-shell spinor-helicity variables for massive particles \cite{Conde:2016vxs,Conde:2016izb,Arkani-Hamed:2017jhn,Aoude:2020onz}. 
Usually, one defines a set of spinors for the momentum $p_{\mu}$ of a massive particle. 
For heavy particles, we instead define spinors for the velocity $v_{\mu}$ such that
\begin{align}
	v_{\mu} (\sigma^{\mu})_{\alpha \dot \alpha} = | v^{I} \rangle_{\alpha} [v_{I} |_{\dot \alpha} \,,
\end{align}
%%%%
%%%%
where $I$ is a massive little group index.
This approach presents several advantages for describing classical scattering.

First, the relevant (classical) expansion is precisely in the interaction scale over the masses of the heavy particles.
The classical expansion is thus manifested, and, in line with effective-field-theory lore, it is useful to make this expansion early on in the calculation. 
Second, the incoming and outgoing particles are described by the same $v^{\mu}$, since this quantity is conserved throughout the classical scattering. 
Defining the spin vector via the Pauli-Lubanski pseudovector, it is thus natural to choose the constant four-velocity for the reference vector, $S^{\mu} = - \frac{1}{2} \epsilon^{\mu\nu\alpha\beta} v_{\nu} J_{\alpha\beta}$.
Consequently, there is no need for the boosting of external states or Hilbert space matching, which stands in contrast to alternative approaches \cite{Guevara:2018wpp,Chung:2018kqs,Guevara:2019fsj,Arkani-Hamed:2019ymq}.
Finally, the spin-supplementary condition, $v_{\mu} S^{\mu} = 0$, is automatically satisfied with this choice of reference vector, enabling us to identify the spin vector appearing in on-shell amplitudes with the classical spin vector.

Throughout the calculation, we employ auxiliary variables $z_{I}$ to absorb any open little group index \cite{Chiodaroli:2021eug}, $| \bm{v} \rangle = | v^{I} \rangle z_{I}$ and $\langle \bar{\bm{v}} | = \langle v^{I} | \bar{z}_{I}$.
In turn, all spin dependence will appear in the combination
\begin{align}
	\frac{\langle \bar{\bm{v}}|^{2s} \{S^{\mu_{1}},\dots,S^{\mu_{n}}\} | \bm{v} \rangle^{2s}}{\langle \bar{\bm{v}}\bm{v} \rangle^{2s}} = m^{n} \mathfrak{a}^{\mu_{1}}\dots\mathfrak{a}^{\mu_{n}} ,
\end{align}
where curly brackets denote symmetrization.
In the scattering amplitudes we will tacitly strip off any overall factor of $\langle \bar{\bm{v}}\bm{v} \rangle^{2s}$ and use the ring radius $\mathfrak{a}^{\mu}$ to keep track of the spin dependence of the two spinning particles.\footnote{See ref.~\cite{Aoude:2021oqj} for a more rigorous treatment of these overall spinor products.}

%%%%%%%%%%%%%%%%%%%%%%%%%%%%%%%%%%%%%%%%%%%%%%%%%%%%%%%
%%%%%%% End Heavy Particles %%%%%%%%%%%%%%%%%%%%%%%%%%%
%%%%%%%%%%%%%%%%%%%%%%%%%%%%%%%%%%%%%%%%%%%%%%%%%%%%%%%

%%%%%%%%%%%%%%%%%%%%%%%%%%%%%%%%%%%%%%%%%%%%%%%%%%%%%%%
%%%%%%% Begin Calculation %%%%%%%%%%%%%%%%%%%%%%%%%%%%%
%%%%%%%%%%%%%%%%%%%%%%%%%%%%%%%%%%%%%%%%%%%%%%%%%%%%%%%

\section{Building blocks for $\cO(G^{2})$ scattering}\label{sec:calculation}

\begin{figure}
    \centering
    \begin{tikzpicture}[scale=1, transform shape]
\usetikzlibrary{decorations.pathmorphing}
\tikzset{snake it/.style={decorate, decoration=snake}}
    \node (A2) at (2.5,-.75) {$+$};
    \path [draw=black] (-1.5,0) -- (1.5,0);
    \path [draw=black] (-1.5,-1.5) -- (1.5,-1.5);
    \path [draw=black] (3.5,0) -- (6.5,0);
    \path [draw=black] (3.5,-1.5) -- (6.5,-1.5);
    \path [draw=black, snake it] (0,-1.5) -- (-1,0);
    \path [draw=black, snake it] (0,-1.5) -- (1,0);
    \path [draw=black, snake it] (5,0) -- (4,-1.5);
    \path [draw=black, snake it] (5,0) -- (6,-1.5);
    \filldraw[fill=white] (-1,0) circle (6pt);
    \filldraw[fill=white] (1,0) circle (6pt);
    \filldraw[fill=white] (0,-1.5) circle (6pt);
    \filldraw[fill=white] (4,-1.5) circle (6pt);
    \filldraw[fill=white] (6,-1.5) circle (6pt);
    \filldraw[fill=white] (5,0) circle (6pt);
\end{tikzpicture}
\caption{The relevant unitarity cuts for the classical dynamics at $\cO{(G^2)}$. The straight lines are black holes while the wavy lines are the exchanged gravitons. All propagators are on-shell, so each vertex represents an on-shell, tree-level scattering amplitude.}
    \label{fig:cuts}
\end{figure}
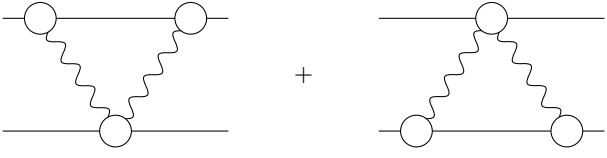

We want to describe the classical scattering of two spinning particles at $\cO(G^2)$.
This is done through a one-loop scattering amplitude, which can be decomposed into a basis of box, triangle, and bubble scalar integrals. For classical physics, we only need the coefficients of the triangle integrals, which we calculate from the triangle unitarity cut shown in \cref{fig:cuts}. 

The triangle cuts are built from tree-level three-point and Compton amplitudes.
Our ambition to produce a second post-Minkowskian (2PM) amplitude relevant to Kerr-black-hole scattering constrains these tree-level amplitudes in the following ways:
\begin{itemize}
    \item The three-point amplitude which describes the interactions between a Kerr black hole and a single graviton is given by the 'minimal' amplitude of ref.~\cite{Arkani-Hamed:2017jhn}; see refs.~\cite{Levi:2015msa,Vines:2017hyw,Guevara:2018wpp,Chung:2018kqs}. The Compton amplitude must factorize onto this amplitude on physical residues.
    \item Kerr-black-hole scattering at low spin orders (up to fourth order in spin) is known to satisfy the so-called black-hole spin-structure assumption \cite{Aoude:2022trd}. Specifically, the spin dependence appears only in the combinations ${(q\cdot\mathfrak{a}_{i})(q\cdot\mathfrak{a}_{j})-q^{2}(\mathfrak{a}_{i}\cdot\mathfrak{a}_{j})}$, ${q^{2}(v_{i}\cdot\mathfrak{a}_{j})}$, and $\epsilon_{\mu\nu\rho\sigma}q^{\mu}v_{1}^{\nu}v_{2}^{\rho}\mathfrak{a}_{i}^{\sigma}$ for $i,j=1,2$ \cite{Damgaard:2019lfh,Bern:2020buy,Kosmopoulos:2021zoq,Chen:2021kxt,Aoude:2022trd,Bern:2022kto,Bautista:2022wjf}. This assumption is equivalent to the shift symmetry of ref.~\cite{Bern:2022kto} for two-massive-particle scattering, and is a manifestation of an analogous shift symmetry at the level of the Compton amplitude \cite{Aoude:2022thd}. We begin by computing the 2PM amplitude which obeys this assumption, and relax it later on.
\end{itemize}
In conjunction, these two points imply that the same-helicity Compton amplitude does not contribute to classical scattering at 2PM.
However, when we relax the second constraint we will need to consider the contributions from same-helicity Compton-amplitude contact terms to the unitarity cuts.

To evaluate the 2PM amplitude in accordance with these two constraints, all we need in addition to the three-point Kerr amplitude (which can be found, e.g., in ref.~\cite{Guevara:2018wpp}) is the shift-symmetric opposite-helicity Compton amplitude of ref.~\cite{Aoude:2022trd}.
For a positive-helicity graviton with momentum $q_{4}^{\mu}$ and a negative-helicity graviton with momentum $q_{3}^{\mu}$ (both outgoing), this amplitude is
\begin{align}
	\label{eq:Compton1}
	\cM_{4} =&  e^{(q_4 - q_3) \cdot \mathfrak{a}} \sum_{n=0}^{2s} \frac{1}{n!} \bar K_{n}  ,
\end{align}
where
\begin{align}
	\label{eq:Compton2}
	K_{n} &= \frac{y^4}{s_{34} t_{13} t_{14}} \left( \frac{t_{14} - t_{13}}{y} w\cdot \mathfrak{a} \right)^{n},   \\
	\bar K_{n} &= \begin{cases}
		K_{n} , & n < 4  \\
		K_{n} + m^2 (w \cdot \mathfrak{a} )^4 d^{(4)}_{0}, & n = 4  \\
		K_{3} L_{n-3} - K_{2} \mathfrak{s}_{2} L_{n-4} &\\
		+ m^2 (w\cdot \mathfrak{a})^{4} \sum_{j=0}^{\floor{(n-4)/2}} d^{(n)}_{j} \mathfrak{s}_{1}^{n-4-2j} \mathfrak{s}^{j}_{2}, & n> 4.
	\end{cases}\nonumber
\end{align}
Here we use $w^{\mu} = [4|\bar \sigma^{\mu}|3\rangle/2$. Also, $\mathfrak{s}_{1} = (q_3 - q_4)\cdot \mathfrak{a}$ and $\mathfrak{s}_{2} = -4 (q_3 \cdot \mathfrak{a})(q_4 \cdot \mathfrak{a}) + s_{34} \mathfrak{a}^2$. 
These enter in the sums
\begin{align}
	L_{m} = \sum_{j=0}^{\floor{m/2}} \binom{m+1}{2j+1} \mathfrak{s}_{1}^{m-2j} (\mathfrak{s}_1^2 - \mathfrak{s}_{2})^{j} .
\end{align}
This amplitude factorizes onto the three-point amplitudes describing Kerr black holes on all three poles, $s_{34}\equiv(q_{3}+q_{4})^{2}$ and $t_{1i}=(p_{1}-q_{i})^{2}-m^{2}$. 
The $d_j^{(n)}$ are unfixed dimensionless coefficients for shift-symmetric contact terms.
The upper bound of the sum in \cref{eq:Compton1} depends on the total quantum spin $s$ of the massive particle.
We take this quantity to infinity to describe classical spinning objects.

%%%%%%%%%%%%%%%%%%%%%%%%%%%%%%%%%%%%%%%%%%%%%%%%%%%%%%%
%%%%%%% End Calculation %%%%%%%%%%%%%%%%%%%%%%%%%%%%%%%
%%%%%%%%%%%%%%%%%%%%%%%%%%%%%%%%%%%%%%%%%%%%%%%%%%%%%%%

%%%%%%%%%%%%%%%%%%%%%%%%%%%%%%%%%%%%%%%%%%%%%%%%%%%%%%%
%%%%%%% Begin Results %%%%%%%%%%%%%%%%%%%%%%%%%%%%%%%%%
%%%%%%%%%%%%%%%%%%%%%%%%%%%%%%%%%%%%%%%%%%%%%%%%%%%%%%%

\section{All-order-in-spin amplitude}\label{sec:result}

The details of the construction of the 2PM amplitude using unitarity can be found, e.g., in refs.~\cite{Bern:2020buy,Chen:2021kxt,Aoude:2022trd}.
The scattering amplitude for two spinning particles at $\cO(G^2)$ is
\begin{align}\label{eq:2PMAmpEvenOddSplit}
	\cM_{2{\rm PM}} = \frac{2 G^2 \pi^2 m^2_1 m^2_2}{\sqrt{-q^2}} \left( \cM^{\rm even}_{2{\rm PM}} + \cM^{\rm odd}_{2{\rm PM}} \right) \,,
\end{align}
which we have split into an even- and an odd-in-spin part.
The even-in-spin part is
\begin{align}
	\label{eq:allSpinEvenResult}
	\cM^{\rm even}_{2{\rm PM}} =& - m_1 (8\hyperref[eq:omega]{\omega}^4 - 8\hyperref[eq:omega]{\omega}^2 + 1) \hyperref[eq:Xkl]{X_{0,0}} %\sum_{n=0}^{\infty} \frac{(-1)^{n+1} 2^{3-2n} }{(\omega^2 - 1)^{n+1} (3/2)_{n-1} } \Vtot^{n} F_{n-1} 
	\nonumber \\ &
	+ m_1 (\hyperref[eq:omega]{\omega} \hyperref[eq:Vij]{V_{12}} - \hyperref[eq:omega]{\omega}^2 \hyperref[eq:Vij]{V_{22}}) (2\hyperref[eq:omega]{\omega}^2 - 1) \hyperref[eq:Xkl]{X_{1,1}} %\sum_{n=0}^{\infty} \frac{(-1)^{n} 2^{3-2n}}{(\omega^2 - 1)^{n+1}(3/2)_{n}} \Vtot^{n} F_{n} 
	\nonumber \\ &
	- m_1 \sum_{n=0}^{\infty} \left( \hyperref[eq:Fn]{F_{2n}} \hyperref[eq:Pn]{P_{n}} + \hyperref[eq:Fn]{F_{2n+1}} \hyperref[eq:Qn]{Q_{n}} \right)
	+ \left( 1 \leftrightarrow 2 \right) \,,
\end{align}
and the odd-in-spin part is
\begin{align}
	\label{eq:allSpinOddResult}
	\cM^{\rm odd}_{2{\rm PM}} =& 
	+i m_1 \hyperref[eq:omega]{\omega} \hyperref[eq:calE]{\cE} (2\hyperref[eq:omega]{\omega}^2 - 1) 2 \hyperref[eq:Xkl]{X_{0,1}} %\sum_{n=0}^{\infty} \frac{(-1)^{n} 2^{4-2n}}{(\omega^2 - 1)^{n+1}(3/2)_{n-1}} \Vtot^n F_{n} 
	\nonumber \\ &
	- i m_1 \hyperref[eq:calE]{\cE} (\hyperref[eq:Vij]{V_{12}} - \hyperref[eq:omega]{\omega} \hyperref[eq:Vij]{V_{22}}) \frac{(8\hyperref[eq:omega]{\omega}^4 - 8 \hyperref[eq:omega]{\omega}^2 + 1)}{8(\hyperref[eq:omega]{\omega}^2 - 1)} \hyperref[eq:Xkl]{X_{1,2}}%\sum_{n=0}^{\infty}\frac{(-1)^{n+1} 2^{-2n}}{(\omega^2-1)^{n+2} (3/2)_{n} } \Vtot^n F_{n+1} 
	\nonumber \\ &
	- i m_1 \sum_{n=0}^{\infty} \left( \hyperref[eq:Fn]{F_{2n}} \hyperref[eq:Rn]{R_{n}} + \hyperref[eq:Fn]{F_{2n+1}} \hyperref[eq:Sn]{S_{n}} \right) 
	+ \left( 1 \leftrightarrow 2 \right) \,,
\end{align}
where
\begin{align}\label{eq:Xkl}
	X_{k,l} = \sum_{n=0}^{\infty} \frac{(-1)^{n+1} 2^{3-2n} }{(\hyperref[eq:omega]{\omega}^2 - 1)^{n+1} (3/2)_{n-1+k} } (\hyperref[eq:Vtot]{\Vtot})^{n} \hyperref[eq:Fn]{F_{n-1+l}}  \,.
\end{align}
We have included a symmetrization where all masses, spins, and velocities of the two heavy particles are swapped, along with changing the sign of the transferred momentum. 
All definitions are given in \cref{sec:definitions}, and nearly all symbols are clickable to jump to their definitions.

If we restrict our attention to the sector with one spinning and one nonspinning particle, we find agreement with the result in ref.~\cite{Aoude:2022thd}. Also, by expanding to eighth order in spin, we find the result in ref.~\cite{Aoude:2022trd}.

While \cref{eq:allSpinEvenResult,eq:allSpinOddResult} encode the all-order-in-spin amplitude, it is useful to study various spin sectors independently to ease the task of searching for structure in the result.
We provide $\cM^{\rm odd}_{\rm 2PM}$ and $\cM^{\rm even}_{\rm 2PM}$ explicitly expanded to $\cO(\mathfrak{a}^{28})$ in the ancillary files \texttt{amplitude2PModdinspinTospin27.mx} and \texttt{amplitude2PMeveninspinTospin28.mx}, respectively.
Note that the overall factor in \cref{eq:2PMAmpEvenOddSplit} has been omitted from the results in those files.

Below we also study the amplitude in some simplifying scenarios. 
We will first consider the probe limit for a spinning particle with up to quadratic orders in spin moving in a Kerr background. Then we will look at the aligned spin case to all orders in the spin for both particles.

\subsection{Probe limit}

We start with the probe limit for particle 2 including up to quadratic orders in its spin and all orders in the spin of the Kerr background.
The 2PM scattering amplitude in the probe limit is
\begin{align}
	\cM_{2{\rm PM}\; {\rm probe}} =& \frac{2 G^2 \pi^2 m^2_1 m^2_2}{\sqrt{-q^2}} \left( \cM^{\rm even}_{2{\rm PM}\; {\rm probe}} + \cM^{\rm odd}_{2{\rm PM}\; {\rm probe}} \right)  \,,
\end{align}
where the even-in-spin part is 
\begin{widetext}
	\begin{align}
		\label{eq:allSpinEvenResultProbe}
		&\cM^{\rm even}_{2{\rm PM}\; {\rm probe}}  =  m_1 \left[3 (5\hyperref[eq:omega]{\omega}^2 - 1) \hyperref[eq:Fa]{F^{(\mathfrak{a}_{1})}_{0}} - \frac{2\hyperref[eq:Vij]{V_{11}} - (\hyperref[eq:omega]{\omega}^2 - 1)\hyperref[eq:Qij]{Q_{11}}}{4} \hyperref[eq:Fa]{F^{(\mathfrak{a}_{1})}_{2}} - (8\hyperref[eq:omega]{\omega}^4 - 8\hyperref[eq:omega]{\omega}^2 + 1) \left(1 - \frac{\hyperref[eq:calEi]{\cE_{2}}^{2}}{2(\hyperref[eq:omega]{\omega}^2 - 1)} \right) \hyperref[eq:Ykl]{Y_{0,0}} \right]
		\nonumber \\ &
		+ m_1 \hyperref[eq:Qij]{Q_{12}} \left[ \left( 4\hyperref[eq:omega]{\omega}^2 - 1 \right) \hyperref[eq:Fa]{F^{(\mathfrak{a}_1)}_{1}} - \frac{1}{2}  ( 8\hyperref[eq:omega]{\omega}^4 - 8\hyperref[eq:omega]{\omega}^2 + 1 )  \hyperref[eq:Ykl]{Y_{0,1}} \right] 
		\nonumber %\\ &
		+
		m_1 \hyperref[eq:omega]{\omega} \hyperref[eq:Vij]{V_{12}} \left[ 4 \hyperref[eq:Fa]{F^{(\mathfrak{a}_1)}_{1}} 
		-\frac{ 8\hyperref[eq:omega]{\omega}^4 - 8\hyperref[eq:omega]{\omega}^2 + 1 }{2(\hyperref[eq:omega]{\omega}^2 - 1)} \hyperref[eq:Ykl]{Y_{0,1}} 
		-\frac{ 4\hyperref[eq:omega]{\omega}^2 - 3 }{4(\hyperref[eq:omega]{\omega}^2 - 1)}\hyperref[eq:Ykl]{Y_{1,1}}
		\right] 
		\nonumber \\ &+
		m_1 \hyperref[eq:Vij]{V_{22}} \left[ \frac{\hyperref[eq:omega]{\omega}^2 + 1}{2(\hyperref[eq:omega]{\omega}^2 - 1)} \hyperref[eq:Fa]{F_{1}^{(\mathfrak{a}_{1})}} - \frac{1}{4(\hyperref[eq:omega]{\omega}^2 - 1)} \hyperref[eq:Vij]{V_{11}} \hyperref[eq:Fa]{F_{2}^{(\mathfrak{a}_{1})}} 
		+  \frac{\hyperref[eq:omega]{\omega}^2 (4\hyperref[eq:omega]{\omega}^2 - 3)}{2(\hyperref[eq:omega]{\omega}^2 - 1)} \hyperref[eq:Ykl]{Y_{0,1}}
		\right] 
		\nonumber %\\ &
		+
		m_1 \hyperref[eq:omega]{\omega} \hyperref[eq:Vij]{V_{12}} \hyperref[eq:Qij]{Q_{12}} \left[ - \frac{1}{2(\hyperref[eq:omega]{\omega}^2 - 1)} \hyperref[eq:Fa]{F_{2}^{(\mathfrak{a}_{1})}} - \frac{4\hyperref[eq:omega]{\omega}^2 - 3}{8(\hyperref[eq:omega]{\omega}^2 - 1)} \hyperref[eq:Ykl]{Y_{1,2}} \right] 
		\nonumber \\ &+
		m_1\hyperref[eq:Qij]{Q_{22}} \left[ \frac{ 15\hyperref[eq:omega]{\omega}^4 - 15\hyperref[eq:omega]{\omega}^2 + 2 }{2(\hyperref[eq:omega]{\omega}^2 - 1)}  \hyperref[eq:Fa]{F_{1}^{(\mathfrak{a}_{1})}} 
		+ \frac{(8\hyperref[eq:omega]{\omega}^4 - 8 \hyperref[eq:omega]{\omega}^2 + 1)\hyperref[eq:Qij]{Q_{11}} -\hyperref[eq:omega]{\omega}^2 \hyperref[eq:Vij]{V_{11}}}{4(\hyperref[eq:omega]{\omega}^2 - 1)} \hyperref[eq:Fa]{F_{2}^{(\mathfrak{a}_{1})}}
		\right]  \,,
	\end{align}
	and the contribution at odd spin orders is 
	%
	\iffalse
	\begin{align}
		\label{eq:allSpinOddResultProbe}
		\cM^{\rm odd}_{2{\rm PM }\; {\rm probe}} =& 
		- i m_1 \omega \cE_{1}
		\left[  \left( 4 + \frac{2\omega^2 - 1}{\omega^2 - 1} Q_{22} - \frac{2\omega^2}{(\omega^2 - 1)^{2}} V_{22} \right) \Fa_1   - 2 (2\omega^2 - 1)  \left( 1 - \frac{\cE_{2}^2}{2(\omega^2 - 1)} \right) Y_{0,1} 
		\right]
		\nonumber \\ &
		-i m_1 V_{12} \cE_{1} \left[ - \frac{1}{2(\omega^2 - 1)} F^{(\mathfrak{a}_{1})}_{2} - \frac{4\omega^2 - 1}{8(\omega^2 - 1)} Y_{1,2}   \right]
		\nonumber \\ &
		-i m_1 \omega \cE_{2} \left[ \frac{11\omega^2 - 5}{\omega^2 - 1} \Fa_{0} + 4 \Fa_{1} + \left( \frac{5}{4} Q_{11} - \frac{1}{2(\omega^2 - 1)} V_{11} \right) \Fa_{2}
		- 4 (2\omega^2 - 1) Y_{0,0} 
		\right]
		\nonumber \\ &
		-i m_1 V_{12} \cE_{2} \left[-\frac{3\omega^2 - 1}{(\omega^2 - 1)^2} \Fa_{1} 
		- \frac{4\omega^2 - 1}{4(\omega^2 - 1)} Y_{1,1}
		\right] \,,
	\end{align}
	\fi
	\begin{align}
		\label{eq:allSpinOddResultProbe}
		&\cM^{\rm odd}_{2{\rm PM }\; {\rm probe}} = 
		- i m_1 \hyperref[eq:omega]{\omega} \hyperref[eq:calEi]{\cE_{1}}
		\left[  \left( 4 + \frac{2\hyperref[eq:omega]{\omega}^2 - 1}{\hyperref[eq:omega]{\omega}^2 - 1} \hyperref[eq:Qij]{Q_{22}} - \frac{2\hyperref[eq:omega]{\omega}^2}{(\hyperref[eq:omega]{\omega}^2 - 1)^{2}} \hyperref[eq:Vij]{V_{22}} \right) \hyperref[eq:Fa]{\Fa_1}   - 2 (2\hyperref[eq:omega]{\omega}^2 - 1)  \left( 1 - \frac{\hyperref[eq:calEi]{\cE_{2}}^2}{2(\hyperref[eq:omega]{\omega}^2 - 1)} \right) \hyperref[eq:Ykl]{Y_{0,1}}
		\right]
		\nonumber \\ &
		-i m_1 \hyperref[eq:Vij]{V_{12}} \hyperref[eq:calEi]{\cE_{1}} \left[ - \frac{1}{2(\hyperref[eq:omega]{\omega}^2 - 1)} \hyperref[eq:Fa]{F^{(\mathfrak{a}_{1})}_{2}} - \frac{4\hyperref[eq:omega]{\omega}^2 - 1}{8(\hyperref[eq:omega]{\omega}^2 - 1)} \hyperref[eq:Ykl]{Y_{1,2}}   \right]
		-i m_1 \hyperref[eq:Vij]{V_{12}} \cE_{2} \left[-\frac{3\hyperref[eq:omega]{\omega}^2 - 1}{(\hyperref[eq:omega]{\omega}^2 - 1)^2}\hyperref[eq:Fa]{\Fa_{1}}
		- \frac{4\hyperref[eq:omega]{\omega}^2 - 1}{4(\hyperref[eq:omega]{\omega}^2 - 1)} \hyperref[eq:Ykl]{Y_{1,1}}
		\right]
		\nonumber \\ &
		-i m_1 \hyperref[eq:omega]{\omega} \hyperref[eq:calEi]{\cE_{2}} \left[ \frac{11\hyperref[eq:omega]{\omega}^2 - 5}{\hyperref[eq:omega]{\omega}^2 - 1} \hyperref[eq:Fa]{\Fa_{0}} + 4 \hyperref[eq:Fa]{\Fa_{1}} + \left( \frac{5}{4} \hyperref[eq:Qij]{Q_{11}} - \frac{1}{2(\hyperref[eq:omega]{\omega}^2 - 1)} \hyperref[eq:Vij]{V_{11}} \right) \hyperref[eq:Fa]{\Fa_{2}}
		- 4 (2\hyperref[eq:omega]{\omega}^2 - 1) \hyperref[eq:Ykl]{Y_{0,0}} 
		\right] \,,
	\end{align}
	where $\Fa_{n} = F_{n}\vert_{\mathfrak{a}_{2} \rightarrow 0} \label{eq:Fa}$ and $Y_{k,l} = X_{k,l}\vert_{\mathfrak{a}_{2} \rightarrow 0} \label{eq:Ykl}$. As expected, the contact terms with unfixed coefficients do not enter in the probe limit at this spin order for the probe particle; they first appear at fourth order in the spin of the probe particle. Therefore, these probe-limit results describe the scattering of Kerr black holes.

\subsection{Aligned spins}
Next, we write down the scattering amplitude for aligned spins. 
This kinematic configuration sets all $V_{ij}$ to $0$.
The aligned-spin scattering amplitude is
\begin{align}
	\label{eq:resultaligned}
\cM_{2{\rm PM\;aligned}} =& \frac{2 G^2 \pi^2 m^2_1 m^2_2}{\sqrt{-q^2}} \left( \cM^{\rm even}_{2{\rm PM\;aligned}} + \cM^{\rm odd}_{2{\rm PM\;aligned}} \right) 
\end{align}
where the even-in-spin part is
\begin{align}
	\label{eq:allSpinEvenResultAligned}
	\cM^{\rm even}_{2{\rm PM\;aligned}} =& m_1 \frac{4(8\hyperref[eq:omega]{\omega}^4 - 8\hyperref[eq:omega]{\omega}^2 + 1)}{\hyperref[eq:omega]{\omega}^2 - 1} \hyperref[eq:Fn]{F_{-1}} 
	%\nonumber \\ &
	- m_1 \sum_{n=0}^{\infty} \left( \hyperref[eq:Fn]{F_{2n}} P^{\rm aligned}_{n} +\hyperref[eq:Fn]{ F_{2n+1}} Q^{\rm aligned}_{n} \right)
	+ \left( 1 \leftrightarrow 2 \right) \,.
\end{align}
We again symmetrize by swapping masses, spins, and velocities of the two heavy particles, along with changing the sign of the transferred momentum.
This part of the amplitude depends on the terms 
\begin{align}
	P^{\rm aligned}_{n} =& 
	\delta_{n0} \left(- (5\hyperref[eq:omega]{\omega}^2 - 1)\left(3 + \frac{\hyperref[eq:Qij]{Q_{22}}}{4} \hyperref[eq:hn]{h^{(2)}_{0}} \right) 	\right) 
	+ \hyperref[eq:Dn]{D_{2n}} + \left(\frac{\hyperref[eq:Qij]{Q_{12}}+\hyperref[eq:Qij]{Q_{22}}}{16} \right)^{2n} \hyperref[eq:CnContactTerm]{C_{2n}} 
		\nonumber \\ &
	+ \left( \hyperref[eq:Qtot]{\Qtot} (\hyperref[eq:omega]{\omega}^2 - 1) \right) \left[ \delta_{n1}  \left( - \frac{1}{4} - \left(\frac{\hyperref[eq:Qij]{Q_{22}}}{16}\right)^{2} ( \hyperref[eq:hn]{h_{1}^{(2)}} - 3 \hyperref[eq:hn]{h^{(4)}_{0}} )   - \left(\frac{\hyperref[eq:Qij]{Q_{22}}}{16}\right)^{3} \hyperref[eq:hn]{h_{1}^{(4)}} \right)
		- \delta_{n2} \left(\left(\frac{\hyperref[eq:Qij]{Q_{22}}}{16} \right)^2 \frac{1}{4} \hyperref[eq:hn]{h_{0}^{(4)}}\right) 
			\right] \,,
\end{align}
and
\begin{align}
	Q^{\rm aligned}_{n} =& 
	\delta_{n0} (\hyperref[eq:Qij]{Q_{12}} + \hyperref[eq:Qij]{Q_{22}}) \left( \frac{3}{4} (5\hyperref[eq:omega]{\omega}^2 - 1) - \frac{\hyperref[eq:Qij]{Q_{22}}}{16} \hyperref[eq:omega]{\omega}^2 \hyperref[eq:hn]{h_{0}^{(1)}} - \left( \frac{\hyperref[eq:Qij]{Q_{22}}}{16} \right)^{2} (\hyperref[eq:omega]{\omega}^2 - 1) \left(  \frac{1}{4} \hyperref[eq:hn]{h_{1}^{(1)}} + 3 \hyperref[eq:hn]{h_{1}^{(3)}} \right) + \left(\frac{\hyperref[eq:Qij]{Q_{22}}}{16} \right)^{3} (\hyperref[eq:omega]{\omega}^2 - 1) \hyperref[eq:hn]{h_{2}^{(3)}} \right)
	\nonumber \\ &
	+ \delta_{n1} (\hyperref[eq:Qij]{Q_{12}} + \hyperref[eq:Qij]{Q_{22}}) \hyperref[eq:Qtot]{\Qtot} (\hyperref[eq:omega]{\omega}^2 - 1) \frac{1}{16} \left(1 - \frac{\hyperref[eq:Qij]{Q_{22}}}{16} \hyperref[eq:hn]{h_{0}^{(1)}} + \left(\frac{\hyperref[eq:Qij]{Q_{22}}}{16}\right)^{2} 4 \hyperref[eq:hn]{h_{1}^{(3)}} \right)
	+\hyperref[eq:Dn]{D_{2n+1}} %+ {\rm  termG}^{(3)}_{n}
	+ \left(\frac{\hyperref[eq:Qij]{Q_{12}}+\hyperref[eq:Qij]{Q_{22}}}{16} \right)^{2n+1} 
	\hyperref[eq:CnContactTerm]{C_{2n+1}} \,.
\end{align}
The odd-in-spin part is
\begin{align}
	\label{eq:allSpinOddResultAligned}
	\cM^{\rm odd}_{2{\rm PM\;aligned}} =& 
	-i m_1 \hyperref[eq:omega]{\omega} \left[ \left( \hyperref[eq:calE]{\cE} \frac{8(2\hyperref[eq:omega]{\omega}^2 - 1)}{\hyperref[eq:omega]{\omega}^2 - 1} - \hyperref[eq:calEi]{\cE_{2}} \frac{5\hyperref[eq:omega]{\omega}^2 - 3}{\hyperref[eq:omega]{\omega}^2 - 1} \right)  \hyperref[eq:Fn]{F_{0}} 
	+ 4 \hyperref[eq:calE]{\cE}  \hyperref[eq:Fn]{F_{1}} 
	+ \left( -  \hyperref[eq:calEi]{\cE_{1}} \frac{3}{2} \hyperref[eq:Qij]{Q_{22}} -  \hyperref[eq:calEi]{\cE_{2}} \frac{3}{4} ( \hyperref[eq:Qij]{Q_{11}} + \hyperref[eq:Qij]{Q_{22}} ) \right) \hyperref[eq:Fn]{F_{2}} \right]
	\nonumber \\ &
	- i m_1 \sum_{n=0}^{\infty} \hyperref[eq:Fn]{F_{n}} \hyperref[eq:En]{E_{n}}  
	+ \left( 1 \leftrightarrow 2 \right) \,.
\end{align}
\end{widetext}
The functions $C_{n}$, $D_{n}$, and $E_{n}$ are given in \cref{sec:definitions}, but with the constraints of aligned-spin scattering imposed. 
Contact terms contribute to the even-in-spin part, but are absent from the odd-in-spin one. This is reminiscent of the contact-term contributions to the $\mathfrak{a}_{1}^{\infty} \times \mathfrak{a}_{2}^{0}$ sector of the scattering, where imposing the black-hole spin-structure assumption barred the appearance of contact-term coefficients at odd spin orders for that sector \cite{Aoude:2022thd}.

In that previous analysis of $\mathfrak{a}_{1}^{\infty} \times \mathfrak{a}_{2}^{0}$ scattering at 2PM, the $\omega \rightarrow \infty$ limit was considered as a potential identifier of Kerr black holes.
With the black-hole spin-structure assumption imposed, certain values of the unfixed Wilson coefficients of even-in-spin contact terms improved the behavior of parts of the amplitude containing $(V_{11})^{n}$ for $n\geq 2$ in this kinematic limit.
The simplifications of the aligned-spin configuration allow us to more easily check consistency of those previously-determined coefficient values in other spin sectors.

Analyzing other spin sectors, we find the $\omega\rightarrow\infty$ limit to be an unreliable handle for fixing the values of the contact-term coefficients.
This is primarily because the coefficient values which improve the $\omega\rightarrow\infty$ behavior for part of the amplitude in one sector are different from the values suggested by another sector.
Our conclusion about the utility of this limit aligns with observations in ref.~\cite{Bautista:2023szu}, which saw that the values of Wilson coefficients matching solutions to the Teukolsky equation did not produce the best high-energy behavior of the 2PM amplitude.

%%%%%%% End Results %%%%%%%%%%%%%%%%%%%%%%%%%%%%%%%%%%%
%%%%%%%%%%%%%%%%%%%%%%%%%%%%%%%%%%%%%%%%%%%%%%%%%%%%%%%

%%%%%%%%%%%%%%%%%%%%%%%%%%%%%%%%%%%%%%%%%%%%%%%%%%%%%%%
%%%%%%% Begin Spin Structure %%%%%%%%%%%%%%%%%%%%%%%%%%
%%%%%%%%%%%%%%%%%%%%%%%%%%%%%%%%%%%%%%%%%%%%%%%%%%%%%%%
\section{Breaking the spin structure}

The results so far are constrained in two ways: the three-point amplitude used expresses the Kerr-black-hole spin-multipole moments, and the Compton amplitude exhibits the spin-shift symmetry.
The latter of these has recently been shown to be in tension with the description of Kerr black holes by the Teukolsky equation \cite{Bautista:2022wjf}, so we allow for its relaxation here.
To do so, we insert the general Compton amplitude
\begin{align}
	\cM_4^{h_{1}h_{2},\rm{gen.}} = \cM_4^{h_{1}h_{2}} + m^2 \bigg( \mathcal{C}^{h_{1}h_{2}} + \mathcal{D}^{h_{1}h_{2}}\bigg) \,
\end{align}
into the cuts.
Here, $\cM_{4}^{h(-h)}$ is the shift-symmetric, opposite-helicity Compton amplitude, including shift-symmetric contact terms, used up until this point, and $\cM^{hh}$ is the factorizable portion of the same-helicity Compton amplitude, which does not contribute to scattering at 2PM for Kerr-black-hole spin-induced multipoles.
$\mathcal{C}^{h_{1}h_{2}}$ and $\mathcal{D}^{h_{1}h_{2}}$ are the most-general contact terms that can be relevant to black-hole scattering at $\cO(G^{2})$ which are, respectively, analytic and non-analytic in the spin vector.\footnote{Refs.~\cite{Bautista:2022wjf,Haddad:2023ylx} referred to these as conservative and dissipative contact terms. The latter variety were first considered in ref.~\cite{Bautista:2022wjf}. We move away from these identifiers---as has ref.~\cite{Bautista:2023szu}---as we understand that the precise relation between these contact terms and classical conservative/dissipative effects is unclear. We thank Fabian Bautista and Justin Vines for discussions about this. }
In particular, the latter contain one factor of $|\mathfrak{a}|\equiv\sqrt{\mathfrak{a}^{2}}$.
These contact terms were constructed in ref.~\cite{Haddad:2023ylx}, and we use a slightly modified form of them here, while also imposing crossing symmetry; see \cref{app:ContactTerms}.
The analytic coefficients are labelled by $\{a,b,c,d,e\}$ and the non-analytic ones are labelled by $\{f,g,p,q,r\}$.

When introducing contact terms, the same-helicity Compton amplitude can produce non-vanishing contributions to the 2PM amplitude.
We must thus consider now the contributions from opposite- and same-helicity Compton amplitudes to the unitarity cut in the most general case.
Note, however, that ref.~\cite{Bautista:2022wjf} demonstrated that the same-helicity Compton amplitude matching solutions to the Teukolsky equation up to sixth order in spin did not require any of the contact terms considered here.

\subsection{Opposite-helicity contact term contributions}

For opposite graviton helicities, crossing symmetry is expressed by the following constraints on the Compton amplitude:
\begin{align}\label{eq:CrossingSymmetryConditionsOH}
    \cM_{4}^{+-}=\cM_{4}^{-+}|_{q_{3}\leftrightarrow q_{4}}=\bar{\cM}_{4}^{-+}|_{\mathfrak{a}\rightarrow-\mathfrak{a}}.
\end{align}
The first of these has the more concrete implications for our purposes.
Specifically, this first condition means that the coefficients appearing in both opposite-helicity sectors must be identical, $x^{-+,(s)}_{i,(l,),j}=x^{+-,(s)}_{i,(l,),j}\equiv x^{(s)}_{i,(l,),j}$ for all coefficients in \cref{eq:AnalyticContactOH,eq:NonAnalyticContactOH}.

The second crossing symmetry condition in \cref{eq:CrossingSymmetryConditionsOH} imposes \cref{eq:OHCoeffConjugates} on the contact-term coefficients.
These relations affect the parity properties of the 2PM amplitude depending on whether the coefficients are taken to be real, complex, or imaginary.
To maintain maximal generality, we do not assume anything about the reality of the coefficients, so these relations do not factor into the result in this section.

With crossing symmetry accounted for, the contribution from the opposite-helicity contact terms to the 2PM amplitude is
\begin{widetext}
    \begin{align}\label{eq:OHContact2PM}
        &\cM_{2\rm PM}^{\rm{OH-contact}}=\frac{\pi^{2}G^{2}}{m_{1}m_{2}q^{4}\sqrt{-q^{2}}}\sum_{s_{1}=0}^{\infty}\sum_{s_{2}=5}^{\infty}\sum_{i=0}^{\floor{(s_{2}-5)/2}}\sum_{j=0}^{s_{2}-5-2i}\sum_{\pm}\sum_{t=-4}^{4}q^{2i}\mathfrak{a}^{2i}_{2}(q\cdot\mathfrak{a}_{2})^{s_{2}-5-2i-j}\left\{\frac{[1+(-1)^{j+t+s_{1}}]}{2}\right. \\
        &\times\left(\mathfrak{a}_{2}^{2}\sum_{l=0}^{i}q^{-2l}\hyperref[eq:ContactH]{\mathcal{H}_{s_{1}}^{j,t,2l}}\left(\frac{1}{4}\hyperref[eq:Qij]{Q_{11}},\hyperref[eq:uvOuterProducts]{z},\hyperref[eq:uvOuterProducts]{x_{1,\pm}},\hyperref[eq:uvOuterProducts]{x_{2,\pm}}\right)\right.\left[\frac{\mathfrak{a}_{2}^{2}}{m_{2}}a^{(s_{2}-1)}_{i,l,j}\left(\hyperref[eq:HelicityVectorFF]{A_{t}^{(p_{2},4)}}+(-1)^{j+s_{1}}\hyperref[eq:HelicityVectorFF]{\bar{A}_{t}^{(p_{2},4)}}\right)+|\mathfrak{a}_{2}|g^{(s_{2}-1)}_{i,l,j}\left(\hyperref[eq:HelicityVectorFF]{A_{t}^{(p_{2},3)}}+(-1)^{j+s_{1}}\hyperref[eq:HelicityVectorFF]{\bar{A}_{t}^{(p_{2},3)}}\right)\right]\notag \\
        &+\left(\frac{2}{m_{2}}\right)^{-j}\hyperref[eq:ContactH]{\mathcal{H}_{s_{1}}^{0,t,j}}\left(\frac{1}{4}\hyperref[eq:Qij]{Q_{11}},\hyperref[eq:uvOuterProducts]{y_{22}},\hyperref[eq:uvOuterProducts]{y_{12,\pm}},1\right)\left[m_{2}\mathfrak{a}_{2}^{2}c^{(s_{2}-1)}_{i,j}\left(\hyperref[eq:HelicityVectorFF]{A_{t,\pm}^{(\mathfrak{a}_{2},2)}}+(-1)^{j+s_{1}}\hyperref[eq:HelicityVectorFF]{\bar{A}_{t,\pm}^{(\mathfrak{a}_{2},2)}}\right)\right.\notag \\
        &\left.\left.+m_{2}^{2}|\mathfrak{a}_{2}|q^{(s_{2}-1)}_{i,j}\left(\hyperref[eq:HelicityVectorFF]{A_{t,\pm}^{(\mathfrak{a}_{2},1)}}+(-1)^{j+s_{1}}\hyperref[eq:HelicityVectorFF]{\bar{A}_{t,\pm}^{(\mathfrak{a}_{2},1)}}\right)+m_{2}^{3}e^{(s_{2}-1)}_{i,j}\left(\hyperref[eq:HelicityVectorFF]{A_{t,\pm}^{(\mathfrak{a}_{2},0)}}+(-1)^{j+s_{1}}\hyperref[eq:HelicityVectorFF]{\bar{A}_{t,\pm}^{(\mathfrak{a}_{2},0)}}\right)\right]\right)-\frac{[1-(-1)^{j+t+s_{1}}]}{2}\notag \\
        &\times\left(\mathfrak{a}_{2}^{4}\sum_{l=0}^{i}q^{-2l}\hyperref[eq:ContactH]{\mathcal{H}_{s_{1}}^{j,t,2l+1}}\left(\frac{1}{4}\hyperref[eq:Qij]{Q_{11}},\hyperref[eq:uvOuterProducts]{z},\hyperref[eq:uvOuterProducts]{x_{1,\pm}},\hyperref[eq:uvOuterProducts]{x_{2,\pm}}\right)\right.\left[b^{(s_{2})}_{i,l,j}\left(\hyperref[eq:HelicityVectorFF]{A_{t}^{(p_{2},3)}}-(-1)^{j+s_{1}}\hyperref[eq:HelicityVectorFF]{\bar{A}_{t}^{(p_{2},3)}}\right)+\frac{|\mathfrak{a}_{2}|}{m_{2}}f^{(s_{2})}_{i,l,j}\left(\hyperref[eq:HelicityVectorFF]{A_{t}^{(p_{2},4)}}-(-1)^{j+s_{1}}\hyperref[eq:HelicityVectorFF]{\bar{A}_{t}^{(p_{2},4)}}\right)\right]\notag \\
        &+\hyperref[eq:ContactH]{\mathcal{H}_{s_{1}}^{j,t,1}}\left(\frac{1}{4}\hyperref[eq:Qij]{Q_{11}},\hyperref[eq:uvOuterProducts]{z},\hyperref[eq:uvOuterProducts]{x_{1,\pm}},\hyperref[eq:uvOuterProducts]{x_{2,\pm}}\right)\left[m_{2}^{2}\mathfrak{a}_{2}^{2}d^{(s_{2})}_{i,j}\left(\hyperref[eq:HelicityVectorFF]{A_{t}^{(p_{2},1)}}-(-1)^{j+s_{1}}\hyperref[eq:HelicityVectorFF]{\bar{A}_{t}^{(p_{2},1)}}\right)\right.\notag \\
        &\left.\left.\left.+m_{2}\mathfrak{a}_{2}^{2}|\mathfrak{a}_{2}|p^{(s_{2})}_{i,j}\left(\hyperref[eq:HelicityVectorFF]{A_{t}^{(p_{2},2)}}-(-1)^{j+s_{1}}\hyperref[eq:HelicityVectorFF]{\bar{A}_{t}^{(p_{2},2)}}\right)+m_{2}^{3}|\mathfrak{a}_{2}|r^{(s_{2})}_{i,j}\left(\hyperref[eq:HelicityVectorFF]{A_{t}^{(p_{2},0)}}-(-1)^{j+s_{1}}\hyperref[eq:HelicityVectorFF]{\bar{A}_{t}^{(p_{2},0)}}\right)\right]\right)\right\}+(1\leftrightarrow2).\notag
    \end{align} 
\end{widetext}
We remind the reader that nearly all symbols are hyperlinked to their definitions in the appendices.
To facilitate a search for structure, we have expanded and collected the contributions up to thirteenth order in the spins of both particles in the files \texttt{OH2PMAnalyticContactTerms.mx} and \texttt{OH2PMNonAnalyticContactTerms.mx}.
The index $s_{i}$ labels the order in the spin-multipole expansion of particle $i$.
All spin dependence of the particle whose propagator is taken on shell is encapsulated in the functions $\mathcal{H}_{n}^{j,t,k}(a,b,c,d)$.

\subsection{Same-helicity contact term contributions}

We move now to the contributions from same-helicity contact terms.
Crossing symmetry again imposes two relations on the same-helicity Compton amplitudes:
\begin{align}
    \cM_{4}^{--}=\bar{\cM}_{4}^{++}|_{\mathfrak{a}\leftrightarrow-\mathfrak{a}},\quad \cM^{++}_{4}=\cM^{++}_{4}|_{q_{3}\leftrightarrow q_{4}}.
\end{align}
The second of these implies that many of the free coefficients must vanish, a condition which is summarized in \cref{eq:SHCoeffVanishing}.
From the first constraint above, we obtain relations between the coefficients of different helicity sectors.
Specifically, $x^{--,(s)}_{i,(l,)j}=(-1)^{s}\bar{x}^{++,(s)}_{i,(l,)j}$ for the analytic-in-spin coefficients $\{a,b,c,d,e\}$ and $x^{--,(s)}_{i,(l,)j}=(-1)^{s+1}\bar{x}^{++,(s)}_{i,(l,)j}$ for the non-analytic coefficients $\{f,g,p,q,r\}$.
Once again, the reality of the coefficients affects the parity properties of the amplitude.
We assume nothing on this front to not sacrifice any generality.

The contribution to the 2PM amplitude from the same-helicity analytic contact terms is
\begin{widetext}
\begin{align}\label{eq:SHAContact2PM}
    &\cM_{\rm2PM}^{\rm SH-contact, A}=\frac{\pi^{2}G^{2}m_{2}^{2}}{m_{1}\sqrt{-q^{2}}}\sum_{s_{2}=4}^{\infty}\sum_{i=0}^{\floor{(s_{2}-4)/2}}\sum_{j=0}^{s_{2}-4-2i}\sum_{\pm}\frac{1+(-1)^{j}}{2}\mathfrak{a}_{2}^{2i}(q\cdot\mathfrak{a}_{2})^{s_{2}-4-2i-j} \\
    &\times \left\{\sum_{l=0}^{i}q^{2l}\hyperref[eq:ContactH]{\mathcal{H}_{0}^{j,0,2(i-l)}}(0,\hyperref[eq:uvOuterProducts]{z},1,\hyperref[eq:uvOuterProducts]{x_{2,\pm}})m_{1}^{4}q^{4}\mathfrak{a}_{2}^{4}\left(a^{--,(s_{2})}_{i,l,j}e^{-q\cdot\mathfrak{a}_{1}}+(-1)^{s_{2}}\bar{a}^{--,(s_{2})}_{i,l,j}e^{q\cdot\mathfrak{a}_{1}}\right)\right.\notag \\
    &+\sum_{t=-2}^{2}\hyperref[eq:ContactH]{\mathcal{H}_{0}^{j,2t,2i}}(0,\hyperref[eq:uvOuterProducts]{z},1,\hyperref[eq:uvOuterProducts]{x_{2,\pm}})\left[m_{1}^{2}\mathfrak{a}_{2}^{2}\left(c^{--,(s_{2})}_{i,j}e^{-q\cdot\mathfrak{a}_{1}}\hyperref[eq:HelicityVectorFF]{B^{(p_{2},2)}_{2t}}+(-1)^{s_{2}}\bar{c}^{--,(s_{2})}_{i,j}e^{q\cdot\mathfrak{a}_{1}}\hyperref[eq:HelicityVectorFF]{\bar{B}^{(p_{2},2)}_{2t}}\right)\right.\notag \\
    &\left.\left.+\frac{1}{q^{4}}\left(e^{--,(s_{2})}_{i,j}e^{-q\cdot\mathfrak{a}_{1}}\hyperref[eq:HelicityVectorFF]{B^{(p_{2},0)}_{2t}}+(-1)^{s_{2}}\bar{e}^{--,(s_{2})}_{i,j}e^{q\cdot\mathfrak{a}_{1}}\hyperref[eq:HelicityVectorFF]{\bar{B}^{(p_{2},0)}_{2t}}\right)\right]\right\}+(1\leftrightarrow2).\notag
\end{align}
The non-analytic same-helicity contact terms enter the 2PM amplitude through
\begin{align}\label{eq:SHNAContact2PM}
    &\cM_{\rm2PM}^{\rm SH-contact, NA}=-\frac{\pi^{2}G^{2}m_{2}^{2}}{m_{1}\sqrt{-q^{2}}}|\mathfrak{a}_{2}|\sum_{s_{2}=5}^{\infty}\sum_{i=0}^{\floor{(s_{2}-5)/2}}\sum_{j=0}^{s_{2}-5-2i}\sum_{\pm}\frac{1-(-1)^{j}}{2}\mathfrak{a}_{2}^{2i}(q\cdot\mathfrak{a}_{2})^{s_{2}-5-2i-j} \\
    &\times\left\{\sum_{l=0}^{i}q^{2l}\hyperref[eq:ContactH]{\mathcal{H}_{0}^{j,0,2(i-l)+1}}(0,\hyperref[eq:uvOuterProducts]{z},1,\hyperref[eq:uvOuterProducts]{x_{2,\pm}})m_{1}^{4}q^{4}\mathfrak{a}_{2}^{4}\left(f_{i,l,j}^{--,(s_{2})}e^{-q\cdot\mathfrak{a}_{1}}-(-1)^{s_{2}}\bar{f}_{i,l,j}^{--,(s_{2})}e^{q\cdot\mathfrak{a}_{1}}\right)\right.\notag \\
    &+\sum_{t=-2}^{2}\hyperref[eq:ContactH]{\mathcal{H}_{0}^{j,2t,2i+1}}(0,\hyperref[eq:uvOuterProducts]{z},1,\hyperref[eq:uvOuterProducts]{x_{2,\pm}})\left[m_{1}^{2}\mathfrak{a}_{2}^{2}\left(p_{i,j}^{--,(s_{2})}e^{-q\cdot\mathfrak{a}_{1}}\hyperref[eq:HelicityVectorFF]{B^{(p_{2},2)}_{2t}}-(-1)^{s_{2}}\bar{p}_{i,j}^{--,(s_{2})}e^{q\cdot\mathfrak{a}_{1}}\hyperref[eq:HelicityVectorFF]{\bar{B}^{(p_{2},2)}_{2t}}\right)\right.\notag \\
    &\left.\left.+\frac{1}{q^{4}}\left(r_{i,j}^{--,(s_{2})}e^{-q\cdot\mathfrak{a}_{1}}\hyperref[eq:HelicityVectorFF]{B^{(p_{2},0)}_{2t}}-(-1)^{s_{2}}\bar{r}_{i,j}^{--,(s_{2})}e^{q\cdot\mathfrak{a}_{1}}\hyperref[eq:HelicityVectorFF]{\bar{B}^{(p_{2},0)}_{2t}}\right)\right]\right\}+(1\leftrightarrow2).\notag
\end{align}
\end{widetext}
As usual, we symmetrize by swapping the masses, spins, and velocities of the two heavy particles, along with changing the sign of the transferred momentum.
The factors of $(1\pm(-1)^{j})/2$ manifest crossing symmetry constraints by removing terms which anyway vanish by \cref{eq:SHCoeffVanishing}.
The spin of the massive particle whose propagator is cut only contributes via exponentials.
This makes it easy to expand these contributions to finite order in the spin of the contact terms, while retaining all spin orders for the other particle.
We have collected the results expanded up to spins $\mathfrak{a}_{1}^\infty \times \mathfrak{a}_{2}^{i\leq28}$ in the files \texttt{SH2PMAnalyticContactTerms.mx} and \texttt{SH2PMNonAnalyticContactTerms.mx}.

\Cref{eq:OHContact2PM,eq:SHAContact2PM,eq:SHNAContact2PM} account for all contact contributions to the 2PM amplitude that are potentially relevant to Kerr-black-hole scattering.
Identifying the appropriate amplitude describing the scattering of two Kerr black holes thus amounts to determining the appropriate values of the free parameters in these equations.

%%%%%%%%%%%%%%%%%%%%%%%%%%%%%%%%%%%%%%%%%%%%%%%%%%%%%%%
%%%%%%% End Spin Structure %%%%%%%%%%%%%%%%%%%%%%%%%%%%
%%%%%%%%%%%%%%%%%%%%%%%%%%%%%%%%%%%%%%%%%%%%%%%%%%%%%%%

%%%%%%%%%%%%%%%%%%%%%%%%%%%%%%%%%%%%%%%%%%%%%%%%%%%%%%%
%%%%%%% Begin Comparison %%%%%%%%%%%%%%%%%%%%%%%%%%%%%%
%%%%%%%%%%%%%%%%%%%%%%%%%%%%%%%%%%%%%%%%%%%%%%%%%%%%%%%

\section{Comparison to literature}\label{sec:Comparisons}

As we have accounted for all contact terms potentially relevant to Kerr-black-hole scattering at 2PM, we expect our result to contain all results present in the literature for Kerr-black-hole scattering at this order.
Indeed, up to cubic order in spin, we agree with refs.~\cite{Damgaard:2019lfh,Bern:2020buy,Kosmopoulos:2021zoq,Chen:2021kxt}.

At quartic, quintic, and sextic orders in spin, the most general 2PM amplitude we can compare with was presented in ref.~\cite{Bautista:2023szu}.
As mentioned above, the same-helicity contact terms matching the Teukolsky equation were shown to all be vanishing up to sixth order in spin in ref.~\cite{Bautista:2022wjf}, so matching to ref.~\cite{Bautista:2023szu} requires that we omit contributions from \cref{eq:SHAContact2PM,eq:SHNAContact2PM}.
Beginning with quartic order in spin, we agree with the amplitude there under the mapping of coefficients
\begin{align}
    a^{4}_{0,0,0}\rightarrow-\frac{1}{16}c_{1,2},&\quad c^{4}_{0,0}\rightarrow\frac{1}{4}(c_{1,1}-2c_{1,2}),\notag \\
    d^{(4)}_{0}+e^{4}_{0,0}\rightarrow-&24(c_{1,0}-c_{1,1}+c_{1,2}), \\
    g^{4}_{0,0,0}\rightarrow0,&\quad q^{4}_{0,0}\rightarrow0.\notag
\end{align}
The parameters in this paper are on the left-hand sides, while those of ref.~\cite{Bautista:2023szu} are on the right-hand sides.
Note that the parameters $d_{0}^{(4)}$ and $e^{4}_{0,0}$ are redundant; we have split them in this manner such that the shift-symmetric result is obtained by setting all coefficients introduced in \cref{app:ContactTerms} to 0. This redundancy can easily be avoided by setting to zero the contact terms in \cref{sec:definitions} when considering the most general case. 

At quintic order in spin, the additional parameter mappings needed to match to ref.~\cite{Bautista:2023szu} are
\begin{align}
    a^{5}_{0,0,0}\rightarrow0,\quad c^{5}_{0,0}&\rightarrow0,\quad e^{5}_{0,0}\rightarrow0,\notag \\
    a^{5}_{0,0,1}+\frac{1}{4}b^{5}_{0,0,0}&\rightarrow\frac{c_{2,2}}{16}+\frac{1}{128}(c_{3,0}-c_{3,1})-\frac{1}{480},\notag \\
    c^{5}_{0,1}+\frac{3}{2}d^{5}_{0,0}\rightarrow-\frac{1}{80}&(20c_{2,1}-40c_{2,2}-30c_{3,0}+15c_{3,1}+8),\notag\\
    g^{5}_{0,0,0}\rightarrow0,&\quad q^{5}_{0,0}\rightarrow0, \\
    r^{5}_{0,0}+\frac{1}{4}q^{5}_{0,1}&\rightarrow-\frac{1}{4}(c_{4,0}-c_{4,1}+c_{4,2}),\notag \\
    \frac{3}{2}g^{5}_{0,0,1}+p^{5}_{0,0}&\rightarrow\frac{1}{16}(c_{4,1}-2c_{4,2}).\notag
\end{align}
We also agree with ref.~\cite{Bautista:2023szu} at sixth order in spin under an appropriate coefficient map, which we omit here for readability.

Above sixth order in spin, we have reproduced our shift-symmetric results from refs.~\cite{Aoude:2022trd,Aoude:2022thd}.

Additionally to these finite-spin results, an $\cO(G^{2})$ amplitude involving all spin orders of both scattering objects in the aligned-spin configuration was recently presented in ref.~\cite{Alessio:2023kgf},
%the validity of which we dispute
with which we do not agree.
The analysis there is based on an interesting attempt to manifest the Newman-Janis shift in a QED Lagrangian, by means of defining a new gauge-covariant derivative which incorporates the classical spin vector in its action on a charged scalar field.
The three-point and Compton amplitudes are then evaluated from the scalar-QED Lagrangian with this gauge-covariant derivative, and subsequently double copied to obtain gravitational amplitudes.

Both the QED and gravitational opposite-helicity Compton amplitudes in ref.~\cite{Alessio:2023kgf} are missing contributions involving $w\cdot\mathfrak{a}$, which is in disagreement with other analyses \cite{Arkani-Hamed:2017jhn,Aoude:2020onz,Bautista:2021wfy,Aoude:2022trd,Saketh:2022wap,Chiodaroli:2021eug,Cangemi:2022bew,Bautista:2022wjf}.
It is implied in ref.~\cite{Alessio:2023kgf} that this discrepancy can be remedied by adding contact terms through the introduction of higher-point operators to the action.
However, terms with $w\cdot\mathfrak{a}$ have non-vanishing residues on the physical factorization channels for QED, QCD, and gravity, even in the classical limit \cite{Aoude:2022trd}.
As such, their contributions to the amplitude cannot be encoded in contact terms, and their absence means that the four-point amplitude of ref.~\cite{Alessio:2023kgf} does not have the correct factorization properties for a Compton amplitude.
This propagates into the $\cO(G^{2})$ amplitude, which cannot be obtained from our results above for any choices of the free parameters.

%%%%%%%%%%%%%%%%%%%%%%%%%%%%%%%%%%%%%%%%%%%%%%%%%%%%%%%
%%%%%%% End Comparison %%%%%%%%%%%%%%%%%%%%%%%%%%%%%%%%
%%%%%%%%%%%%%%%%%%%%%%%%%%%%%%%%%%%%%%%%%%%%%%%%%%%%%%%

%%%%%%%%%%%%%%%%%%%%%%%%%%%%%%%%%%%%%%%%%%%%%%%%%%%%%%%
%%%%%%% Begin Conclusion %%%%%%%%%%%%%%%%%%%%%%%%%%%%%%
%%%%%%%%%%%%%%%%%%%%%%%%%%%%%%%%%%%%%%%%%%%%%%%%%%%%%%%

\section{Conclusion}\label{sec:conclusion}

With the aim of providing the data needed to pinpoint the amplitude describing the scattering of Kerr black holes at $\cO(G^{2})$, we have computed the most general 2PM amplitude consistent with known properties of Kerr-black-hole scattering.
Specifically, the totality of our results contain only two externally-imposed principles.
The first is that the scattering is constructed from the Kerr-black-hole linear-in-curvature spin-induced multipoles \cite{Levi:2015msa,Vines:2017hyw}.
Said otherwise, the factorization properties of the amplitudes used here are all consistent with the known Kerr-black-hole three-point amplitudes \cite{Arkani-Hamed:2017jhn,Guevara:2018wpp,Chung:2018kqs}.
The second is that the general set of contact terms considered have coefficients which can be made dimensionless through rescalings by factors of the mass only.
As was argued in ref.~\cite{Haddad:2023ylx}, any other contact terms would only be relevant to black-hole scattering at higher PM orders.
Our computation made use of the synergy between HPET and on-shell methods, which allows for the convenient extraction of classical physics \cite{Aoude:2020onz}.
On top of this, the one-loop classical amplitude was derived using unitarity cuts.

What remains unknown is which values of the unfixed coefficients describe Kerr-black-hole scattering at this PM order.
Prescriptions for those coefficients appearing up to sixth order in spin were put forth recently in ref.~\cite{Bautista:2022wjf} by comparing to solutions to the Teukolsky equation.
However, an amplitudes-based understanding of Kerr black holes remains desirable, as it may present a handle for understanding the structure of all spin-multipole orders instead of uncovering one spin order at a time.
Moreover, an ideal scenario would see such an image of black-hole amplitudes be extendable to higher PM orders.

At leading order in Newton's constant, the amplitudes describing Kerr black holes stand out due to their simplicity compared to those relevant for more general objects \cite{Arkani-Hamed:2017jhn,Haddad:2020que,Aoude:2020ygw,Aoude:2021oqj,Haddad:2023ylx}.
Extending this to higher orders, while it is not clear what principles one should adhere to to identify the appropriate Kerr amplitudes, one might expect special structures to appear when the Kerr values for the Wilson coefficients are imposed.
Going the other way, the result in this paper allows for a search for the coefficient values which endow the amplitude with special properties.
A potential starting point in this search for structure is to map the Kerr-black-hole conjectures of ref.~\cite{Levi:2022rrq} onto the unfixed coefficients here and subsequently study the implications for the amplitude.

The all-spin amplitude at finite spin orders can be analyzed with the ancillary \texttt{2PM\_All\_Spins.nb} notebook.

%%%%%%%%%%%%%%%%%%%%%%%%%%%%%%%%%%%%%%%%%%%%%%%%%%%%%%%
%%%%%%% End Conclusion %%%%%%%%%%%%%%%%%%%%%%%%%%%%%%%%
%%%%%%%%%%%%%%%%%%%%%%%%%%%%%%%%%%%%%%%%%%%%%%%%%%%%%%%

\subsection*{Acknowledgments}

We thank Francesco Alessio, Fabian Bautista, Clifford Cheung, Julio Parra-Martinez, and Justin Vines for helpful discussions.
We are also grateful to Fabian Bautista for sharing a preliminary version of ref.~\cite{Bautista:2023szu}. 
R.A. is supported by the F.R.S.-FNRS project no. 40005600 and the FSR Program of UCLouvain.
K. H. is supported by the Knut and Alice Wallenberg Foundation under grants KAW 2018.0116 (From Scattering Amplitudes to Gravitational Waves) and KAW 2018.0162.
A. H. is supported by the DOE under award number DE-SC0011632 and by the Walter Burke Institute for Theoretical Physics.

\appendix

%%%%%%%%%%%%%%%%%%%%%%%%%%%%%%%%%%%%%%%%%%%%%%%%%%%%%%%
%%%%%%% Begin Definitions %%%%%%%%%%%%%%%%%%%%%%%%%%%%%
%%%%%%%%%%%%%%%%%%%%%%%%%%%%%%%%%%%%%%%%%%%%%%%%%%%%%%%

\section{Definitions}\label{sec:definitions}

Here we define all notation used in the all-order-in-spin result above.
To begin, we summarize the kinematic variables out of which all other notation is composed.

\subsection{Four-vectors}

We consider the scattering of two objects with initial momenta $p_{i}^{\mu}=m_{i}v_{i}^{\mu}$ and classical spin vectors $\mathfrak{a}_{i}^{\mu}$ for $i=1,2$.
The $v_{i}^{\mu}$ represent the four-velocities of the objects, and satisfy $v_{i}^{2}=1$.
After the scattering, a four-momentum $q^{\mu}$ is transferred from object 2 to object 1.
Some portions of the result depend on sums of the spins or velocities:
\begin{align}
	\mathfrak{a}^{\mu} \equiv \mathfrak{a}_{1}^{\mu} + \mathfrak{a}_{2}^{\mu} ,\quad v^{\mu}\equiv v_{1}^{\mu}+v_{2}^{\mu}.
\end{align}
The former should not be confused with the spin in eqs.~\eqref{eq:Compton1} and \eqref{eq:Compton2}, which describes the Compton amplitude for one spinning particle.

\subsection{Lorentz invariants}

The four-vectors above enter the results through the following Lorentz invariants:
\begin{align}
    \omega &\equiv v_{1}\cdot v_{2} ,\label{eq:omega} \\
	Q_{ij} &\equiv (q\cdot \mathfrak{a}_{i}) (q \cdot \mathfrak{a}_{j}) - q^2 (\mathfrak{a}_{i} \cdot \mathfrak{a}_{j})\label{eq:Qij} , \\
	V_{ij} &\equiv q^2 (v \cdot \mathfrak{a}_{i}) (v \cdot \mathfrak{a}_{j})\label{eq:Vij} , \\
	\cE_{i} &\equiv \epsilon_{\mu\nu\rho\sigma} q^{\mu} v_{1}^{\nu} v_{2}^{\rho} \mathfrak{a}_{i}^{\sigma} ,\label{eq:calEi} \\
	\cE &\equiv \cE_{1} + \cE_{2} ,\label{eq:calE}
\end{align}
The spin-supplementary conditions (SSCs) $v_{i}\cdot\mathfrak{a}_{i}=0$ help to slightly reduce the number of invariants.
Common combinations of these are
\begin{align}
	\Qtot &= Q_{11} + Q_{22} + 2 Q_{12} + V_{22} ,\label{eq:Qtot} \\
	\Vtot &= V_{11} - 2 \omega V_{12} + \omega^2 V_{22}\label{eq:Vtot} .
\end{align}
Note that $\cE^2 = \Vtot - (\omega^2 - 1) \Qtot$ and $\cE^2_{i} = V_{ii} - (\omega^2 - 1) Q_{ii}$.

\subsection{Hypergeometric functions}

Much of the spin dependence of the amplitude is captured by the hypergeometric function
\begin{align}
	F_{j} &= \frac{1}{\Gamma[j+1]} \;_{0}F_{1}(j+1; \Qtot/4)\label{eq:Fn} \,.
\end{align}
We also define
\begin{widetext}
\begin{align}\label{eq:Bnj}
	B_{n,j}[b] = \sum_{\underset{\rm even/odd}{m=n-j}}^{\infty} \theta(m) b_{m} \left[ \binom{m}{\tfrac{m+n-j}{2}} + \binom{m}{\tfrac{m-n-j}{2}} \right] 2^{n+m} \left( \frac{\hyperref[eq:Qij]{Q_{22}} + \hyperref[eq:Vij]{V_{22}}}{16} \right)^{\tfrac{m-n+j}{2}} \,,
\end{align}
where $\theta(m)$ is the Heaviside function and the sum over $m$ is for even/odd integers when $n-j$ is even/odd.
Also,
\begin{align}
	H_{m}^{(k)} = 
	\sum_{j=0}^{\infty} \frac{(-1)^{m}(2j+1-m)^{1-k}}{(m+2j+1+k)(m+2j+2+k)} \frac{(\hyperref[eq:Qij]{Q_{22}})^{j}}{\Gamma[m+1]\Gamma[2j+2]} \,,
\end{align}
which can be expressed in terms of hypergeometric functions.
We also use the short-hand notation 
\begin{align}
	\label{eq:hn}
	h^{(1)}_{n} =& 2^{-2n-1} \hyperref[eq:Bnj]{B_{2n+1,0}[H^{(0)}]} \, , \quad
	h^{(2)}_{n} = 2^{-2n-1} \hyperref[eq:Bnj]{B_{2n,0}[H^{(0)}]} \, , \nonumber \\
	h^{(3)}_{n} =& -2^{-2n-1} \hyperref[eq:Bnj]{B_{2n,0}[H^{(1)}]} \, , \quad
	h^{(4)}_{n} = 2^{-2n-1} \hyperref[eq:Bnj]{B_{2n+1,0}[H^{(1)}]} \,.
\end{align}

\subsection{Shift-symmetric contact terms}

The shift-symmetric contact terms enter the even-in-spin part of the $\cO(G^{2})$ amplitude through
\begin{align}
	\label{eq:CnContactTerm}
	C_{n} &= (-1)^{n} \left[
		- \frac{ 3 \hyperref[eq:Qij]{Q_{22}^2} + 30 \hyperref[eq:Qij]{Q_{22}} \hyperref[eq:Vij]{V_{22}} + 35 \hyperref[eq:Vij]{V_{22}^{2}}}{64}\hyperref[eq:Bnj]{B_{n,0}[\bar c]}
		+ (\hyperref[eq:Qij]{Q_{22}} + 7 \hyperref[eq:Vij]{V_{22}}) \hyperref[eq:Bnj]{B_{n,2}[\bar c]} 
		- 4 \hyperref[eq:Bnj]{B_{n,4}[\bar c]}
	\right] \left( 1 - \frac{\delta_{n0}}{2} \right)  \,, \\
		%\bar c_{n} &= \sum_{k=0}^{\infty} \sum_{j=k}^{\floor{n/2}+k} \frac{d^{(n + 2k + 4)}_{j}}{(n + 2k + 4)!} \binom{j}{k} (-\hyperref[eq:Qij]{Q_{22}})^{k} \,, \\
		\bar c_{n} &= \frac{2}{3} \delta_{n0} + \sum_{k=0}^{\infty} \sum_{j=k}^{\floor{n/2}+k} \frac{d^{(n + 2k + 4)}_{j}}{(n + 2k + 4)!} \binom{j}{k} (-\hyperref[eq:Qij]{Q_{22}})^{k} \,.
\end{align}
%
%where we shift $d^{(4)}_{0} \rightarrow d^{(4)}_{0} +  16$.

For odd spins, the shift-symmetric contact terms are grouped as
\begin{align}
	\tilde C_{2n+j} =& (-1)^{j+1} \left(  \frac{(v_{1} \cdot \mathfrak{a}_{2}) (\hyperref[eq:Qij]{Q_{22}} + 3 \hyperref[eq:Vij]{V_{22}})}{m_{1}}  \right) \left[  \hyperref[eq:G6indices]{G(\mathfrak{a}_{2},\mathfrak{a},\mathfrak{a}_{2}, \epsilon(qp_1\mathfrak{a}_{2}\cdot),2n+j,1)} Z^{(1-j,-j)}_{n} 
	\right. \nonumber \\ & \left.\qquad\qquad + (1 - \delta_{2n+j,0} ) \hyperref[eq:G6indices]{G(\mathfrak{a},\mathfrak{a}_{2},\mathfrak{a}, \epsilon(qp_1\mathfrak{a}_{2}\cdot),2n+j-1,1)} Z^{(1-j,1-j)}_{n} \right]
	\nonumber \\ &
	+ (-1)^{j+1} \left( -  \frac{8 (v_{1} \cdot \mathfrak{a}_{2}) }{m_{1}}  \right) \left[  \hyperref[eq:G6indices]{G(\mathfrak{a}_{2},\mathfrak{a}, \epsilon(qp_1\mathfrak{a}_{2}\cdot),\mathfrak{a}_{2}, 2n+j,1)} \hyperref[eq:At1t2]{A(\mathfrak{a}_{2},\mathfrak{a}_{2})} Z^{(1-j,-j)}_{n} 
	\right. \nonumber \\ & \left. \qquad\qquad+ (1 - \delta_{2n+j,0} ) 
	\hyperref[eq:G6indices]{G(\mathfrak{a},\mathfrak{a}_{2},\epsilon(qp_1\mathfrak{a}_{2}\cdot),\mathfrak{a}_{2},2n+j,1)} Z^{(1-j,1-j)}_{n} \right]
	\nonumber \\ &
	+ (-1)^{j+1} \left( -  \frac{8 (v_{1} \cdot \mathfrak{a}_{2}) }{m_{1}}  \right) \left[  \hyperref[eq:G6indices]{G(\mathfrak{a}_{2},\mathfrak{a},\mathfrak{a}_{2}, \epsilon(qp_1\mathfrak{a}_{2}\cdot), 2n+j,1)}
	 \hyperref[eq:At1t2]{A(\mathfrak{a}_{2},\mathfrak{a}_{2})^{2}} Z^{(1-j,1-j)}_{n} 
	\right. \nonumber \\ & \left. \qquad\qquad+ (1 - \delta_{2n+j,0} ) 
	\hyperref[eq:G6indices]{G(\mathfrak{a},\mathfrak{a}_{2}, \mathfrak{a},\epsilon(qp_1\mathfrak{a}_{2}\cdot),2n+j-1,1)} Z^{(1-j,2-j)}_{n} \right] \label{eq:C2nTilde}
\end{align}
where
\begin{align}
	Z_{n}^{(k,j)} =& \sum_{m=0}^{\infty} \binom{2m+k}{m-n+l} 2^{2m+2n+1} \left( \frac{\hyperref[eq:Qij]{Q_{22}} + \hyperref[eq:Vij]{V_{22}}}{16} \right)^{m-n+l} \bar c_{2m+k} \,.
\end{align}
These functions appear in the all-order-in-spin result below.

\subsection{Other functions}

The all-order-in-spin result is expressed through the following functions. 
We need
\begin{align}\label{eq:At1t2}
	A(t_1,t_2) =& \frac{1}{16} \left[q^2 (v_1\cdot t_1) (v_1 \cdot t_2) + (q \cdot t_1) (q \cdot t_2) - q^2 (t_1 \cdot t_2) \right], \\
	G_{n,l}(t_1,t_2) =& \sum_{k=0}^{\floor{n/2}} \binom{n}{2k+l} \left[A(t_1,t_2) \right]^{n-2k-l} \left[ \frac{q^2}{256} \epsilon(qv_1t_1t_2)^2 \right]^{k} \,, \\
	G(t_1,t_2,n) =& G_{n,0}(t_1,t_2) \,, \label{eq:G3indices}\\
	G(t_1,t_2,t_3,t_4,n_1,n_2) =& G_{n_1,0}(t_1,t_2) G_{n_2,0}(t_3,t_4) + G_{n_1,1}(t_1,t_2)G_{n_2,1}(t_3,t_4) \left(\frac{q^2}{256} \epsilon(qv_1t_1t_2)\epsilon(qv_1t_3t_4)\right)  \,, \label{eq:G6indices} \\
	G(t_1,t_2,t_3,t_4,t_5,t_6,n_1,n_2,n_3) =& 
	G(t_1,t_2,n_1)G(t_3,t_4,t_5,t_6,n_2,n_3) + 
	G(t_3,t_4,n_2)G(t_1,t_2,t_5,t_6,n_1,n_3) 
	\nonumber \\ &+
	G(t_4,t_5,n_3)G(t_1,t_2,t_3,t_4,n_1,n_2) 
	- 2 G(t_1,t_2,n_1) G(t_3,t_4,n_2) G(t_5,t_6,n_3) \,,\label{eq:G9indices}
\end{align}
and
\begin{align}\label{eq:Dn}
	D_{n} =& \frac{1}{8} \left( (5\hyperref[eq:omega]{\omega}^2 - 1) \hyperref[eq:Qij]{Q_{22}} + 5 (7\hyperref[eq:omega]{\omega}^2 - 1) \hyperref[eq:Vij]{V_{22}} \right) \hyperref[eq:kn]{k_{n,0,0}^{(0)}} \left(1 - \frac{\delta_{n0}}{2} \right)
	- 4 (3\hyperref[eq:omega]{\omega}^2 - 1) \hyperref[eq:kn]{k_{n,2,0}^{(0)}}
	- 32 \hyperref[eq:omega]{\omega} \frac{(v_1\cdot \mathfrak{a}_{2})}{m_2} \hyperref[eq:kn]{k_{n,1,1}^{(0)}}
	\nonumber \\ &
	- \frac{4 (\hyperref[eq:Qij]{Q_{22}} + 3 \hyperref[eq:Vij]{V_{22}})}{q^2 m^2_2} \hyperref[eq:kn]{k_{n,0,2}^{(0)}}
	+ \frac{32}{q^2 m^2_2} \hyperref[eq:kn]{k_{n,2,2}^{(0)}}
	%\nonumber \\ &
	- \frac{1}{2} (\hyperref[eq:Qij]{Q_{22}} + 13 \hyperref[eq:Vij]{V_{22}} ) (\hyperref[eq:omega]{\omega}^2 - 1) \hyperref[eq:kn]{k^{(1)}_{n,1,0}}  - 128 \hyperref[eq:omega]{\omega} \frac{(v_1\cdot \mathfrak{a}_{2})}{m_2} \hyperref[eq:kn]{k^{\prime(1)}_{n,0,1}}  + \frac{128}{q^2 m_2^2} \hyperref[eq:kn]{k^{\prime(1)}_{n,0,2}} 
	\nonumber \\ &
	+ 8 (\hyperref[eq:Qij]{Q_{22}} + 3 \hyperref[eq:Vij]{V_{22}})\hyperref[eq:omega]{\omega} \frac{(v_2\cdot \mathfrak{a}_{2})}{m_2} \hyperref[eq:kn]{k^{(1)}_{n,0,1}}  - \frac{8(\hyperref[eq:Qij]{Q_{22}} + 13 \hyperref[eq:Vij]{V_{22}})}{q^2 m^2_2} \hyperref[eq:knprime]{k^{\prime(1)}_{n,2,2}} 
	\nonumber \\ &
	+ 8 (\hyperref[eq:Qij]{\omega}^2 - 1) \hyperref[eq:kn]{k^{(1)}_{n,3,0}} - 64 \hyperref[eq:omega]{\omega} \frac{(v_1\cdot \mathfrak{a}_{2})}{m_2} \hyperref[eq:kn]{k^{(1)}_{n,2,1}}  - \frac{4(\hyperref[eq:Qij]{Q_{22}} + 13 \hyperref[eq:Vij]{V_{22}})}{q^2 m_2^2}\hyperref[eq:kn]{ k^{(1)}_{n,1,2}}
	+ \frac{64}{q^2 m_2^2} \hyperref[eq:kn]{k^{(1)}_{n,3,2}}  \,,
\end{align}
as well as
\begin{align}
	E_{2n+j} =&
	\left( -  \frac{12 \hyperref[eq:omega]{\omega} (v_1 \cdot \mathfrak{a}_{2})}{m^2_1 m_2} \right)
	\left( \hyperref[eq:g2n]{g^{(j+1)}_{2n+j,0,0}}[e_{\rm tot}] \right)
	\nonumber \\ &
	+ \left(   \frac{64 }{m^2_1 m^2_2 q^2} \right)
	\left( \hyperref[eq:G6indices]{G(\mathfrak{a}_{2},\mathfrak{a},e_{\rm tot},p_{2},2n+j,1)}  \hyperref[eq:At1t2]{A(\mathfrak{a}_{2},\mathfrak{a}_{2})} 2^{2n+2j-1} \hyperref[eq:hn]{h^{(j+1)}_{n+j}} +
	\hyperref[eq:G6indices]{G(\mathfrak{a},\mathfrak{a}_{2},e_{\rm tot},p_2,2n+j,1)}2^{2n+2j-1} \hyperref[eq:hn]{h^{(j+1)}_{n+j-1}} \right)
	\nonumber \\ &
	+ \left(   \frac{64 }{m^2_1 m^2_2q^2} \right)
	\left(
	\hyperref[eq:g2n]{g^{(j+1)}_{2n+j,1,1}}[e_{\rm tot}]	
	+ \delta_{n,1-j} \hyperref[eq:G6indices]{G(\mathfrak{a},p_2,\mathfrak{a},e_{\rm tot},1,1)} \hyperref[eq:At1t2]{A(\mathfrak{a}_{2},\mathfrak{a}_{2})}2^{2n+2j-1} \hyperref[eq:hn]{h^{(j+1)}_{j}}
	\right)
	\nonumber \\ &
	+ \left( - \frac{64 \hyperref[eq:omega]{\omega}}{m_1 m_2 q^2} \right) \left( 
	\hyperref[eq:g2nprime]{g_{n,2,1}^{\prime(j+3)}}
	+ \delta_{n,1-j} \hyperref[eq:G6indices]{G(\mathfrak{a},p_2,\mathfrak{a},e,1,1)} \hyperref[eq:At1t2]{A(\mathfrak{a}_{2},\mathfrak{a}_{2})^{2}} 2^{2n} \hyperref[eq:hn]{h_{1}^{(j+3)}}
	\right. \nonumber \\ & \left. \qquad\qquad\qquad\qquad
	+ \delta_{n,1}\delta_{j1} \hyperref[eq:G9indices]{G(\mathfrak{a},\mathfrak{a}_{2},\mathfrak{a},p_2,\mathfrak{a},e,1,1,1)} \hyperref[eq:At1t2]{A(\mathfrak{a}_{2},\mathfrak{a}_{2})} 2^{2n} \hyperref[eq:hn]{h_{0}^{(j+3)}}
	\right)
	\nonumber \\ &
	+ \left( - \frac{192 (v_1\cdot \mathfrak{a}_{2})}{m_1 m_2^2 q^2} \right) \left( 
	\hyperref[eq:g2n]{g_{n,1,2}^{\prime (j+3)}}
	+ \delta_{n,1-j} \hyperref[eq:G6indices]{G(\mathfrak{a},p_2,\mathfrak{a}_2,e,2,1)} \hyperref[eq:At1t2]{A(\mathfrak{a}_{2},\mathfrak{a}_{2})} 2^{2n} \hyperref[eq:hn]{h_{1}^{(j+3)}}
    \right. \nonumber \\ & \left. \qquad\qquad\qquad\qquad
	+ \delta_{n,1}\delta_{j1} \hyperref[eq:G6indices]{G(\mathfrak{a},p_2,\mathfrak{a},e,2,1)} \hyperref[eq:At1t2]{A(\mathfrak{a}_{2},\mathfrak{a}_{2})} 2^{2n} \hyperref[eq:hn]{h_{0}^{(j+3)}}
	\right)
	\nonumber \\ &
	+ \left( 8  \hyperref[eq:omega]{\omega} \cE_2 \right) \left(
	\hyperref[eq:G3indices]{G(\mathfrak{a}_{2}, \mathfrak{a},2n+j)}
	\hyperref[eq:At1t2]{A(\mathfrak{a}_2,\mathfrak{a}_2)^2} 2^{2n} \hyperref[eq:hn]{h_{n+1}^{(j+3)}}
	+ \hyperref[eq:G3indices]{G(\mathfrak{a}, \mathfrak{a}_{2},2n+j)} 2^{2n} \hyperref[eq:hn]{h_{n-1}^{(j+3)}}
	\right. \nonumber \\ & \left. \qquad\qquad\qquad\qquad
	+ \delta_{n0}\delta_{j1} \hyperref[eq:G3indices]{G(\mathfrak{a}, \mathfrak{a}_{2},1)}   \hyperref[eq:At1t2]{A(\mathfrak{a}_{2},\mathfrak{a}_{2})} 2^{2n} \hyperref[eq:hn]{h_{0}^{(j+3)}}
	\right)
	\nonumber \\ &
	+ \left( \frac{8 \omega (\hyperref[eq:Qij]{Q_{22}} + 13 \hyperref[eq:Vij]{V_{22}})}{m_1 m_2 q^2} \right) \left( 
	\hyperref[eq:g2n]{g_{n,0,1}^{\prime(j+3)}}
	+ \delta_{n0}\delta_{j1} \hyperref[eq:G6indices]{G(\mathfrak{a}_{2},p_2,\mathfrak{a},e,1,1)}2^{2n} \hyperref[eq:hn]{h_{0}^{(j+3)}}
	\right)
	\nonumber \\ &
	+ \left(-  \frac{16  (v_1\cdot \mathfrak{a}_{2})^{3}}{m_1 m_2^2} \right) \left(
	\hyperref[eq:G9indices]{G(\mathfrak{a}_{2},\mathfrak{a},\mathfrak{a}_{2},p_2,\mathfrak{a}_{2},e^{\prime},2n+j,1,1)} 2^{2n} \hyperref[eq:hn]{h_{n+1}^{(j+3)}}
	+ \hyperref[eq:G9indices]{G(\mathfrak{a},\mathfrak{a}_{2},\mathfrak{a},p_2,\mathfrak{a},e^{\prime},2n+j-2,1,1)} 2^{2n} \hyperref[eq:hn]{h_{n-1}^{(j+3)}}
	\right. \nonumber \\ & \left. \qquad\qquad\qquad\qquad
	+ \delta_{n0}\delta_{j1} \hyperref[eq:G6indices]{G(\mathfrak{a}_{2},p_{2},\mathfrak{a},e^{\prime},1,1)} 2^{2n} \hyperref[eq:hn]{h_{0}^{(j+3)}}
	\right)
	\nonumber \\ &
	+ \left( -  \frac{24 (v_1\cdot \mathfrak{a}_2) (\hyperref[eq:omega]{\omega}^2 - 1)}{m_1} \right) \left( 
	\hyperref[eq:g2n]{g_{n,1,0}^{\prime(j+3)}}
	+ \delta_{n0} \delta_{j1} \hyperref[eq:G3indices]{G(\mathfrak{a},e,1)} \hyperref[eq:At1t2]{A(\mathfrak{a}_{2},\mathfrak{a}_{2})} 2^{2n} \hyperref[eq:hn]{h^{(j+3)}_{0}}
	\right)
	\nonumber \\ &
	+\left( -  \frac{3}{4} (\hyperref[eq:Qij]{Q_{22}} + 7 \hyperref[eq:Vij]{V_{22}}) \hyperref[eq:omega]{\omega} \hyperref[eq:calEi]{\cE_{2}} \right) \left( \hyperref[eq:G3indices]{G(\mathfrak{a}_2,\mathfrak{a},2n+j)} 2^{2n} \hyperref[eq:hn]{h_{n}^{(j+3)}}(2-\delta_{n0}\delta_{j0}) \right) \,,\label{eq:En}
\end{align}
with $e_{\rm tot}=m_1 m_2 \omega \epsilon(qp_1\mathfrak{a}_{2}\cdot) - (p_1\cdot \mathfrak{a}_{2}) \epsilon(qp_1p_{2}\cdot)$, $e=\epsilon(qp_1\mathfrak{a}_2\cdot)$, and $e^{\prime}=\epsilon(qp_1p_2\cdot)$.
Also, 
%%%
\begin{subequations}
\begin{align}
	k_{n,k,l}^{(j_2)} =& \hyperref[eq:G6indices]{G(\mathfrak{a}_{2},\mathfrak{a},\mathfrak{a}_{2},p_2,n,l)} \hyperref[eq:At1t2]{A(\mathfrak{a}_{2},\mathfrak{a}_{2})^{k}} 2^{-k-l} 
	\hyperref[eq:Bnj]{B_{n+k+l,0}[H^{(j_2)}]}
	\nonumber \\ &
	+ \hyperref[eq:G6indices]{G(\mathfrak{a},\mathfrak{a}_2,\mathfrak{a},p_2,n-l,l)} 2^{k+l} \hyperref[eq:Bnj]{B_{n-k-l,0}[H^{(j_2)}]} \theta(n-k-l) \,,
	\label{eq:kn}\\ 
	k_{n,k,l}^{\prime(j_2)} =& \hyperref[eq:G6indices]{G(\mathfrak{a}_{2},\mathfrak{a},\mathfrak{a}_{2},p_2,n,l)}  2^{-k} \hyperref[eq:Bnj]{B_{n+k-1,0}[H^{(j_2)}]}
	\nonumber  \\&
	+ \hyperref[eq:G6indices]{G(\mathfrak{a},\mathfrak{a}_2,\mathfrak{a},p_2,n-l,l)} \hyperref[eq:At1t2]{A(\mathfrak{a}_{2},\mathfrak{a}_{2})^{l+1-k}} 
	2^{k-2} \hyperref[eq:Bnj]{B_{n+1-k,0}[H^{(j_2)}]} \theta(n-l) \, 
	\label{eq:knprime} \\ 
	g^{(i)}_{2n+j,k,l}[\cE] =& \hyperref[eq:G9indices]{G(\mathfrak{a}_{2},\mathfrak{a},\mathfrak{a}_{2},p_2,\mathfrak{a}_{2},\cE,2n+j,l,1)} \hyperref[eq:At1t2]{A(\mathfrak{a}_{2},\mathfrak{a}_{2})^{k}} 2^{2n+2j} 
	\hyperref[eq:hn]{h_{n+j+\tfrac{k+l}{2}}^{(i)}}
	\nonumber \\ &
	+ \hyperref[eq:G9indices]{G(\mathfrak{a},\mathfrak{a}_2,\mathfrak{a},p_2,\mathfrak{a},\cE,2n+j-l-1,l,1)} 2^{2n+2j} \hyperref[eq:hn]{h_{n+j-\tfrac{k+l}{2}-1}^{(i)}} \,,
	\label{eq:g2n} \\ 
	g_{n,k,l}^{\prime(i,j_1,j_2)} =& 
	\hyperref[eq:G9indices]{G(\mathfrak{a}_{2},\mathfrak{a},\mathfrak{a}_{2},p_2,\mathfrak{a}_{2},\cE,2n+j_1,l,1)} \hyperref[eq:At1t2]{A(\mathfrak{a}_{2},\mathfrak{a}_{2})^{k}} 
	2^{2n} \hyperref[eq:hn]{h_{n+\tfrac{k+l+1}{2}}^{(i)}}
	\nonumber \\ &
	+ 
	\hyperref[eq:G9indices]{G(\mathfrak{a},\mathfrak{a}_2,\mathfrak{a},p_2,\mathfrak{a},\cE,2n+j_1-l-1,l,1)} 2^{2n} \hyperref[eq:hn]{h_{n-\tfrac{k+l+1}{2}}^{(i)}}  \,.	\label{eq:g2nprime} 
\end{align}
\end{subequations}
These functions are the building blocks of the $\cO(G^2)$ amplitude.

\subsection{All-order-in-spin result}

In the even-in-spin part of the all-order-in-spin amplitude in eq.~\eqref{eq:allSpinEvenResult}, we introduced
\begin{align}\label{eq:Pn}
	P_{n} =& \delta_{n0} \left( 3 - 15 \hyperref[eq:omega]{\omega}^2 \right) 
		\nonumber \\ &
		+ \delta_{n1} \left(2 \hyperref[eq:Vtot]{\Vtot} - \hyperref[eq:Qtot]{\Qtot} (\hyperref[eq:omega]{\omega}^2 - 1) \right)\left( \frac{1}{4} +  \hyperref[eq:At1t2]{A(\mathfrak{a}_2,\mathfrak{a}_2)}^2 \hyperref[eq:hn]{h_{1}^{(2)}} - \frac{1}{16}(\hyperref[eq:Qij]{Q_{22}} + 13 \hyperref[eq:Vij]{V_{22}}) \hyperref[eq:At1t2]{A(\mathfrak{a}_{2},\mathfrak{a}_{2})} \hyperref[eq:hn]{h^{(4)}_{0}}
		+ \hyperref[eq:At1t2]{A(\mathfrak{a}_2,\mathfrak{a}_{2})}^{3} \hyperref[eq:hn]{h_{1}^{(4)}}
		\right)
		\nonumber \\ &
		+  
		\delta_{n1}\left( 16 (\hyperref[eq:omega]{\omega}^2 - 1) \hyperref[eq:G3indices]{G(\mathfrak{a},\mathfrak{a}_{2},2)} - 128 \hyperref[eq:omega]{\omega} \frac{(v_1\cdot \mathfrak{a}_{2})}{m_2} \hyperref[eq:G6indices]{G(\mathfrak{a},\mathfrak{a}_{2},\mathfrak{a},p_2,1,1)} \right) \hyperref[eq:At1t2]{A(\mathfrak{a}_2,\mathfrak{a}_2)}
	    \; 2 \hyperref[eq:hn]{h_{0}^{(4)}}
		\nonumber \\ &
		+ \delta_{n2} \left(\frac{2^{10}}{q^2 m_2^2} \hyperref[eq:G6indices]{G(\mathfrak{a},\mathfrak{a}_{2},\mathfrak{a},p_2,2,2)}\hyperref[eq:At1t2]{A(\mathfrak{a}_2,\mathfrak{a}_{2})}  \hyperref[eq:hn]{h_{0}^{(4)}} \right)
		%\nonumber \\ &
		+ \hyperref[eq:Dn]{D_{2n}} %+ {\rm  termG}^{(4)}_{n}
		+ \hyperref[eq:G3indices]{G(\mathfrak{a},\mathfrak{a}_{2},2n)}\hyperref[eq:Cn]{C_{2n}} \,,
\end{align}
and
\begin{align}\label{eq:Qn}
	Q_{n} =& 
	\frac{3}{4} \delta_{n0} \left( (5\hyperref[eq:omega]{\omega}^2 - 1) (\hyperref[eq:Qij]{Q_{12}} +\hyperref[eq:Qij]{ Q_{22}}) - \frac{14}{3} \hyperref[eq:omega]{\omega} \hyperref[eq:Vij]{V_{12}} + \frac{29\hyperref[eq:omega]{\omega}^2 - 3}{3} \hyperref[eq:Vij]{V_{22}} \right)
	\nonumber \\ &
	+ \delta_{n0} \left(\frac{1}{16} (\hyperref[eq:omega]{\omega} \hyperref[eq:Vij]{V_{12}} (5 \hyperref[eq:Qij]{Q_{22}} + 7 \hyperref[eq:Vij]{V_{22}}) - \hyperref[eq:omega]{\omega}^2 \hyperref[eq:Qij]{Q_{22}} (\hyperref[eq:Qij]{Q_{12}} + \hyperref[eq:Qij]{Q_{22}}) - \hyperref[eq:Vij]{V_{22}} (\hyperref[eq:Qij]{Q_{12}} + \hyperref[eq:Qij]{Q_{22}} + \hyperref[eq:Vij]{V_{22}} + \hyperref[eq:omega]{\omega}^2 (6 \hyperref[eq:Qij]{Q_{22}} + 7 \hyperref[eq:Vij]{V_{22}}))  ) \right) \hyperref[eq:hn]{h^{(1)}_{0}}
	\nonumber \\ &
	+ \delta_{n0} \left(\frac{64}{q^2 m^2_2} \hyperref[eq:G6indices]{G(\mathfrak{a},p_{2},\mathfrak{a}_2,p_2,1,1)} \hyperref[eq:At1t2]{A(\mathfrak{a}_{2},\mathfrak{a}_{2})} \left( \hyperref[eq:At1t2]{A(\mathfrak{a}_{2},\mathfrak{a}_{2})} \hyperref[eq:hn]{h^{(1)}_{1}} + \frac{\hyperref[eq:Qij]{Q_{22}} + 13 \hyperref[eq:Vij]{V_{22}}}{4} \hyperref[eq:hn]{h^{(3)}_{1}} - 4 \hyperref[eq:At1t2]{A(\mathfrak{a}_{2},\mathfrak{a}_{2})}^{2} \hyperref[eq:hn]{h^{(3)}_{2}} \right) \right)
	\nonumber \\ &
	- \delta_{n1} \frac{128}{m^2_2 q^2} \left( 2 \hyperref[eq:G6indices]{G(\mathfrak{a}_{1},\mathfrak{a}_{2},\mathfrak{a}_{1},p_2,1,2)} + 2 \hyperref[eq:At1t2]{A(\mathfrak{a}_{2},\mathfrak{a}_{2})} (3 \hyperref[eq:G3indices]{G(\mathfrak{a}_{1},p_2,2)}+
	\hyperref[eq:G3indices]{G(\mathfrak{a}_{2},p_2,2)} 
	+ 3 \hyperref[eq:G6indices]{G(\mathfrak{a}_{1},p_2,\mathfrak{a}_{2},p_2,1,1)} ) ) \right)
	\nonumber \\ &
	+ \delta_{n1} \left(\frac{256}{q^2 m^2_2} \hyperref[eq:G6indices]{G(\mathfrak{a},\mathfrak{a}_{2},\mathfrak{a},p_2,1,2)}  \left( \hyperref[eq:At1t2]{A(\mathfrak{a}_{2},\mathfrak{a}_{2})} \hyperref[eq:hn]{h^{(1)}_{0}}  - 4\hyperref[eq:At1t2]{A(\mathfrak{a}_{2},\mathfrak{a}_{2})}^{2}   \hyperref[eq:hn]{h_{1}^{(3)}}
	\right) \right)
	\nonumber \\ &
	- 16 (\hyperref[eq:omega]{\omega}^2 - 1) 
	\delta_{n0}\hyperref[eq:G3indices]{G(\mathfrak{a},\mathfrak{a}_2,1)}\hyperref[eq:At1t2]{A(\mathfrak{a}_2,\mathfrak{a}_2)}^2
	 \; 2 \hyperref[eq:hn]{h_{1}^{(3)}}
	\nonumber \\ &
	+ 128 \hyperref[eq:omega]{\omega} \frac{(v_1\cdot \mathfrak{a}_{2})}{m_2} 
	\delta_{n0}\hyperref[eq:G3indices]{G(\mathfrak{a},p_2,1)}
	\hyperref[eq:At1t2]{A(\mathfrak{a}_2,\mathfrak{a}_2)}^2
	 \; 2 \hyperref[eq:hn]{h_{1}^{(3)}}
	\nonumber \\ &
	+ \hyperref[eq:Dn]{D_{2n+1}}
	+ \hyperref[eq:G3indices]{G(\mathfrak{a},\mathfrak{a}_{2},2n+1)} 
	\hyperref[eq:CnContactTerm]{C_{2n+1}} \,.
\end{align}
In the odd-in-spin part of the all-order-in-spin amplitude in eq.~\eqref{eq:allSpinOddResult}, we used
\begin{align}\label{eq:Rn}
	R_{n} =& \delta_{n0} \left(- \hyperref[eq:omega]{\omega} \hyperref[eq:calEi]{\cE_{2}} \frac{5\hyperref[eq:omega]{\omega}^2 - 3}{\hyperref[eq:omega]{\omega}^2 - 1} \right) 
	\nonumber \\ &
	+ \delta_{n1} \left( -  \hyperref[eq:calEi]{\cE_{1}} \left(\frac{1}{2} \hyperref[eq:omega]{\omega} (3 \hyperref[eq:Qij]{Q_{22}} + 4 \hyperref[eq:Vij]{V_{22}}) - \frac{4\hyperref[eq:omega]{\omega}^2 - 1}{2(\hyperref[eq:omega]{\omega}^2 - 1)} \hyperref[eq:Vij]{V_{12}} \right) 
	- \hyperref[eq:calEi]{\cE_{2}} \left( \frac{1}{4} \hyperref[eq:omega]{\omega} ( 3 \hyperref[eq:Qij]{Q_{11}} + 3\hyperref[eq:Qij]{Q_{22}} + 5 \hyperref[eq:Vij]{V_{22}} ) - \frac{1}{2} \hyperref[eq:Vij]{V_{12}} - \frac{3}{2(\hyperref[eq:omega]{\omega}^2 - 1)} \hyperref[eq:omega]{\omega} \hyperref[eq:Vij]{V_{11}} \right)  \right) 
	\nonumber \\ &
+ \hyperref[eq:En]{E_{2n}} + \hyperref[eq:C2nTilde]{\tilde{C}_{2n}}
\end{align}
and
\begin{align}\label{eq:Sn}
	S_{n} =& \delta_{n0} \left(  4\hyperref[eq:omega]{\omega} \hyperref[eq:calE]{\cE} \right) 
+ \hyperref[eq:En]{E_{2n+1}} +  \hyperref[eq:C2nTilde]{\tilde{C}_{2n+1}} \, .
\end{align}
These expressions, together with the other definitions in this appendix, constitute the all-order-in-spin amplitude at $\cO(G^2)$.

%%%%%%%%%%%%%%%%%%%%%%%%%%%%%%%%%%%%%%%%%%%%%%%%%%%%%%%
%%%%%%% End Definitions %%%%%%%%%%%%%%%%%%%%%%%%%%%%%%%
%%%%%%%%%%%%%%%%%%%%%%%%%%%%%%%%%%%%%%%%%%%%%%%%%%%%%%%

%%%%%%%%%%%%%%%%%%%%%%%%%%%%%%%%%%%%%%%%%%%%%%%%%%%%%%%
%%%%%%% Begin most-general contact terms %%%%%%%%%%%%%%
%%%%%%%%%%%%%%%%%%%%%%%%%%%%%%%%%%%%%%%%%%%%%%%%%%%%%%%

\section{Most-general contact terms relevant at 2PM}\label{app:ContactTerms}

The most general set of contact deformations of the Compton amplitude that are potentially relevant to black-hole scattering at 2PM was constructed in ref.~\cite{Haddad:2023ylx}.
In the computation above we have included these contact terms in a slightly modified form relative to that reference.
Specifically, we have introduced $Q_{34}^{\pm,\mu}=q_{4}^{\mu}\pm q_{3}^{\mu}$. The analytic-in-spin contact terms in the opposite-helicity sector are
\begin{align}\label{eq:AnalyticContactOH}
    &m^{2}\mathcal{C}^{-+}=\frac{y^{4}}{m^{2}}\sum_{s=4}^{\infty}\sum_{i=0}^{\floor{(s-4)/2}}\sum_{j=0}^{s-4-2i}\sum_{l=0}^{i}a^{-+,(s)}_{i,l,j}\mathfrak{a}^{4+2i}s_{34}^{i-l}(Q_{34}^{-}\cdot\mathfrak{a})^{j}(Q_{34}^{+}\cdot\mathfrak{a})^{s-4-2i-j}\frac{(t_{14}-t_{13})^{2l}}{m^{2l}}\notag \\
    &+\frac{y^{3}}{m}(w\cdot\mathfrak{a})\sum_{s=5}^{\infty}\sum_{i=0}^{\floor{(s-5)/2}}\sum_{j=0}^{s-5-2i}\sum_{l=0}^{i}b^{-+,(s)}_{i,l,j}\mathfrak{a}^{4+2i}(Q_{34}^{-}\cdot\mathfrak{a})^{j}(Q_{34}^{+}\cdot\mathfrak{a})^{s-5-2i-j}s_{34}^{i-l}\frac{(t_{14}-t_{13})^{2l+1}}{m^{2l+1}}\notag \\
    &+y^{2}(w\cdot\mathfrak{a})^{2}\sum_{s=4}^{\infty}\sum_{i=0}^{\floor{(s-4)/2}}\sum_{j=0}^{s-4-2i}c^{-+,(s)}_{i,j}\mathfrak{a}^{2+2i}(Q_{34}^{-}\cdot\mathfrak{a})^{j}(Q_{34}^{+}\cdot\mathfrak{a})^{s-4-2i-j}s_{34}^{i}\notag \\
    &+m\frac{(t_{14}-t_{13})}{m}y(w\cdot\mathfrak{a})^{3}\sum_{s=5}^{\infty}\sum_{i=0}^{\floor{(s-5)/2}}\sum_{j=0}^{s-5-2i}d^{-+,(s)}_{i,j}\mathfrak{a}^{2+2i}(Q_{34}^{-}\cdot\mathfrak{a})^{j}(Q_{34}^{+}\cdot\mathfrak{a})^{s-5-2i-j}s_{34}^{i}\notag \\
    &+m^{2}(w\cdot\mathfrak{a})^{4}\sum_{s=4}^{\infty}\sum_{i=0}^{\floor{(s-4)/2}}\sum_{j=0}^{s-4-2i}e^{-+,(s)}_{i,j}\mathfrak{a}^{2i}(Q_{34}^{-}\cdot\mathfrak{a})^{j}(Q_{34}^{+}\cdot\mathfrak{a})^{s-4-2i-j}s_{34}^{i}.
\end{align}
The opposite-helicity Compton amplitude can also be deformed by contact terms which are not analytic in the spin vector:
\begin{align}\label{eq:NonAnalyticContactOH}
    &m^{2}\mathcal{D}^{-+}=|\mathfrak{a}|\frac{y^{4}}{m^{2}}\sum_{s=5}^{\infty}\sum_{i=0}^{\floor{(s-5)/2}}\sum_{j=0}^{s-5-2i}\sum_{l=0}^{i}f_{i,l,j}^{-+,(s)}s_{34}^{i-l}\mathfrak{a}^{4+2i}(Q_{34}^{+}\cdot\mathfrak{a})^{s-5-2i-j}(Q_{34}^{-}\cdot\mathfrak{a})^{j}\frac{(t_{14}-t_{13})^{2l+1}}{m^{2l+1}}\notag \\
    &+|\mathfrak{a}|\frac{y^{3}}{m}(w\cdot\mathfrak{a})\sum_{s=4}^{\infty}\sum_{i=0}^{\floor{(s-4)/2}}\sum_{j=0}^{s-4-2i}\sum_{l=0}^{i}g_{i,l,j}^{-+,(s)}s_{34}^{i-l}\mathfrak{a}^{2+2i}(Q_{34}^{+}\cdot\mathfrak{a})^{s-4-2i-j}(Q_{34}^{-}\cdot\mathfrak{a})^{j}\frac{(t_{14}-t_{13})^{2l}}{m^{2l}}\notag \\
    &+|\mathfrak{a}|y^{2}(w\cdot\mathfrak{a})^{2}\sum_{s=5}^{\infty}\sum_{i=0}^{\floor{(s-5)/2}}\sum_{j=0}^{s-5-2i}p_{i,j}^{-+,(s)}s_{34}^{i}\mathfrak{a}^{2+2i}(Q_{34}^{+}\cdot\mathfrak{a})^{s-5-2i-j}(Q_{34}^{-}\cdot\mathfrak{a})^{j}\frac{(t_{14}-t_{13})}{m}\notag \\
    &+m|\mathfrak{a}|y(w\cdot\mathfrak{a})^{3}\sum_{s=4}^{\infty}\sum_{i=0}^{\floor{(s-4)/2}}\sum_{j=0}^{s-4-2i}q_{i,j}^{-+,(s)}s_{34}^{i}\mathfrak{a}^{2i}(Q_{34}^{+}\cdot\mathfrak{a})^{s-4-2i-j}(Q_{34}^{-}\cdot\mathfrak{a})^{j}\notag \\
    &+m^{2}|\mathfrak{a}|(w\cdot\mathfrak{a})^{4}\sum_{s=5}^{\infty}\sum_{i=0}^{\floor{(s-5)/2}}\sum_{j=0}^{s-5-2i}r_{i,j}^{-+,(s)}s_{34}^{i}\mathfrak{a}^{2i}(Q_{34}^{+}\cdot\mathfrak{a})^{s-5-2i-j}(Q_{34}^{-}\cdot\mathfrak{a})^{j}\frac{(t_{14}-t_{13})}{m}.
\end{align}
A consequence of crossing symmetry is that the coefficients are the same for both opposite-helicity amplitudes, $x^{-+,(s)}_{i,(l,),j}=x^{+-,(s)}_{i,(l,)j}\equiv x^{(s)}_{i,(l,)j}$.
Moreover, crossing symmetry relates these coefficients to their complex conjugates:
\begin{align}
    a^{(s)}_{i,l,j}=(-1)^{s+j}\bar{a}^{(s)}_{i,l,j},&\quad b^{(s)}_{i,l,j}=(-1)^{s+j+1}\bar{b}^{(s)}_{i,l,j},\label{eq:OHCoeffConjugates} \\
    c^{(s)}_{i,j}=(-1)^{s+j}\bar{c}^{(s)}_{i,j},\quad d^{(s)}_{i,j}=&(-1)^{s+j+1}\bar{d}^{(s)}_{i,j},\quad e^{(s)}_{i,j}=(-1)^{s+j}\bar{e}^{(s)}_{i,j},\notag \\
    f^{(s)}_{i,l,j}=(-1)^{s+j}\bar{f}^{(s)}_{i,l,j},&\quad g^{(s)}_{i,l,j}=(-1)^{s+j+1}\bar{g}^{(s)}_{i,l,j}, \\
    p^{(s)}_{i,j}=(-1)^{s+j}\bar{p}^{(s)}_{i,j},\quad q^{(s)}_{i,j}&=(-1)^{s+j+1}\bar{q}^{(s)}_{i,j},\quad r^{(s)}_{i,j}=(-1)^{s+j}\bar{r}^{(s)}_{i,j}.\notag
\end{align}
While these relations were not needed for writing the all-spin contribution to the 2PM amplitude, they have implications for the parity properties of the amplitude.

In the same-helicity (two-plus) case, the analytic contact terms are
\begin{align}\label{eq:AnalyticContactSH}
    &m^{2}\mathcal{C}^{++}=\frac{y^{4}_{++}}{m^{2}}\sum_{s=4}^{\infty}\sum_{i=0}^{\floor{(s-4)/2}}\sum_{j=0}^{s-4-2i}\sum_{l=0}^{i}a^{++,(s)}_{i,l,j}\mathfrak{a}^{4+2i}(Q_{34}^{-}\cdot\mathfrak{a})^{j}(Q_{34}^{+}\cdot\mathfrak{a})^{s-4-2i-j}s_{34}^{l}\frac{(t_{14}-t_{13})^{2(i-l)}}{m^{2(i-l)}}\notag \\
    &+\frac{y^{3}_{++}}{m}(w_{++}\cdot\mathfrak{a})\sum_{s=5}^{\infty}\sum_{i=0}^{\floor{(s-5)/2}}\sum_{j=0}^{s-5-2i}\sum_{l=0}^{i}b^{++,(s)}_{i,l,j}\mathfrak{a}^{4+2i}(Q_{34}^{-}\cdot\mathfrak{a})^{j}(Q_{34}^{+}\cdot\mathfrak{a})^{s-5-2i-j}s_{34}^{l}\frac{(t_{14}-t_{13})^{2(i-l)+1}}{m^{2(i-l)+1}}\notag \\
    &+y^{2}_{++}(w_{++}\cdot\mathfrak{a})^{2}\sum_{s=4}^{\infty}\sum_{i=0}^{\floor{(s-4)/2}}\sum_{j=0}^{s-4-2i}c^{++,(s)}_{i,j}\mathfrak{a}^{2+2i}(Q_{34}^{-}\cdot\mathfrak{a})^{j}(Q_{34}^{+}\cdot\mathfrak{a})^{s-4-2i-j}\frac{(t_{14}-t_{13})^{2i}}{m^{2i}}\notag \\
    &+my_{++}(w_{++}\cdot\mathfrak{a})^{3}\sum_{s=5}^{\infty}\sum_{i=0}^{\floor{(s-5)/2}}\sum_{j=0}^{s-5-2i}d^{++,(s)}_{i,j}\mathfrak{a}^{2+2i}(Q_{34}^{-}\cdot\mathfrak{a})^{j}(Q_{34}^{+}\cdot\mathfrak{a})^{s-5-2i-j}\frac{(t_{14}-t_{13})^{2i+1}}{m^{2i+1}}\notag \\
    &+m^{2}(w_{++}\cdot\mathfrak{a})^{4}\sum_{s=4}^{\infty}\sum_{i=0}^{\floor{(s-4)/2}}\sum_{j=0}^{s-4-2i}e^{++,(s)}_{i,j}\mathfrak{a}^{2i}(Q_{34}^{-}\cdot\mathfrak{a})^{j}(Q_{34}^{+}\cdot\mathfrak{a})^{s-4-2i-j}\frac{(t_{14}-t_{13})^{2i}}{m^{2i}},
\end{align}
whereas the non-analytic deformations are
\begin{align}\label{eq:NonAnalyticContactSH}
    &m^{2}\mathcal{D}^{++}=|\mathfrak{a}|\frac{y_{++}^{4}}{m^{2}}\sum_{s=5}^{\infty}\sum_{i=0}^{\floor{(s-5)/2}}\sum_{j=0}^{s-5-2i}\sum_{l=0}^{i}f_{i,l,j}^{++,(s)}s_{34}^{l}\mathfrak{a}^{4+2i}(Q_{34}^{+}\cdot\mathfrak{a})^{s-5-2i-j}(Q_{34}^{-}\cdot\mathfrak{a})^{j}\frac{(t_{14}-t_{13})^{2(i-l)+1}}{m^{2(i-l)+1}}\notag \\
    &+|\mathfrak{a}|\frac{y_{++}^{3}}{m}(w_{++}\cdot\mathfrak{a})\sum_{s=4}^{\infty}\sum_{i=0}^{\floor{(s-4)/2}}\sum_{j=0}^{s-4-2i}\sum_{l=0}^{i}g_{i,l,j}^{++,(s)}s_{34}^{l}\mathfrak{a}^{2+2i}(Q_{34}^{+}\cdot\mathfrak{a})^{s-4-2i-j}(Q_{34}^{-}\cdot\mathfrak{a})^{j}\frac{(t_{14}-t_{13})^{2(i-l)}}{m^{2(i-l)}}\notag \\
    &+|\mathfrak{a}|y_{++}^{2}(w_{++}\cdot\mathfrak{a})^{2}\sum_{s=5}^{\infty}\sum_{i=0}^{\floor{(s-5)/2}}\sum_{j=0}^{s-5-2i}p_{i,j}^{++,(s)}\mathfrak{a}^{2+2i}(Q_{34}^{+}\cdot\mathfrak{a})^{s-5-2i-j}(Q_{34}^{-}\cdot\mathfrak{a})^{j}\frac{(t_{14}-t_{13})^{2i+1}}{m^{2i+1}}\notag \\
    &+m|\mathfrak{a}|y_{++}(w_{++}\cdot\mathfrak{a})^{3}\sum_{s=4}^{\infty}\sum_{i=0}^{\floor{(s-4)/2}}\sum_{j=0}^{s-4-2i}q_{i,j}^{++,(s)}\mathfrak{a}^{2i}(Q_{34}^{+}\cdot\mathfrak{a})^{s-4-2i-j}(Q_{34}^{-}\cdot\mathfrak{a})^{j}\frac{(t_{14}-t_{13})^{2i}}{m^{2i}}\notag \\
    &+m^{2}|\mathfrak{a}|(w_{++}\cdot\mathfrak{a})^{4}\sum_{s=5}^{\infty}\sum_{i=0}^{\floor{(s-5)/2}}\sum_{j=0}^{s-5-2i}r_{i,j}^{++,(s)}\mathfrak{a}^{2i}(Q_{34}^{+}\cdot\mathfrak{a})^{s-5-2i-j}(Q_{34}^{-}\cdot\mathfrak{a})^{j}\frac{(t_{14}-t_{13})^{2i+1}}{m^{2i+1}}.
\end{align}
The helicity weights of the gravitons for the same-helicity configuration are carried by $w_{++}^{\mu}\equiv[4|p_{1}\sigma^{\mu}|3]/2m$ and $y_{++}\equiv 2p_{1}\cdot w_{++}.$

Crossing symmetry again relates the coefficients between both same-helicity configurations through $x^{--,(s)}_{i,(l,)j}=(-1)^{s}\bar{x}^{++,(s)}_{i,(l,)j}$ in the conservative sector and $x^{--,(s)}_{i,(l,)j}=(-1)^{s+1}\bar{x}^{++,(s)}_{i,(l,)j}$ in the dissipative sector.
Further consequences of crossing symmetry are
\begin{align}
    a^{hh,(s)}_{i,l,j}=(-1)^{j}a^{hh,(s)}_{i,l,j}&\quad b^{hh,(s)}_{i,l,j}=(-1)^{j+1}b^{hh,(s)}_{i,l,j},\label{eq:SHCoeffVanishing} \\
    c^{hh,(s)}_{i,j}=(-1)^{j}c^{hh,(s)}_{i,j},\quad d^{hh,(s)}_{i,j}&=(-1)^{j+1}d^{hh,(s)}_{i,j},\quad e^{hh,(s)}_{i,j}=(-1)^{j}e^{hh,(s)}_{i,j},\notag \\
    f^{hh,(s)}_{i,l,j}=(-1)^{j+1}f^{hh,(s)}_{i,l,j},&\quad g^{hh,(s)}_{i,l,j}=(-1)^{j}g^{hh,(s)}_{i,l,j},\notag \\
    p^{hh,(s)}_{i,j}=(-1)^{j+1}p^{hh,(s)}_{i,j},\quad q^{hh,(s)}_{i,j}&=(-1)^{j}q^{hh,(s)}_{i,j},\quad r^{hh,(s)}_{i,j}=(-1)^{j+1}r^{hh,(s)}_{i,j},\notag
\end{align}
which impose the vanishing of many of the coefficients for certain values of $j$.

All coefficients in all helicity sectors have been made dimensionless by introducing appropriate factors of the mass.

In the general-contact-term section of the 2PM amplitude, the spin dependence of the particle whose massive propagator is taken on shell is encapsulated in the function
\begin{align}\label{eq:ContactH}
    &\mathcal{H}_{n}^{j,t,k}(a,b,c,d)=\left(\frac{2}{m_{2}}\right)^{k}\frac{1}{n!}\frac{2-\delta_{0,n}}{2}{}_{0}F_{1}(;n+1;a)b^{(k-j-n)/2}\bar{d}^{j} \\
    &\quad\times\left[c^{n}{{k}\choose{(j+k-t-n)/2}}{}_{2}F_{1}\left(-j,\frac{t+n-j-k}{2};\frac{2+k-j+n+t}{2};\frac{d}{\bar{d}}\right)\right.\notag \\
    &\quad\left.+\bar{c}^{n}{{k}\choose{(j+k-t+n)/2}}{}_{2}F_{1}\left(-j,\frac{t-n-j-k}{2};\frac{2+k-j-n+t}{2};\frac{d}{\bar{d}}\right)\right],\notag
\end{align}
where a bar over a symbol represents complex conjugation.

When the tree-level Compton amplitudes are multiplied together in the cut, the helicity vectors $w^{\mu}$, $w_{++}^{\mu}$, and their conjugates are also multiplied.
This gives emergence to the quantities
\begin{align}
    L_{yy}\equiv q^{2}m_{1}m_{2}\omega,&\quad E_{yy}^{\mu}\equiv2\epsilon^{\mu\nu\alpha\beta}q_{\nu}p_{1\alpha}p_{2\beta}, \\
    L_{yw}\equiv\frac{1}{2}q^{2}(p_{1}\cdot\mathfrak{a}_{2}),&\quad E_{yw}^{\mu}\equiv\epsilon^{\mu\nu\alpha\beta}q_{\nu}p_{1\alpha}\mathfrak{a}_{2\beta}, \\
    W^{\mu}=\frac{m_{1}}{m_{2}}[p_{2}^{\mu}(q&\cdot\mathfrak{a}_{2})+i\epsilon^{\mu\nu\alpha\beta}q_{\nu}\mathfrak{a}_{2\alpha}p_{2\beta}].
\end{align}

The form factors in arising from the contact terms are most succinctly defined in terms of the vectors
\begin{align}\label{eq:LoopSolutionuv}
    u^{\mu}_{\pm}=\langle Q_{\pm}|\gamma^{\mu}|P_{\pm}],&\quad V^{\mu}_{\pm}=\pm\frac{\sqrt{-q^{2}}}{64m_{1}}[Q_{\pm}|\gamma^{\mu}|P_{\pm}\rangle, \\
    P_{\pm}^{\mu}=p_{1}^{\mu}\pm\frac{m_{1}}{\sqrt{-q^{2}}}q^{\mu},&\quad Q_{\pm}^{\mu}=q^{\mu}\mp\frac{\sqrt{-q^{2}}}{m_{1}}p_{1}^{\mu},
\end{align}
which appear in the solutions for the loop momentum which satisfy the triangle-cut conditions $l^{2}=(l+q)^{2}=p_{1}\cdot l=0$.
In terms of these, the form factors are
\begin{align}\label{eq:HelicityVectorFF}
    A^{(x,n)}_{t}&=\left(2u_{\pm}\cdot x\right)^{-t/2}\left(2V_{\pm}\cdot x\right)^{t/2}A^{(n)}_{t}, \\
    B^{(x,2)}_{2r}&=\left(2u_{\pm}\cdot x\right)^{-r}\left(2V_{\pm}\cdot x\right)^{r}B^{(2)}_{2r,\mu\nu}W^{\mu}W^{\nu}, \\
    B^{(x,0)}_{2r}&=\left(2u_{\pm}\cdot x\right)^{-r}\left(2V_{\pm}\cdot x\right)^{r}B^{(0)}_{2r,\mu\nu\rho\tau}W^{\mu}W^{\nu}W^{\rho}W^{\tau},
\end{align}
for $0\leq n\leq4$, $-4\leq t\leq4$, $-2\leq r\leq 2$, where
\begin{align}
    A^{(4)}_{-4}&=\left(V_{\pm}\cdot E_{yy}\right)^{4}, \\
    A^{(4)}_{-3}&=4iL_{yy}\left(V_{\pm}\cdot E_{yy}\right)^{3}, \\
    A^{(4)}_{-2}&=2\left(V_{\pm}\cdot E_{yy}\right)^{2}\left[-3L_{yy}^{2}+2\left(u_{\pm}\cdot E_{yy}\right)\left(V_{\pm}\cdot E_{yy}\right)\right], \\
    A^{(4)}_{-1}&=-4iL_{yy}\left(V_{\pm}\cdot E_{yy}\right)\left[L_{yy}^{2}-3\left(u_{\pm}\cdot E_{yy}\right)\left(V_{\pm}\cdot E_{yy}\right)\right], \\
    A^{(4)}_{0}&=L_{yy}^{4}-12L_{yy}^{2}\left(u_{\pm}\cdot E_{yy}\right)\left(V_{\pm}\cdot E_{yy}\right)+6\left(u_{\pm}\cdot E_{yy}\right)^{2}\left(V_{\pm}\cdot E_{yy}\right)^{2}, \\
    A^{(3)}_{-4}&=\left(V_{\pm}\cdot E_{yy}\right)^{3}\left(V_{\pm}\cdot E_{yw}\right), \\
    A^{(3)}_{-3}&=i\left(V_{\pm}\cdot E_{yy}\right)^{2}\left[3L_{yy}\left(V_{\pm}\cdot E_{yw}\right)+L_{yw}\left(V_{\pm}\cdot E_{yy}\right)\right], \\
    A^{(3)}_{-2}&=\left(V_{\pm}\cdot E_{yy}\right)\left[\left(V_{\pm}\cdot E_{yy}\right)\left(-3L_{yy}L_{yw}+\left(u_{\pm}\cdot E_{yw}\right)\left(V_{\pm}\cdot E_{yy}\right)\right)\right.\notag \\
    &\qquad\left.-3\left(V_{\pm}\cdot E_{yw}\right)\left(L_{yy}^{2}-\left(u_{\pm}\cdot E_{yy}\right)\left(V_{\pm}\cdot E_{yy}\right)\right)\right], \\
    A^{(3)}_{-1}&=-i\left[-3\left(V_{\pm}\cdot E_{yy}\right)\left[-L_{yw}L_{yy}^{2}+L_{yy}\left(u_{\pm}\cdot E_{yw}\right)\left(V_{\pm}\cdot E_{yy}\right)+L_{yw}\left(u_{\pm}\cdot E_{yy}\right)\left(V_{\pm}\cdot E_{yy}\right)\right]\right.\\
    &\qquad\left.+\left(V_{\pm}\cdot E_{yw}\right)\left[L_{yy}^{3}-6L_{yy}\left(u_{\pm}\cdot E_{yy}\right)\left(V_{\pm}\cdot E_{yy}\right)\right]\right], \\
    A^{(3)}_{0}&=L_{yy}^{2}\left[L_{yw}L_{yy}-3\left(V_{\pm}\cdot E_{yy}\right)\left(u_{\pm}\cdot E_{yw}\right)\right]+3\left(u_{\pm}\cdot E_{yy}\right)^{2}\left(V_{\pm}\cdot E_{yy}\right)\left(V_{\pm}\cdot E_{yw}\right) \\
    &\qquad-3\left(u_{\pm}\cdot E_{yy}\right)\left[2L_{yw}L_{yy}\left(V_{\pm}\cdot E_{yy}\right)-\left(V_{\pm}\cdot E_{yy}\right)^{2}\left(u_{\pm}\cdot E_{yw}\right)+L_{yy}^{2}\left(V_{\pm}\cdot E_{yw}\right)\right], \\
    A^{(2)}_{-4}&=\left(V_{\pm}\cdot E_{yy}\right)^{2}\left(V_{\pm}\cdot E_{yw}\right)^{2}, \\
    A^{(2)}_{-3}&=2i\left(V_{\pm}\cdot E_{yy}\right)\left(V_{\pm}\cdot E_{yw}\right)\left[L_{yy}\left(V_{\pm}\cdot E_{yw}\right)+L_{yw}\left(V_{\pm}\cdot E_{yy}\right)\right], \\
    A^{(2)}_{-2}&=-L_{yw}^{2}\left(V_{\pm}\cdot E_{yy}\right)^{2}+2\left(V_{\pm}\cdot E_{yw}\right)\left(V_{\pm}\cdot E_{yy}\right)\left[-2L_{yy}L_{yw}+\left(u_{\pm}\cdot E_{yw}\right)\left(V_{\pm}\cdot E_{yy}\right)\right] \\
    &\qquad-\left(V_{\pm}\cdot E_{yw}\right)^{2}\left[L_{yy}^{2}-2\left(V_{\pm}\cdot E_{yy}\right)\left(u_{\pm}\cdot E_{yy}\right)\right], \\
    A^{(2)}_{-1}&=-2i\left[-L_{yy}\left(V_{\pm}\cdot E_{yw}\right)^{2}\left(u_{\pm}\cdot E_{yy}\right)+L_{yw}\left(V_{\pm}\cdot E_{yy}\right)\left[L_{yw}L_{yy}-\left(u_{\pm}\cdot E_{yw}\right)\left(V_{\pm}\cdot E_{yy}\right)\right]\right. \\
    &\qquad\left.+\left(V_{\pm}\cdot E_{yw}\right)\left[-2L_{yy}\left(u_{\pm}\cdot E_{yw}\right)\left(V_{\pm}\cdot E_{yy}\right)+L_{yw}\left[L_{yy}^{2}-2\left(u_{\pm}\cdot E_{yy}\right)\left(V_{\pm}\cdot E_{yy}\right)\right]\right]\right], \\
    A^{(2)}_{0}&=L_{yw}^{2}L_{yy}^{2}-4L_{yw}L_{yy}\left[\left(V_{\pm}\cdot E_{yw}\right)\left(u_{\pm}\cdot E_{yy}\right)+\left(u_{\pm}\cdot E_{yw}\right)\left(V_{\pm}\cdot E_{yy}\right)\right] \\
    &\quad-2L_{yw}^{2}\left(u_{\pm}\cdot E_{yy}\right)\left(V_{\pm}\cdot E_{yy}\right)-2L_{yy}^{2}\left(u_{\pm}\cdot E_{yw}\right)\left(V_{\pm}\cdot E_{yw}\right) \\
    &\quad+\left(V_{\pm}\cdot E_{yw}\right)^{2}\left(u_{\pm}\cdot E_{yy}\right)^{2}+\left(u_{\pm}\cdot E_{yw}\right)^{2}\left(V_{\pm}\cdot E_{yy}\right)^{2}+4\left(u_{\pm}\cdot E_{yw}\right)\left(V_{\pm}\cdot E_{yw}\right)\left(u_{\pm}\cdot E_{yy}\right)\left(V_{\pm}\cdot E_{yy}\right), \\
    B^{(2)}_{-4,\mu\nu}&=0,\quad B^{(2)}_{-2,\mu\nu}=V_{\pm,\mu}V_{\pm,\nu},\quad B^{(2)}_{0,\mu\nu}=2u_{\pm,\mu}V_{\pm,\nu}, \\
    B^{(0)}_{-4,\mu\nu\rho\tau}&=V_{\pm,\mu}V_{\pm,\nu}V_{\pm,\rho}V_{\pm,\tau},\quad
    B^{(0)}_{-2,\mu\nu\rho\tau}=4u_{\pm,\mu}V_{\pm,\nu}V_{\pm,\rho}V_{\pm,\tau},\quad B^{(0)}_{0,\mu\nu\rho\tau}=6u_{\pm,\mu}u_{\pm,\nu}V_{\pm,\rho}V_{\pm,\tau}.
\end{align}
Those not written explicitly can be generated from these using $A^{(n)}_{t}=A^{(4-n)}_{t}|_{L_{yy}\leftrightarrow L_{yw},E_{yy}\leftrightarrow E_{yw}}$ and $X^{(n)}_{t}=X^{(n)}_{-t}|_{u\leftrightarrow V}$ for $X=A,B$.
Since $u^{\mu}_{\pm}$ and $V^{\mu}_{\pm}$ are related by complex conjugation, we point out that $\bar{X}^{(n)}_{t}=X^{(n)}_{t}|_{i\rightarrow-i}$ as opposed to being the full complex conjugate.
These quantities can be written in terms of the outer products
\begin{align}\label{eq:uvOuterProducts}
    x_{1,\pm}&\equiv4(u_{\pm}\cdot\mathfrak{a}_{1})(V_{\pm}\cdot p_{2})=-\frac{1}{4m_{1}}\left[m_{1}q^{2}(p_{2}\cdot\mathfrak{a}_{1})\pm i\sqrt{-q^{2}}\epsilon^{\mu\nu\alpha\beta}q_{\mu}\mathfrak{a}_{1\nu}p_{1\alpha}p_{2\beta}\right]+\cO(\hbar^{2}), \\
    x_{2,\pm}&\equiv4(u_{\pm}\cdot\mathfrak{a}_{2})(V_{\pm}\cdot p_{2})=\frac{1}{4m_{1}}\left[m_{2}\omega q^{2}(p_{1}\cdot\mathfrak{a}_{2})\mp i\sqrt{-q^{2}}\epsilon^{\mu\nu\alpha\beta}q_{\mu}\mathfrak{a}_{2\nu}p_{1\alpha}p_{2\beta}\right]+\cO(\hbar^{2}), \\
    y_{12,\pm}&\equiv4(u_{\pm}\cdot\mathfrak{a}_{1})(V_{\pm}\cdot\mathfrak{a}_{2})=\frac{1}{4m_{1}}\left[m_{1}Q_{12}\pm i\sqrt{-q^{2}}\epsilon^{\mu\nu\alpha\beta}q_{\mu}\mathfrak{a}_{1\nu}\mathfrak{a}_{2\alpha}p_{1\beta}\right], \\
    y_{22}&\equiv4(u_{\pm}\cdot\mathfrak{a}_{2})(V_{\pm}\cdot\mathfrak{a}_{2})=\frac{1}{4}\left(Q_{22}+V_{22}\right), \\
    z&\equiv4(u_{\pm}\cdot p_{2})(V_{\pm}\cdot p_{2})=\frac{1}{4}q^{2}m_{2}^{2}(\omega^{2}-1)+\cO(\hbar^{4}),
\end{align}
such that the vectors $u_{\pm}^{\mu}$ and $V_{\pm}^{\mu}$ are eliminated in favor of the physical four-vectors describing the scattering.

\end{widetext}

\bibliography{BibliographyKerrScattering}

\begin{thebibliography}{71}%
\makeatletter
\providecommand \@ifxundefined [1]{%
 \@ifx{#1\undefined}
}%
\providecommand \@ifnum [1]{%
 \ifnum #1\expandafter \@firstoftwo
 \else \expandafter \@secondoftwo
 \fi
}%
\providecommand \@ifx [1]{%
 \ifx #1\expandafter \@firstoftwo
 \else \expandafter \@secondoftwo
 \fi
}%
\providecommand \natexlab [1]{#1}%
\providecommand \enquote  [1]{``#1''}%
\providecommand \bibnamefont  [1]{#1}%
\providecommand \bibfnamefont [1]{#1}%
\providecommand \citenamefont [1]{#1}%
\providecommand \href@noop [0]{\@secondoftwo}%
\providecommand \href [0]{\begingroup \@sanitize@url \@href}%
\providecommand \@href[1]{\@@startlink{#1}\@@href}%
\providecommand \@@href[1]{\endgroup#1\@@endlink}%
\providecommand \@sanitize@url [0]{\catcode `\\12\catcode `\$12\catcode
  `\&12\catcode `\#12\catcode `\^12\catcode `\_12\catcode `\%12\relax}%
\providecommand \@@startlink[1]{}%
\providecommand \@@endlink[0]{}%
\providecommand \url  [0]{\begingroup\@sanitize@url \@url }%
\providecommand \@url [1]{\endgroup\@href {#1}{\urlprefix }}%
\providecommand \urlprefix  [0]{URL }%
\providecommand \Eprint [0]{\href }%
\providecommand \doibase [0]{http://dx.doi.org/}%
\providecommand \selectlanguage [0]{\@gobble}%
\providecommand \bibinfo  [0]{\@secondoftwo}%
\providecommand \bibfield  [0]{\@secondoftwo}%
\providecommand \translation [1]{[#1]}%
\providecommand \BibitemOpen [0]{}%
\providecommand \bibitemStop [0]{}%
\providecommand \bibitemNoStop [0]{.\EOS\space}%
\providecommand \EOS [0]{\spacefactor3000\relax}%
\providecommand \BibitemShut  [1]{\csname bibitem#1\endcsname}%
\let\auto@bib@innerbib\@empty
%</preamble>
\bibitem [{\citenamefont {Israel}(1967)}]{PhysRev.164.1776}%
  \BibitemOpen
  \bibfield  {author} {\bibinfo {author} {\bibfnamefont {W.}~\bibnamefont
  {Israel}},\ }\href {\doibase 10.1103/PhysRev.164.1776} {\bibfield  {journal}
  {\bibinfo  {journal} {Phys. Rev.}\ }\textbf {\bibinfo {volume} {164}},\
  \bibinfo {pages} {1776} (\bibinfo {year} {1967})}\BibitemShut {NoStop}%
\bibitem [{\citenamefont {Carter}(1971)}]{PhysRevLett.26.331}%
  \BibitemOpen
  \bibfield  {author} {\bibinfo {author} {\bibfnamefont {B.}~\bibnamefont
  {Carter}},\ }\href {\doibase 10.1103/PhysRevLett.26.331} {\bibfield
  {journal} {\bibinfo  {journal} {Phys. Rev. Lett.}\ }\textbf {\bibinfo
  {volume} {26}},\ \bibinfo {pages} {331} (\bibinfo {year} {1971})}\BibitemShut
  {NoStop}%
\bibitem [{\citenamefont {Hansen}(1974)}]{Hansen:1974zz}%
  \BibitemOpen
  \bibfield  {author} {\bibinfo {author} {\bibfnamefont {R.~O.}\ \bibnamefont
  {Hansen}},\ }\href {\doibase 10.1063/1.1666501} {\bibfield  {journal}
  {\bibinfo  {journal} {J. Math. Phys.}\ }\textbf {\bibinfo {volume} {15}},\
  \bibinfo {pages} {46} (\bibinfo {year} {1974})}\BibitemShut {NoStop}%
\bibitem [{\citenamefont {Akiyama}\ \emph {et~al.}(2019)\citenamefont {Akiyama}
  \emph {et~al.}}]{EventHorizonTelescope:2019dse}%
  \BibitemOpen
  \bibfield  {author} {\bibinfo {author} {\bibfnamefont {K.}~\bibnamefont
  {Akiyama}} \emph {et~al.} (\bibinfo {collaboration} {Event Horizon
  Telescope}),\ }\href {\doibase 10.3847/2041-8213/ab0ec7} {\bibfield
  {journal} {\bibinfo  {journal} {Astrophys. J. Lett.}\ }\textbf {\bibinfo
  {volume} {875}},\ \bibinfo {pages} {L1} (\bibinfo {year} {2019})},\ \Eprint
  {http://arxiv.org/abs/1906.11238} {arXiv:1906.11238 [astro-ph.GA]}
  \BibitemShut {NoStop}%
\bibitem [{\citenamefont {Abbott}\ \emph {et~al.}(2016)\citenamefont {Abbott}
  \emph {et~al.}}]{LIGOScientific:2016aoc}%
  \BibitemOpen
  \bibfield  {author} {\bibinfo {author} {\bibfnamefont {B.~P.}\ \bibnamefont
  {Abbott}} \emph {et~al.} (\bibinfo {collaboration} {LIGO Scientific,
  Virgo}),\ }\href {\doibase 10.1103/PhysRevLett.116.061102} {\bibfield
  {journal} {\bibinfo  {journal} {Phys. Rev. Lett.}\ }\textbf {\bibinfo
  {volume} {116}},\ \bibinfo {pages} {061102} (\bibinfo {year} {2016})},\
  \Eprint {http://arxiv.org/abs/1602.03837} {arXiv:1602.03837 [gr-qc]}
  \BibitemShut {NoStop}%
\bibitem [{\citenamefont {Cachazo}\ and\ \citenamefont
  {Guevara}(2020)}]{Cachazo:2017jef}%
  \BibitemOpen
  \bibfield  {author} {\bibinfo {author} {\bibfnamefont {F.}~\bibnamefont
  {Cachazo}}\ and\ \bibinfo {author} {\bibfnamefont {A.}~\bibnamefont
  {Guevara}},\ }\href {\doibase 10.1007/JHEP02(2020)181} {\bibfield  {journal}
  {\bibinfo  {journal} {JHEP}\ }\textbf {\bibinfo {volume} {02}},\ \bibinfo
  {pages} {181} (\bibinfo {year} {2020})},\ \Eprint
  {http://arxiv.org/abs/1705.10262} {arXiv:1705.10262 [hep-th]} \BibitemShut
  {NoStop}%
\bibitem [{\citenamefont {Vines}(2018)}]{Vines:2017hyw}%
  \BibitemOpen
  \bibfield  {author} {\bibinfo {author} {\bibfnamefont {J.}~\bibnamefont
  {Vines}},\ }\href {\doibase 10.1088/1361-6382/aaa3a8} {\bibfield  {journal}
  {\bibinfo  {journal} {Class. Quant. Grav.}\ }\textbf {\bibinfo {volume}
  {35}},\ \bibinfo {pages} {084002} (\bibinfo {year} {2018})},\ \Eprint
  {http://arxiv.org/abs/1709.06016} {arXiv:1709.06016 [gr-qc]} \BibitemShut
  {NoStop}%
\bibitem [{\citenamefont {Guevara}\ \emph
  {et~al.}(2019{\natexlab{a}})\citenamefont {Guevara}, \citenamefont
  {Ochirov},\ and\ \citenamefont {Vines}}]{Guevara:2018wpp}%
  \BibitemOpen
  \bibfield  {author} {\bibinfo {author} {\bibfnamefont {A.}~\bibnamefont
  {Guevara}}, \bibinfo {author} {\bibfnamefont {A.}~\bibnamefont {Ochirov}}, \
  and\ \bibinfo {author} {\bibfnamefont {J.}~\bibnamefont {Vines}},\ }\href
  {\doibase 10.1007/JHEP09(2019)056} {\bibfield  {journal} {\bibinfo  {journal}
  {JHEP}\ }\textbf {\bibinfo {volume} {09}},\ \bibinfo {pages} {056} (\bibinfo
  {year} {2019}{\natexlab{a}})},\ \Eprint {http://arxiv.org/abs/1812.06895}
  {arXiv:1812.06895 [hep-th]} \BibitemShut {NoStop}%
\bibitem [{\citenamefont {Chung}\ \emph {et~al.}(2019)\citenamefont {Chung},
  \citenamefont {Huang}, \citenamefont {Kim},\ and\ \citenamefont
  {Lee}}]{Chung:2018kqs}%
  \BibitemOpen
  \bibfield  {author} {\bibinfo {author} {\bibfnamefont {M.-Z.}\ \bibnamefont
  {Chung}}, \bibinfo {author} {\bibfnamefont {Y.-T.}\ \bibnamefont {Huang}},
  \bibinfo {author} {\bibfnamefont {J.-W.}\ \bibnamefont {Kim}}, \ and\
  \bibinfo {author} {\bibfnamefont {S.}~\bibnamefont {Lee}},\ }\href {\doibase
  10.1007/JHEP04(2019)156} {\bibfield  {journal} {\bibinfo  {journal} {JHEP}\
  }\textbf {\bibinfo {volume} {04}},\ \bibinfo {pages} {156} (\bibinfo {year}
  {2019})},\ \Eprint {http://arxiv.org/abs/1812.08752} {arXiv:1812.08752
  [hep-th]} \BibitemShut {NoStop}%
\bibitem [{\citenamefont {Kosower}\ \emph {et~al.}(2019)\citenamefont
  {Kosower}, \citenamefont {Maybee},\ and\ \citenamefont
  {O'Connell}}]{Kosower:2018adc}%
  \BibitemOpen
  \bibfield  {author} {\bibinfo {author} {\bibfnamefont {D.~A.}\ \bibnamefont
  {Kosower}}, \bibinfo {author} {\bibfnamefont {B.}~\bibnamefont {Maybee}}, \
  and\ \bibinfo {author} {\bibfnamefont {D.}~\bibnamefont {O'Connell}},\ }\href
  {\doibase 10.1007/JHEP02(2019)137} {\bibfield  {journal} {\bibinfo  {journal}
  {JHEP}\ }\textbf {\bibinfo {volume} {02}},\ \bibinfo {pages} {137} (\bibinfo
  {year} {2019})},\ \Eprint {http://arxiv.org/abs/1811.10950} {arXiv:1811.10950
  [hep-th]} \BibitemShut {NoStop}%
\bibitem [{\citenamefont {Vines}\ \emph {et~al.}(2019)\citenamefont {Vines},
  \citenamefont {Steinhoff},\ and\ \citenamefont {Buonanno}}]{Vines:2018gqi}%
  \BibitemOpen
  \bibfield  {author} {\bibinfo {author} {\bibfnamefont {J.}~\bibnamefont
  {Vines}}, \bibinfo {author} {\bibfnamefont {J.}~\bibnamefont {Steinhoff}}, \
  and\ \bibinfo {author} {\bibfnamefont {A.}~\bibnamefont {Buonanno}},\ }\href
  {\doibase 10.1103/PhysRevD.99.064054} {\bibfield  {journal} {\bibinfo
  {journal} {Phys. Rev. D}\ }\textbf {\bibinfo {volume} {99}},\ \bibinfo
  {pages} {064054} (\bibinfo {year} {2019})},\ \Eprint
  {http://arxiv.org/abs/1812.00956} {arXiv:1812.00956 [gr-qc]} \BibitemShut
  {NoStop}%
\bibitem [{\citenamefont {Koemans~Collado}\ \emph {et~al.}(2019)\citenamefont
  {Koemans~Collado}, \citenamefont {Di~Vecchia},\ and\ \citenamefont
  {Russo}}]{KoemansCollado:2019ggb}%
  \BibitemOpen
  \bibfield  {author} {\bibinfo {author} {\bibfnamefont {A.}~\bibnamefont
  {Koemans~Collado}}, \bibinfo {author} {\bibfnamefont {P.}~\bibnamefont
  {Di~Vecchia}}, \ and\ \bibinfo {author} {\bibfnamefont {R.}~\bibnamefont
  {Russo}},\ }\href {\doibase 10.1103/PhysRevD.100.066028} {\bibfield
  {journal} {\bibinfo  {journal} {Phys. Rev. D}\ }\textbf {\bibinfo {volume}
  {100}},\ \bibinfo {pages} {066028} (\bibinfo {year} {2019})},\ \Eprint
  {http://arxiv.org/abs/1904.02667} {arXiv:1904.02667 [hep-th]} \BibitemShut
  {NoStop}%
\bibitem [{\citenamefont {Maybee}\ \emph {et~al.}(2019)\citenamefont {Maybee},
  \citenamefont {O'Connell},\ and\ \citenamefont {Vines}}]{Maybee:2019jus}%
  \BibitemOpen
  \bibfield  {author} {\bibinfo {author} {\bibfnamefont {B.}~\bibnamefont
  {Maybee}}, \bibinfo {author} {\bibfnamefont {D.}~\bibnamefont {O'Connell}}, \
  and\ \bibinfo {author} {\bibfnamefont {J.}~\bibnamefont {Vines}},\ }\href
  {\doibase 10.1007/JHEP12(2019)156} {\bibfield  {journal} {\bibinfo  {journal}
  {JHEP}\ }\textbf {\bibinfo {volume} {12}},\ \bibinfo {pages} {156} (\bibinfo
  {year} {2019})},\ \Eprint {http://arxiv.org/abs/1906.09260} {arXiv:1906.09260
  [hep-th]} \BibitemShut {NoStop}%
\bibitem [{\citenamefont {Guevara}\ \emph
  {et~al.}(2019{\natexlab{b}})\citenamefont {Guevara}, \citenamefont
  {Ochirov},\ and\ \citenamefont {Vines}}]{Guevara:2019fsj}%
  \BibitemOpen
  \bibfield  {author} {\bibinfo {author} {\bibfnamefont {A.}~\bibnamefont
  {Guevara}}, \bibinfo {author} {\bibfnamefont {A.}~\bibnamefont {Ochirov}}, \
  and\ \bibinfo {author} {\bibfnamefont {J.}~\bibnamefont {Vines}},\ }\href
  {\doibase 10.1103/PhysRevD.100.104024} {\bibfield  {journal} {\bibinfo
  {journal} {Phys. Rev. D}\ }\textbf {\bibinfo {volume} {100}},\ \bibinfo
  {pages} {104024} (\bibinfo {year} {2019}{\natexlab{b}})},\ \Eprint
  {http://arxiv.org/abs/1906.10071} {arXiv:1906.10071 [hep-th]} \BibitemShut
  {NoStop}%
\bibitem [{\citenamefont {Siemonsen}\ and\ \citenamefont
  {Vines}(2020)}]{Siemonsen:2019dsu}%
  \BibitemOpen
  \bibfield  {author} {\bibinfo {author} {\bibfnamefont {N.}~\bibnamefont
  {Siemonsen}}\ and\ \bibinfo {author} {\bibfnamefont {J.}~\bibnamefont
  {Vines}},\ }\href {\doibase 10.1103/PhysRevD.101.064066} {\bibfield
  {journal} {\bibinfo  {journal} {Phys. Rev. D}\ }\textbf {\bibinfo {volume}
  {101}},\ \bibinfo {pages} {064066} (\bibinfo {year} {2020})},\ \Eprint
  {http://arxiv.org/abs/1909.07361} {arXiv:1909.07361 [gr-qc]} \BibitemShut
  {NoStop}%
\bibitem [{\citenamefont {Arkani-Hamed}\ \emph {et~al.}(2020)\citenamefont
  {Arkani-Hamed}, \citenamefont {Huang},\ and\ \citenamefont
  {O'Connell}}]{Arkani-Hamed:2019ymq}%
  \BibitemOpen
  \bibfield  {author} {\bibinfo {author} {\bibfnamefont {N.}~\bibnamefont
  {Arkani-Hamed}}, \bibinfo {author} {\bibfnamefont {Y.-t.}\ \bibnamefont
  {Huang}}, \ and\ \bibinfo {author} {\bibfnamefont {D.}~\bibnamefont
  {O'Connell}},\ }\href {\doibase 10.1007/JHEP01(2020)046} {\bibfield
  {journal} {\bibinfo  {journal} {JHEP}\ }\textbf {\bibinfo {volume} {01}},\
  \bibinfo {pages} {046} (\bibinfo {year} {2020})},\ \Eprint
  {http://arxiv.org/abs/1906.10100} {arXiv:1906.10100 [hep-th]} \BibitemShut
  {NoStop}%
\bibitem [{\citenamefont {Damgaard}\ \emph {et~al.}(2019)\citenamefont
  {Damgaard}, \citenamefont {Haddad},\ and\ \citenamefont
  {Helset}}]{Damgaard:2019lfh}%
  \BibitemOpen
  \bibfield  {author} {\bibinfo {author} {\bibfnamefont {P.~H.}\ \bibnamefont
  {Damgaard}}, \bibinfo {author} {\bibfnamefont {K.}~\bibnamefont {Haddad}}, \
  and\ \bibinfo {author} {\bibfnamefont {A.}~\bibnamefont {Helset}},\ }\href
  {\doibase 10.1007/JHEP11(2019)070} {\bibfield  {journal} {\bibinfo  {journal}
  {JHEP}\ }\textbf {\bibinfo {volume} {11}},\ \bibinfo {pages} {070} (\bibinfo
  {year} {2019})},\ \Eprint {http://arxiv.org/abs/1908.10308} {arXiv:1908.10308
  [hep-ph]} \BibitemShut {NoStop}%
\bibitem [{\citenamefont {Aoude}\ \emph {et~al.}(2020)\citenamefont {Aoude},
  \citenamefont {Haddad},\ and\ \citenamefont {Helset}}]{Aoude:2020onz}%
  \BibitemOpen
  \bibfield  {author} {\bibinfo {author} {\bibfnamefont {R.}~\bibnamefont
  {Aoude}}, \bibinfo {author} {\bibfnamefont {K.}~\bibnamefont {Haddad}}, \
  and\ \bibinfo {author} {\bibfnamefont {A.}~\bibnamefont {Helset}},\ }\href
  {\doibase 10.1007/JHEP05(2020)051} {\bibfield  {journal} {\bibinfo  {journal}
  {JHEP}\ }\textbf {\bibinfo {volume} {05}},\ \bibinfo {pages} {051} (\bibinfo
  {year} {2020})},\ \Eprint {http://arxiv.org/abs/2001.09164} {arXiv:2001.09164
  [hep-th]} \BibitemShut {NoStop}%
\bibitem [{\citenamefont {Chung}\ \emph {et~al.}(2020)\citenamefont {Chung},
  \citenamefont {Huang}, \citenamefont {Kim},\ and\ \citenamefont
  {Lee}}]{Chung:2020rrz}%
  \BibitemOpen
  \bibfield  {author} {\bibinfo {author} {\bibfnamefont {M.-Z.}\ \bibnamefont
  {Chung}}, \bibinfo {author} {\bibfnamefont {Y.-t.}\ \bibnamefont {Huang}},
  \bibinfo {author} {\bibfnamefont {J.-W.}\ \bibnamefont {Kim}}, \ and\
  \bibinfo {author} {\bibfnamefont {S.}~\bibnamefont {Lee}},\ }\href {\doibase
  10.1007/JHEP05(2020)105} {\bibfield  {journal} {\bibinfo  {journal} {JHEP}\
  }\textbf {\bibinfo {volume} {05}},\ \bibinfo {pages} {105} (\bibinfo {year}
  {2020})},\ \Eprint {http://arxiv.org/abs/2003.06600} {arXiv:2003.06600
  [hep-th]} \BibitemShut {NoStop}%
\bibitem [{\citenamefont {Guevara}\ \emph {et~al.}(2021)\citenamefont
  {Guevara}, \citenamefont {Maybee}, \citenamefont {Ochirov}, \citenamefont
  {O'connell},\ and\ \citenamefont {Vines}}]{Guevara:2020xjx}%
  \BibitemOpen
  \bibfield  {author} {\bibinfo {author} {\bibfnamefont {A.}~\bibnamefont
  {Guevara}}, \bibinfo {author} {\bibfnamefont {B.}~\bibnamefont {Maybee}},
  \bibinfo {author} {\bibfnamefont {A.}~\bibnamefont {Ochirov}}, \bibinfo
  {author} {\bibfnamefont {D.}~\bibnamefont {O'connell}}, \ and\ \bibinfo
  {author} {\bibfnamefont {J.}~\bibnamefont {Vines}},\ }\href {\doibase
  10.1007/JHEP03(2021)201} {\bibfield  {journal} {\bibinfo  {journal} {JHEP}\
  }\textbf {\bibinfo {volume} {03}},\ \bibinfo {pages} {201} (\bibinfo {year}
  {2021})},\ \Eprint {http://arxiv.org/abs/2012.11570} {arXiv:2012.11570
  [hep-th]} \BibitemShut {NoStop}%
\bibitem [{\citenamefont {Kosmopoulos}\ and\ \citenamefont
  {Luna}(2021)}]{Kosmopoulos:2021zoq}%
  \BibitemOpen
  \bibfield  {author} {\bibinfo {author} {\bibfnamefont {D.}~\bibnamefont
  {Kosmopoulos}}\ and\ \bibinfo {author} {\bibfnamefont {A.}~\bibnamefont
  {Luna}},\ }\href {\doibase 10.1007/JHEP07(2021)037} {\bibfield  {journal}
  {\bibinfo  {journal} {JHEP}\ }\textbf {\bibinfo {volume} {07}},\ \bibinfo
  {pages} {037} (\bibinfo {year} {2021})},\ \Eprint
  {http://arxiv.org/abs/2102.10137} {arXiv:2102.10137 [hep-th]} \BibitemShut
  {NoStop}%
\bibitem [{\citenamefont {Bautista}\ \emph {et~al.}(2021)\citenamefont
  {Bautista}, \citenamefont {Guevara}, \citenamefont {Kavanagh},\ and\
  \citenamefont {Vines}}]{Bautista:2021wfy}%
  \BibitemOpen
  \bibfield  {author} {\bibinfo {author} {\bibfnamefont {Y.~F.}\ \bibnamefont
  {Bautista}}, \bibinfo {author} {\bibfnamefont {A.}~\bibnamefont {Guevara}},
  \bibinfo {author} {\bibfnamefont {C.}~\bibnamefont {Kavanagh}}, \ and\
  \bibinfo {author} {\bibfnamefont {J.}~\bibnamefont {Vines}},\ }\href@noop {}
  {\  (\bibinfo {year} {2021})},\ \Eprint {http://arxiv.org/abs/2107.10179}
  {arXiv:2107.10179 [hep-th]} \BibitemShut {NoStop}%
\bibitem [{\citenamefont {Haddad}(2022)}]{Haddad:2021znf}%
  \BibitemOpen
  \bibfield  {author} {\bibinfo {author} {\bibfnamefont {K.}~\bibnamefont
  {Haddad}},\ }\href {\doibase 10.1103/PhysRevD.105.026004} {\bibfield
  {journal} {\bibinfo  {journal} {Phys. Rev. D}\ }\textbf {\bibinfo {volume}
  {105}},\ \bibinfo {pages} {026004} (\bibinfo {year} {2022})},\ \Eprint
  {http://arxiv.org/abs/2109.04427} {arXiv:2109.04427 [hep-th]} \BibitemShut
  {NoStop}%
\bibitem [{\citenamefont {Herrmann}\ \emph
  {et~al.}(2021{\natexlab{a}})\citenamefont {Herrmann}, \citenamefont
  {Parra-Martinez}, \citenamefont {Ruf},\ and\ \citenamefont
  {Zeng}}]{Herrmann:2021lqe}%
  \BibitemOpen
  \bibfield  {author} {\bibinfo {author} {\bibfnamefont {E.}~\bibnamefont
  {Herrmann}}, \bibinfo {author} {\bibfnamefont {J.}~\bibnamefont
  {Parra-Martinez}}, \bibinfo {author} {\bibfnamefont {M.~S.}\ \bibnamefont
  {Ruf}}, \ and\ \bibinfo {author} {\bibfnamefont {M.}~\bibnamefont {Zeng}},\
  }\href {\doibase 10.1103/PhysRevLett.126.201602} {\bibfield  {journal}
  {\bibinfo  {journal} {Phys. Rev. Lett.}\ }\textbf {\bibinfo {volume} {126}},\
  \bibinfo {pages} {201602} (\bibinfo {year} {2021}{\natexlab{a}})},\ \Eprint
  {http://arxiv.org/abs/2101.07255} {arXiv:2101.07255 [hep-th]} \BibitemShut
  {NoStop}%
\bibitem [{\citenamefont {Herrmann}\ \emph
  {et~al.}(2021{\natexlab{b}})\citenamefont {Herrmann}, \citenamefont
  {Parra-Martinez}, \citenamefont {Ruf},\ and\ \citenamefont
  {Zeng}}]{Herrmann:2021tct}%
  \BibitemOpen
  \bibfield  {author} {\bibinfo {author} {\bibfnamefont {E.}~\bibnamefont
  {Herrmann}}, \bibinfo {author} {\bibfnamefont {J.}~\bibnamefont
  {Parra-Martinez}}, \bibinfo {author} {\bibfnamefont {M.~S.}\ \bibnamefont
  {Ruf}}, \ and\ \bibinfo {author} {\bibfnamefont {M.}~\bibnamefont {Zeng}},\
  }\href {\doibase 10.1007/JHEP10(2021)148} {\bibfield  {journal} {\bibinfo
  {journal} {JHEP}\ }\textbf {\bibinfo {volume} {10}},\ \bibinfo {pages} {148}
  (\bibinfo {year} {2021}{\natexlab{b}})},\ \Eprint
  {http://arxiv.org/abs/2104.03957} {arXiv:2104.03957 [hep-th]} \BibitemShut
  {NoStop}%
\bibitem [{\citenamefont {Chiodaroli}\ \emph {et~al.}(2022)\citenamefont
  {Chiodaroli}, \citenamefont {Johansson},\ and\ \citenamefont
  {Pichini}}]{Chiodaroli:2021eug}%
  \BibitemOpen
  \bibfield  {author} {\bibinfo {author} {\bibfnamefont {M.}~\bibnamefont
  {Chiodaroli}}, \bibinfo {author} {\bibfnamefont {H.}~\bibnamefont
  {Johansson}}, \ and\ \bibinfo {author} {\bibfnamefont {P.}~\bibnamefont
  {Pichini}},\ }\href {\doibase 10.1007/JHEP02(2022)156} {\bibfield  {journal}
  {\bibinfo  {journal} {JHEP}\ }\textbf {\bibinfo {volume} {02}},\ \bibinfo
  {pages} {156} (\bibinfo {year} {2022})},\ \Eprint
  {http://arxiv.org/abs/2107.14779} {arXiv:2107.14779 [hep-th]} \BibitemShut
  {NoStop}%
\bibitem [{\citenamefont {Brandhuber}\ \emph {et~al.}(2021)\citenamefont
  {Brandhuber}, \citenamefont {Chen}, \citenamefont {Travaglini},\ and\
  \citenamefont {Wen}}]{Brandhuber:2021eyq}%
  \BibitemOpen
  \bibfield  {author} {\bibinfo {author} {\bibfnamefont {A.}~\bibnamefont
  {Brandhuber}}, \bibinfo {author} {\bibfnamefont {G.}~\bibnamefont {Chen}},
  \bibinfo {author} {\bibfnamefont {G.}~\bibnamefont {Travaglini}}, \ and\
  \bibinfo {author} {\bibfnamefont {C.}~\bibnamefont {Wen}},\ }\href {\doibase
  10.1007/JHEP10(2021)118} {\bibfield  {journal} {\bibinfo  {journal} {JHEP}\
  }\textbf {\bibinfo {volume} {10}},\ \bibinfo {pages} {118} (\bibinfo {year}
  {2021})},\ \Eprint {http://arxiv.org/abs/2108.04216} {arXiv:2108.04216
  [hep-th]} \BibitemShut {NoStop}%
\bibitem [{\citenamefont {Alessio}\ and\ \citenamefont
  {Di~Vecchia}(2022)}]{Alessio:2022kwv}%
  \BibitemOpen
  \bibfield  {author} {\bibinfo {author} {\bibfnamefont {F.}~\bibnamefont
  {Alessio}}\ and\ \bibinfo {author} {\bibfnamefont {P.}~\bibnamefont
  {Di~Vecchia}},\ }\href {\doibase 10.1016/j.physletb.2022.137258} {\bibfield
  {journal} {\bibinfo  {journal} {Phys. Lett. B}\ }\textbf {\bibinfo {volume}
  {832}},\ \bibinfo {pages} {137258} (\bibinfo {year} {2022})},\ \Eprint
  {http://arxiv.org/abs/2203.13272} {arXiv:2203.13272 [hep-th]} \BibitemShut
  {NoStop}%
\bibitem [{\citenamefont {Aoude}\ \emph
  {et~al.}(2022{\natexlab{a}})\citenamefont {Aoude}, \citenamefont {Haddad},\
  and\ \citenamefont {Helset}}]{Aoude:2022trd}%
  \BibitemOpen
  \bibfield  {author} {\bibinfo {author} {\bibfnamefont {R.}~\bibnamefont
  {Aoude}}, \bibinfo {author} {\bibfnamefont {K.}~\bibnamefont {Haddad}}, \
  and\ \bibinfo {author} {\bibfnamefont {A.}~\bibnamefont {Helset}},\ }\href
  {\doibase 10.1007/JHEP07(2022)072} {\bibfield  {journal} {\bibinfo  {journal}
  {JHEP}\ }\textbf {\bibinfo {volume} {07}},\ \bibinfo {pages} {072} (\bibinfo
  {year} {2022}{\natexlab{a}})},\ \Eprint {http://arxiv.org/abs/2203.06197}
  {arXiv:2203.06197 [hep-th]} \BibitemShut {NoStop}%
\bibitem [{\citenamefont {Aoude}\ \emph
  {et~al.}(2022{\natexlab{b}})\citenamefont {Aoude}, \citenamefont {Haddad},\
  and\ \citenamefont {Helset}}]{Aoude:2022thd}%
  \BibitemOpen
  \bibfield  {author} {\bibinfo {author} {\bibfnamefont {R.}~\bibnamefont
  {Aoude}}, \bibinfo {author} {\bibfnamefont {K.}~\bibnamefont {Haddad}}, \
  and\ \bibinfo {author} {\bibfnamefont {A.}~\bibnamefont {Helset}},\ }\href
  {\doibase 10.1103/PhysRevLett.129.141102} {\bibfield  {journal} {\bibinfo
  {journal} {Phys. Rev. Lett.}\ }\textbf {\bibinfo {volume} {129}},\ \bibinfo
  {pages} {141102} (\bibinfo {year} {2022}{\natexlab{b}})},\ \Eprint
  {http://arxiv.org/abs/2205.02809} {arXiv:2205.02809 [hep-th]} \BibitemShut
  {NoStop}%
\bibitem [{\citenamefont {Febres~Cordero}\ \emph {et~al.}(2023)\citenamefont
  {Febres~Cordero}, \citenamefont {Kraus}, \citenamefont {Lin}, \citenamefont
  {Ruf},\ and\ \citenamefont {Zeng}}]{FebresCordero:2022jts}%
  \BibitemOpen
  \bibfield  {author} {\bibinfo {author} {\bibfnamefont {F.}~\bibnamefont
  {Febres~Cordero}}, \bibinfo {author} {\bibfnamefont {M.}~\bibnamefont
  {Kraus}}, \bibinfo {author} {\bibfnamefont {G.}~\bibnamefont {Lin}}, \bibinfo
  {author} {\bibfnamefont {M.~S.}\ \bibnamefont {Ruf}}, \ and\ \bibinfo
  {author} {\bibfnamefont {M.}~\bibnamefont {Zeng}},\ }\href {\doibase
  10.1103/PhysRevLett.130.021601} {\bibfield  {journal} {\bibinfo  {journal}
  {Phys. Rev. Lett.}\ }\textbf {\bibinfo {volume} {130}},\ \bibinfo {pages}
  {021601} (\bibinfo {year} {2023})},\ \Eprint
  {http://arxiv.org/abs/2205.07357} {arXiv:2205.07357 [hep-th]} \BibitemShut
  {NoStop}%
\bibitem [{\citenamefont {Bellazzini}\ \emph {et~al.}(2022)\citenamefont
  {Bellazzini}, \citenamefont {Isabella},\ and\ \citenamefont
  {Riva}}]{Bellazzini:2022wzv}%
  \BibitemOpen
  \bibfield  {author} {\bibinfo {author} {\bibfnamefont {B.}~\bibnamefont
  {Bellazzini}}, \bibinfo {author} {\bibfnamefont {G.}~\bibnamefont
  {Isabella}}, \ and\ \bibinfo {author} {\bibfnamefont {M.~M.}\ \bibnamefont
  {Riva}},\ }\href@noop {} {\  (\bibinfo {year} {2022})},\ \Eprint
  {http://arxiv.org/abs/2211.00085} {arXiv:2211.00085 [hep-th]} \BibitemShut
  {NoStop}%
\bibitem [{\citenamefont {Bautista}\ \emph {et~al.}(2022)\citenamefont
  {Bautista}, \citenamefont {Guevara}, \citenamefont {Kavanagh},\ and\
  \citenamefont {Vinese}}]{Bautista:2022wjf}%
  \BibitemOpen
  \bibfield  {author} {\bibinfo {author} {\bibfnamefont {Y.~F.}\ \bibnamefont
  {Bautista}}, \bibinfo {author} {\bibfnamefont {A.}~\bibnamefont {Guevara}},
  \bibinfo {author} {\bibfnamefont {C.}~\bibnamefont {Kavanagh}}, \ and\
  \bibinfo {author} {\bibfnamefont {J.}~\bibnamefont {Vinese}},\ }\href@noop {}
  {\  (\bibinfo {year} {2022})},\ \Eprint {http://arxiv.org/abs/2212.07965}
  {arXiv:2212.07965 [hep-th]} \BibitemShut {NoStop}%
\bibitem [{\citenamefont {Cangemi}\ \emph {et~al.}(2022)\citenamefont
  {Cangemi}, \citenamefont {Chiodaroli}, \citenamefont {Johansson},
  \citenamefont {Ochirov}, \citenamefont {Pichini},\ and\ \citenamefont
  {Skvortsov}}]{Cangemi:2022bew}%
  \BibitemOpen
  \bibfield  {author} {\bibinfo {author} {\bibfnamefont {L.}~\bibnamefont
  {Cangemi}}, \bibinfo {author} {\bibfnamefont {M.}~\bibnamefont {Chiodaroli}},
  \bibinfo {author} {\bibfnamefont {H.}~\bibnamefont {Johansson}}, \bibinfo
  {author} {\bibfnamefont {A.}~\bibnamefont {Ochirov}}, \bibinfo {author}
  {\bibfnamefont {P.}~\bibnamefont {Pichini}}, \ and\ \bibinfo {author}
  {\bibfnamefont {E.}~\bibnamefont {Skvortsov}},\ }\href@noop {} {\  (\bibinfo
  {year} {2022})},\ \Eprint {http://arxiv.org/abs/2212.06120} {arXiv:2212.06120
  [hep-th]} \BibitemShut {NoStop}%
\bibitem [{\citenamefont {Bjerrum-Bohr}\ \emph {et~al.}(2023)\citenamefont
  {Bjerrum-Bohr}, \citenamefont {Chen},\ and\ \citenamefont
  {Skowronek}}]{Bjerrum-Bohr:2023jau}%
  \BibitemOpen
  \bibfield  {author} {\bibinfo {author} {\bibfnamefont {N.~E.~J.}\
  \bibnamefont {Bjerrum-Bohr}}, \bibinfo {author} {\bibfnamefont
  {G.}~\bibnamefont {Chen}}, \ and\ \bibinfo {author} {\bibfnamefont
  {M.}~\bibnamefont {Skowronek}},\ }\href@noop {} {\  (\bibinfo {year}
  {2023})},\ \Eprint {http://arxiv.org/abs/2302.00498} {arXiv:2302.00498
  [hep-th]} \BibitemShut {NoStop}%
\bibitem [{\citenamefont {Brandhuber}\ \emph {et~al.}(2023)\citenamefont
  {Brandhuber}, \citenamefont {Brown}, \citenamefont {Chen}, \citenamefont
  {De~Angelis}, \citenamefont {Gowdy},\ and\ \citenamefont
  {Travaglini}}]{Brandhuber:2023hhy}%
  \BibitemOpen
  \bibfield  {author} {\bibinfo {author} {\bibfnamefont {A.}~\bibnamefont
  {Brandhuber}}, \bibinfo {author} {\bibfnamefont {G.~R.}\ \bibnamefont
  {Brown}}, \bibinfo {author} {\bibfnamefont {G.}~\bibnamefont {Chen}},
  \bibinfo {author} {\bibfnamefont {S.}~\bibnamefont {De~Angelis}}, \bibinfo
  {author} {\bibfnamefont {J.}~\bibnamefont {Gowdy}}, \ and\ \bibinfo {author}
  {\bibfnamefont {G.}~\bibnamefont {Travaglini}},\ }\href@noop {} {\  (\bibinfo
  {year} {2023})},\ \Eprint {http://arxiv.org/abs/2303.06111} {arXiv:2303.06111
  [hep-th]} \BibitemShut {NoStop}%
\bibitem [{\citenamefont {Herderschee}\ \emph {et~al.}(2023)\citenamefont
  {Herderschee}, \citenamefont {Roiban},\ and\ \citenamefont
  {Teng}}]{Herderschee:2023fxh}%
  \BibitemOpen
  \bibfield  {author} {\bibinfo {author} {\bibfnamefont {A.}~\bibnamefont
  {Herderschee}}, \bibinfo {author} {\bibfnamefont {R.}~\bibnamefont {Roiban}},
  \ and\ \bibinfo {author} {\bibfnamefont {F.}~\bibnamefont {Teng}},\
  }\href@noop {} {\  (\bibinfo {year} {2023})},\ \Eprint
  {http://arxiv.org/abs/2303.06112} {arXiv:2303.06112 [hep-th]} \BibitemShut
  {NoStop}%
\bibitem [{\citenamefont {Elkhidir}\ \emph {et~al.}(2023)\citenamefont
  {Elkhidir}, \citenamefont {O'Connell}, \citenamefont {Sergola},\ and\
  \citenamefont {Vazquez-Holm}}]{Elkhidir:2023dco}%
  \BibitemOpen
  \bibfield  {author} {\bibinfo {author} {\bibfnamefont {A.}~\bibnamefont
  {Elkhidir}}, \bibinfo {author} {\bibfnamefont {D.}~\bibnamefont {O'Connell}},
  \bibinfo {author} {\bibfnamefont {M.}~\bibnamefont {Sergola}}, \ and\
  \bibinfo {author} {\bibfnamefont {I.~A.}\ \bibnamefont {Vazquez-Holm}},\
  }\href@noop {} {\  (\bibinfo {year} {2023})},\ \Eprint
  {http://arxiv.org/abs/2303.06211} {arXiv:2303.06211 [hep-th]} \BibitemShut
  {NoStop}%
\bibitem [{\citenamefont {Georgoudis}\ \emph {et~al.}(2023)\citenamefont
  {Georgoudis}, \citenamefont {Heissenberg},\ and\ \citenamefont
  {Vazquez-Holm}}]{Georgoudis:2023lgf}%
  \BibitemOpen
  \bibfield  {author} {\bibinfo {author} {\bibfnamefont {A.}~\bibnamefont
  {Georgoudis}}, \bibinfo {author} {\bibfnamefont {C.}~\bibnamefont
  {Heissenberg}}, \ and\ \bibinfo {author} {\bibfnamefont {I.}~\bibnamefont
  {Vazquez-Holm}},\ }\href@noop {} {\  (\bibinfo {year} {2023})},\ \Eprint
  {http://arxiv.org/abs/2303.07006} {arXiv:2303.07006 [hep-th]} \BibitemShut
  {NoStop}%
\bibitem [{\citenamefont {Cheung}\ \emph {et~al.}(2018)\citenamefont {Cheung},
  \citenamefont {Rothstein},\ and\ \citenamefont {Solon}}]{Cheung:2018wkq}%
  \BibitemOpen
  \bibfield  {author} {\bibinfo {author} {\bibfnamefont {C.}~\bibnamefont
  {Cheung}}, \bibinfo {author} {\bibfnamefont {I.~Z.}\ \bibnamefont
  {Rothstein}}, \ and\ \bibinfo {author} {\bibfnamefont {M.~P.}\ \bibnamefont
  {Solon}},\ }\href {\doibase 10.1103/PhysRevLett.121.251101} {\bibfield
  {journal} {\bibinfo  {journal} {Phys. Rev. Lett.}\ }\textbf {\bibinfo
  {volume} {121}},\ \bibinfo {pages} {251101} (\bibinfo {year} {2018})},\
  \Eprint {http://arxiv.org/abs/1808.02489} {arXiv:1808.02489 [hep-th]}
  \BibitemShut {NoStop}%
\bibitem [{\citenamefont {Bern}\ \emph
  {et~al.}(2019{\natexlab{a}})\citenamefont {Bern}, \citenamefont {Cheung},
  \citenamefont {Roiban}, \citenamefont {Shen}, \citenamefont {Solon},\ and\
  \citenamefont {Zeng}}]{Bern:2019nnu}%
  \BibitemOpen
  \bibfield  {author} {\bibinfo {author} {\bibfnamefont {Z.}~\bibnamefont
  {Bern}}, \bibinfo {author} {\bibfnamefont {C.}~\bibnamefont {Cheung}},
  \bibinfo {author} {\bibfnamefont {R.}~\bibnamefont {Roiban}}, \bibinfo
  {author} {\bibfnamefont {C.-H.}\ \bibnamefont {Shen}}, \bibinfo {author}
  {\bibfnamefont {M.~P.}\ \bibnamefont {Solon}}, \ and\ \bibinfo {author}
  {\bibfnamefont {M.}~\bibnamefont {Zeng}},\ }\href {\doibase
  10.1103/PhysRevLett.122.201603} {\bibfield  {journal} {\bibinfo  {journal}
  {Phys. Rev. Lett.}\ }\textbf {\bibinfo {volume} {122}},\ \bibinfo {pages}
  {201603} (\bibinfo {year} {2019}{\natexlab{a}})},\ \Eprint
  {http://arxiv.org/abs/1901.04424} {arXiv:1901.04424 [hep-th]} \BibitemShut
  {NoStop}%
\bibitem [{\citenamefont {Cristofoli}\ \emph {et~al.}(2019)\citenamefont
  {Cristofoli}, \citenamefont {Bjerrum-Bohr}, \citenamefont {Damgaard},\ and\
  \citenamefont {Vanhove}}]{Cristofoli:2019neg}%
  \BibitemOpen
  \bibfield  {author} {\bibinfo {author} {\bibfnamefont {A.}~\bibnamefont
  {Cristofoli}}, \bibinfo {author} {\bibfnamefont {N.~E.~J.}\ \bibnamefont
  {Bjerrum-Bohr}}, \bibinfo {author} {\bibfnamefont {P.~H.}\ \bibnamefont
  {Damgaard}}, \ and\ \bibinfo {author} {\bibfnamefont {P.}~\bibnamefont
  {Vanhove}},\ }\href {\doibase 10.1103/PhysRevD.100.084040} {\bibfield
  {journal} {\bibinfo  {journal} {Phys. Rev. D}\ }\textbf {\bibinfo {volume}
  {100}},\ \bibinfo {pages} {084040} (\bibinfo {year} {2019})},\ \Eprint
  {http://arxiv.org/abs/1906.01579} {arXiv:1906.01579 [hep-th]} \BibitemShut
  {NoStop}%
\bibitem [{\citenamefont {Bern}\ \emph
  {et~al.}(2019{\natexlab{b}})\citenamefont {Bern}, \citenamefont {Cheung},
  \citenamefont {Roiban}, \citenamefont {Shen}, \citenamefont {Solon},\ and\
  \citenamefont {Zeng}}]{Bern:2019crd}%
  \BibitemOpen
  \bibfield  {author} {\bibinfo {author} {\bibfnamefont {Z.}~\bibnamefont
  {Bern}}, \bibinfo {author} {\bibfnamefont {C.}~\bibnamefont {Cheung}},
  \bibinfo {author} {\bibfnamefont {R.}~\bibnamefont {Roiban}}, \bibinfo
  {author} {\bibfnamefont {C.-H.}\ \bibnamefont {Shen}}, \bibinfo {author}
  {\bibfnamefont {M.~P.}\ \bibnamefont {Solon}}, \ and\ \bibinfo {author}
  {\bibfnamefont {M.}~\bibnamefont {Zeng}},\ }\href {\doibase
  10.1007/JHEP10(2019)206} {\bibfield  {journal} {\bibinfo  {journal} {JHEP}\
  }\textbf {\bibinfo {volume} {10}},\ \bibinfo {pages} {206} (\bibinfo {year}
  {2019}{\natexlab{b}})},\ \Eprint {http://arxiv.org/abs/1908.01493}
  {arXiv:1908.01493 [hep-th]} \BibitemShut {NoStop}%
\bibitem [{\citenamefont {Bern}\ \emph {et~al.}(2021)\citenamefont {Bern},
  \citenamefont {Luna}, \citenamefont {Roiban}, \citenamefont {Shen},\ and\
  \citenamefont {Zeng}}]{Bern:2020buy}%
  \BibitemOpen
  \bibfield  {author} {\bibinfo {author} {\bibfnamefont {Z.}~\bibnamefont
  {Bern}}, \bibinfo {author} {\bibfnamefont {A.}~\bibnamefont {Luna}}, \bibinfo
  {author} {\bibfnamefont {R.}~\bibnamefont {Roiban}}, \bibinfo {author}
  {\bibfnamefont {C.-H.}\ \bibnamefont {Shen}}, \ and\ \bibinfo {author}
  {\bibfnamefont {M.}~\bibnamefont {Zeng}},\ }\href {\doibase
  10.1103/PhysRevD.104.065014} {\bibfield  {journal} {\bibinfo  {journal}
  {Phys. Rev. D}\ }\textbf {\bibinfo {volume} {104}},\ \bibinfo {pages}
  {065014} (\bibinfo {year} {2021})},\ \Eprint
  {http://arxiv.org/abs/2005.03071} {arXiv:2005.03071 [hep-th]} \BibitemShut
  {NoStop}%
\bibitem [{\citenamefont {Chen}\ \emph {et~al.}(2022)\citenamefont {Chen},
  \citenamefont {Chung}, \citenamefont {Huang},\ and\ \citenamefont
  {Kim}}]{Chen:2021kxt}%
  \BibitemOpen
  \bibfield  {author} {\bibinfo {author} {\bibfnamefont {W.-M.}\ \bibnamefont
  {Chen}}, \bibinfo {author} {\bibfnamefont {M.-Z.}\ \bibnamefont {Chung}},
  \bibinfo {author} {\bibfnamefont {Y.-t.}\ \bibnamefont {Huang}}, \ and\
  \bibinfo {author} {\bibfnamefont {J.-W.}\ \bibnamefont {Kim}},\ }\href
  {\doibase 10.1007/JHEP08(2022)148} {\bibfield  {journal} {\bibinfo  {journal}
  {JHEP}\ }\textbf {\bibinfo {volume} {08}},\ \bibinfo {pages} {148} (\bibinfo
  {year} {2022})},\ \Eprint {http://arxiv.org/abs/2111.13639} {arXiv:2111.13639
  [hep-th]} \BibitemShut {NoStop}%
\bibitem [{\citenamefont {Bern}\ \emph
  {et~al.}(2022{\natexlab{a}})\citenamefont {Bern}, \citenamefont
  {Parra-Martinez}, \citenamefont {Roiban}, \citenamefont {Ruf}, \citenamefont
  {Shen}, \citenamefont {Solon},\ and\ \citenamefont {Zeng}}]{Bern:2021yeh}%
  \BibitemOpen
  \bibfield  {author} {\bibinfo {author} {\bibfnamefont {Z.}~\bibnamefont
  {Bern}}, \bibinfo {author} {\bibfnamefont {J.}~\bibnamefont
  {Parra-Martinez}}, \bibinfo {author} {\bibfnamefont {R.}~\bibnamefont
  {Roiban}}, \bibinfo {author} {\bibfnamefont {M.~S.}\ \bibnamefont {Ruf}},
  \bibinfo {author} {\bibfnamefont {C.-H.}\ \bibnamefont {Shen}}, \bibinfo
  {author} {\bibfnamefont {M.~P.}\ \bibnamefont {Solon}}, \ and\ \bibinfo
  {author} {\bibfnamefont {M.}~\bibnamefont {Zeng}},\ }\href {\doibase
  10.1103/PhysRevLett.128.161103} {\bibfield  {journal} {\bibinfo  {journal}
  {Phys. Rev. Lett.}\ }\textbf {\bibinfo {volume} {128}},\ \bibinfo {pages}
  {161103} (\bibinfo {year} {2022}{\natexlab{a}})},\ \Eprint
  {http://arxiv.org/abs/2112.10750} {arXiv:2112.10750 [hep-th]} \BibitemShut
  {NoStop}%
\bibitem [{\citenamefont {Bern}\ \emph
  {et~al.}(2022{\natexlab{b}})\citenamefont {Bern}, \citenamefont
  {Kosmopoulos}, \citenamefont {Luna}, \citenamefont {Roiban},\ and\
  \citenamefont {Teng}}]{Bern:2022kto}%
  \BibitemOpen
  \bibfield  {author} {\bibinfo {author} {\bibfnamefont {Z.}~\bibnamefont
  {Bern}}, \bibinfo {author} {\bibfnamefont {D.}~\bibnamefont {Kosmopoulos}},
  \bibinfo {author} {\bibfnamefont {A.}~\bibnamefont {Luna}}, \bibinfo {author}
  {\bibfnamefont {R.}~\bibnamefont {Roiban}}, \ and\ \bibinfo {author}
  {\bibfnamefont {F.}~\bibnamefont {Teng}},\ }\href@noop {} {\  (\bibinfo
  {year} {2022}{\natexlab{b}})},\ \Eprint {http://arxiv.org/abs/2203.06202}
  {arXiv:2203.06202 [hep-th]} \BibitemShut {NoStop}%
\bibitem [{\citenamefont {Bautista}(2023)}]{Bautista:2023szu}%
  \BibitemOpen
  \bibfield  {author} {\bibinfo {author} {\bibfnamefont {Y.~F.}\ \bibnamefont
  {Bautista}},\ }\href@noop {} {\  (\bibinfo {year} {2023})},\ \Eprint
  {http://arxiv.org/abs/2304.04287} {arXiv:2304.04287 [hep-th]} \BibitemShut
  {NoStop}%
\bibitem [{\citenamefont {Barack}\ \emph {et~al.}(2023)\citenamefont {Barack}
  \emph {et~al.}}]{Barack:2023oqp}%
  \BibitemOpen
  \bibfield  {author} {\bibinfo {author} {\bibfnamefont {L.}~\bibnamefont
  {Barack}} \emph {et~al.},\ }\href@noop {} {\  (\bibinfo {year} {2023})},\
  \Eprint {http://arxiv.org/abs/2304.09200} {arXiv:2304.09200 [hep-th]}
  \BibitemShut {NoStop}%
\bibitem [{\citenamefont {Arkani-Hamed}\ \emph {et~al.}(2021)\citenamefont
  {Arkani-Hamed}, \citenamefont {Huang},\ and\ \citenamefont
  {Huang}}]{Arkani-Hamed:2017jhn}%
  \BibitemOpen
  \bibfield  {author} {\bibinfo {author} {\bibfnamefont {N.}~\bibnamefont
  {Arkani-Hamed}}, \bibinfo {author} {\bibfnamefont {T.-C.}\ \bibnamefont
  {Huang}}, \ and\ \bibinfo {author} {\bibfnamefont {Y.-t.}\ \bibnamefont
  {Huang}},\ }\href {\doibase 10.1007/JHEP11(2021)070} {\bibfield  {journal}
  {\bibinfo  {journal} {JHEP}\ }\textbf {\bibinfo {volume} {11}},\ \bibinfo
  {pages} {070} (\bibinfo {year} {2021})},\ \Eprint
  {http://arxiv.org/abs/1709.04891} {arXiv:1709.04891 [hep-th]} \BibitemShut
  {NoStop}%
\bibitem [{\citenamefont {Britto}\ \emph
  {et~al.}(2005{\natexlab{a}})\citenamefont {Britto}, \citenamefont {Cachazo},\
  and\ \citenamefont {Feng}}]{Britto:2004ap}%
  \BibitemOpen
  \bibfield  {author} {\bibinfo {author} {\bibfnamefont {R.}~\bibnamefont
  {Britto}}, \bibinfo {author} {\bibfnamefont {F.}~\bibnamefont {Cachazo}}, \
  and\ \bibinfo {author} {\bibfnamefont {B.}~\bibnamefont {Feng}},\ }\href
  {\doibase 10.1016/j.nuclphysb.2005.02.030} {\bibfield  {journal} {\bibinfo
  {journal} {Nucl. Phys. B}\ }\textbf {\bibinfo {volume} {715}},\ \bibinfo
  {pages} {499} (\bibinfo {year} {2005}{\natexlab{a}})},\ \Eprint
  {http://arxiv.org/abs/hep-th/0412308} {arXiv:hep-th/0412308} \BibitemShut
  {NoStop}%
\bibitem [{\citenamefont {Britto}\ \emph
  {et~al.}(2005{\natexlab{b}})\citenamefont {Britto}, \citenamefont {Cachazo},
  \citenamefont {Feng},\ and\ \citenamefont {Witten}}]{Britto:2005fq}%
  \BibitemOpen
  \bibfield  {author} {\bibinfo {author} {\bibfnamefont {R.}~\bibnamefont
  {Britto}}, \bibinfo {author} {\bibfnamefont {F.}~\bibnamefont {Cachazo}},
  \bibinfo {author} {\bibfnamefont {B.}~\bibnamefont {Feng}}, \ and\ \bibinfo
  {author} {\bibfnamefont {E.}~\bibnamefont {Witten}},\ }\href {\doibase
  10.1103/PhysRevLett.94.181602} {\bibfield  {journal} {\bibinfo  {journal}
  {Phys. Rev. Lett.}\ }\textbf {\bibinfo {volume} {94}},\ \bibinfo {pages}
  {181602} (\bibinfo {year} {2005}{\natexlab{b}})},\ \Eprint
  {http://arxiv.org/abs/hep-th/0501052} {arXiv:hep-th/0501052} \BibitemShut
  {NoStop}%
\bibitem [{\citenamefont {Falkowski}\ and\ \citenamefont
  {Machado}(2021)}]{Falkowski:2020aso}%
  \BibitemOpen
  \bibfield  {author} {\bibinfo {author} {\bibfnamefont {A.}~\bibnamefont
  {Falkowski}}\ and\ \bibinfo {author} {\bibfnamefont {C.~S.}\ \bibnamefont
  {Machado}},\ }\href {\doibase 10.1007/JHEP05(2021)238} {\bibfield  {journal}
  {\bibinfo  {journal} {JHEP}\ }\textbf {\bibinfo {volume} {05}},\ \bibinfo
  {pages} {238} (\bibinfo {year} {2021})},\ \Eprint
  {http://arxiv.org/abs/2005.08981} {arXiv:2005.08981 [hep-th]} \BibitemShut
  {NoStop}%
\bibitem [{\citenamefont {Haddad}(2023)}]{Haddad:2023ylx}%
  \BibitemOpen
  \bibfield  {author} {\bibinfo {author} {\bibfnamefont {K.}~\bibnamefont
  {Haddad}},\ }\href@noop {} {\  (\bibinfo {year} {2023})},\ \Eprint
  {http://arxiv.org/abs/2303.02624} {arXiv:2303.02624 [hep-th]} \BibitemShut
  {NoStop}%
\bibitem [{\citenamefont {Johansson}\ and\ \citenamefont
  {Ochirov}(2019)}]{Johansson:2019dnu}%
  \BibitemOpen
  \bibfield  {author} {\bibinfo {author} {\bibfnamefont {H.}~\bibnamefont
  {Johansson}}\ and\ \bibinfo {author} {\bibfnamefont {A.}~\bibnamefont
  {Ochirov}},\ }\href {\doibase 10.1007/JHEP09(2019)040} {\bibfield  {journal}
  {\bibinfo  {journal} {JHEP}\ }\textbf {\bibinfo {volume} {09}},\ \bibinfo
  {pages} {040} (\bibinfo {year} {2019})},\ \Eprint
  {http://arxiv.org/abs/1906.12292} {arXiv:1906.12292 [hep-th]} \BibitemShut
  {NoStop}%
\bibitem [{\citenamefont {Teukolsky}(1973)}]{Teukolsky:1973ha}%
  \BibitemOpen
  \bibfield  {author} {\bibinfo {author} {\bibfnamefont {S.~A.}\ \bibnamefont
  {Teukolsky}},\ }\href {\doibase 10.1086/152444} {\bibfield  {journal}
  {\bibinfo  {journal} {Astrophys. J.}\ }\textbf {\bibinfo {volume} {185}},\
  \bibinfo {pages} {635} (\bibinfo {year} {1973})}\BibitemShut {NoStop}%
\bibitem [{\citenamefont {Dolan}(2008)}]{Dolan:2008kf}%
  \BibitemOpen
  \bibfield  {author} {\bibinfo {author} {\bibfnamefont {S.~R.}\ \bibnamefont
  {Dolan}},\ }\href {\doibase 10.1088/0264-9381/25/23/235002} {\bibfield
  {journal} {\bibinfo  {journal} {Class. Quant. Grav.}\ }\textbf {\bibinfo
  {volume} {25}},\ \bibinfo {pages} {235002} (\bibinfo {year} {2008})},\
  \Eprint {http://arxiv.org/abs/0801.3805} {arXiv:0801.3805 [gr-qc]}
  \BibitemShut {NoStop}%
\bibitem [{\citenamefont {Georgi}(1990)}]{Georgi:1990um}%
  \BibitemOpen
  \bibfield  {author} {\bibinfo {author} {\bibfnamefont {H.}~\bibnamefont
  {Georgi}},\ }\href {\doibase 10.1016/0370-2693(90)91128-X} {\bibfield
  {journal} {\bibinfo  {journal} {Phys. Lett. B}\ }\textbf {\bibinfo {volume}
  {240}},\ \bibinfo {pages} {447} (\bibinfo {year} {1990})}\BibitemShut
  {NoStop}%
\bibitem [{\citenamefont {Luke}\ and\ \citenamefont
  {Manohar}(1992)}]{Luke:1992cs}%
  \BibitemOpen
  \bibfield  {author} {\bibinfo {author} {\bibfnamefont {M.~E.}\ \bibnamefont
  {Luke}}\ and\ \bibinfo {author} {\bibfnamefont {A.~V.}\ \bibnamefont
  {Manohar}},\ }\href {\doibase 10.1016/0370-2693(92)91786-9} {\bibfield
  {journal} {\bibinfo  {journal} {Phys. Lett. B}\ }\textbf {\bibinfo {volume}
  {286}},\ \bibinfo {pages} {348} (\bibinfo {year} {1992})},\ \Eprint
  {http://arxiv.org/abs/hep-ph/9205228} {arXiv:hep-ph/9205228} \BibitemShut
  {NoStop}%
\bibitem [{\citenamefont {Manohar}\ and\ \citenamefont
  {Wise}(2000)}]{Manohar:2000dt}%
  \BibitemOpen
  \bibfield  {author} {\bibinfo {author} {\bibfnamefont {A.~V.}\ \bibnamefont
  {Manohar}}\ and\ \bibinfo {author} {\bibfnamefont {M.~B.}\ \bibnamefont
  {Wise}},\ }\href@noop {} {\emph {\bibinfo {title} {{Heavy quark physics}}}},\
  Vol.~\bibinfo {volume} {10}\ (\bibinfo {year} {2000})\BibitemShut {NoStop}%
\bibitem [{\citenamefont {Heinonen}\ \emph {et~al.}(2012)\citenamefont
  {Heinonen}, \citenamefont {Hill},\ and\ \citenamefont
  {Solon}}]{Heinonen:2012km}%
  \BibitemOpen
  \bibfield  {author} {\bibinfo {author} {\bibfnamefont {J.}~\bibnamefont
  {Heinonen}}, \bibinfo {author} {\bibfnamefont {R.~J.}\ \bibnamefont {Hill}},
  \ and\ \bibinfo {author} {\bibfnamefont {M.~P.}\ \bibnamefont {Solon}},\
  }\href {\doibase 10.1103/PhysRevD.86.094020} {\bibfield  {journal} {\bibinfo
  {journal} {Phys. Rev. D}\ }\textbf {\bibinfo {volume} {86}},\ \bibinfo
  {pages} {094020} (\bibinfo {year} {2012})},\ \Eprint
  {http://arxiv.org/abs/1208.0601} {arXiv:1208.0601 [hep-ph]} \BibitemShut
  {NoStop}%
\bibitem [{\citenamefont {Brivio}\ and\ \citenamefont
  {Trott}(2019)}]{Brivio:2017vri}%
  \BibitemOpen
  \bibfield  {author} {\bibinfo {author} {\bibfnamefont {I.}~\bibnamefont
  {Brivio}}\ and\ \bibinfo {author} {\bibfnamefont {M.}~\bibnamefont {Trott}},\
  }\href {\doibase 10.1016/j.physrep.2018.11.002} {\bibfield  {journal}
  {\bibinfo  {journal} {Phys. Rept.}\ }\textbf {\bibinfo {volume} {793}},\
  \bibinfo {pages} {1} (\bibinfo {year} {2019})},\ \Eprint
  {http://arxiv.org/abs/1706.08945} {arXiv:1706.08945 [hep-ph]} \BibitemShut
  {NoStop}%
\bibitem [{\citenamefont {Conde}\ and\ \citenamefont
  {Marzolla}(2016)}]{Conde:2016vxs}%
  \BibitemOpen
  \bibfield  {author} {\bibinfo {author} {\bibfnamefont {E.}~\bibnamefont
  {Conde}}\ and\ \bibinfo {author} {\bibfnamefont {A.}~\bibnamefont
  {Marzolla}},\ }\href {\doibase 10.1007/JHEP09(2016)041} {\bibfield  {journal}
  {\bibinfo  {journal} {JHEP}\ }\textbf {\bibinfo {volume} {09}},\ \bibinfo
  {pages} {041} (\bibinfo {year} {2016})},\ \Eprint
  {http://arxiv.org/abs/1601.08113} {arXiv:1601.08113 [hep-th]} \BibitemShut
  {NoStop}%
\bibitem [{\citenamefont {Conde}\ \emph {et~al.}(2016)\citenamefont {Conde},
  \citenamefont {Joung},\ and\ \citenamefont {Mkrtchyan}}]{Conde:2016izb}%
  \BibitemOpen
  \bibfield  {author} {\bibinfo {author} {\bibfnamefont {E.}~\bibnamefont
  {Conde}}, \bibinfo {author} {\bibfnamefont {E.}~\bibnamefont {Joung}}, \ and\
  \bibinfo {author} {\bibfnamefont {K.}~\bibnamefont {Mkrtchyan}},\ }\href
  {\doibase 10.1007/JHEP08(2016)040} {\bibfield  {journal} {\bibinfo  {journal}
  {JHEP}\ }\textbf {\bibinfo {volume} {08}},\ \bibinfo {pages} {040} (\bibinfo
  {year} {2016})},\ \Eprint {http://arxiv.org/abs/1605.07402} {arXiv:1605.07402
  [hep-th]} \BibitemShut {NoStop}%
\bibitem [{\citenamefont {Aoude}\ and\ \citenamefont
  {Ochirov}(2021)}]{Aoude:2021oqj}%
  \BibitemOpen
  \bibfield  {author} {\bibinfo {author} {\bibfnamefont {R.}~\bibnamefont
  {Aoude}}\ and\ \bibinfo {author} {\bibfnamefont {A.}~\bibnamefont
  {Ochirov}},\ }\href {\doibase 10.1007/JHEP10(2021)008} {\bibfield  {journal}
  {\bibinfo  {journal} {JHEP}\ }\textbf {\bibinfo {volume} {10}},\ \bibinfo
  {pages} {008} (\bibinfo {year} {2021})},\ \Eprint
  {http://arxiv.org/abs/2108.01649} {arXiv:2108.01649 [hep-th]} \BibitemShut
  {NoStop}%
\bibitem [{\citenamefont {Levi}\ and\ \citenamefont
  {Steinhoff}(2015)}]{Levi:2015msa}%
  \BibitemOpen
  \bibfield  {author} {\bibinfo {author} {\bibfnamefont {M.}~\bibnamefont
  {Levi}}\ and\ \bibinfo {author} {\bibfnamefont {J.}~\bibnamefont
  {Steinhoff}},\ }\href {\doibase 10.1007/JHEP09(2015)219} {\bibfield
  {journal} {\bibinfo  {journal} {JHEP}\ }\textbf {\bibinfo {volume} {09}},\
  \bibinfo {pages} {219} (\bibinfo {year} {2015})},\ \Eprint
  {http://arxiv.org/abs/1501.04956} {arXiv:1501.04956 [gr-qc]} \BibitemShut
  {NoStop}%
\bibitem [{\citenamefont {Alessio}(2023)}]{Alessio:2023kgf}%
  \BibitemOpen
  \bibfield  {author} {\bibinfo {author} {\bibfnamefont {F.}~\bibnamefont
  {Alessio}},\ }\href@noop {} {\  (\bibinfo {year} {2023})},\ \Eprint
  {http://arxiv.org/abs/2303.12784} {arXiv:2303.12784 [hep-th]} \BibitemShut
  {NoStop}%
\bibitem [{\citenamefont {Saketh}\ and\ \citenamefont
  {Vines}(2022)}]{Saketh:2022wap}%
  \BibitemOpen
  \bibfield  {author} {\bibinfo {author} {\bibfnamefont {M.~V.~S.}\
  \bibnamefont {Saketh}}\ and\ \bibinfo {author} {\bibfnamefont
  {J.}~\bibnamefont {Vines}},\ }\href@noop {} {\  (\bibinfo {year} {2022})},\
  \Eprint {http://arxiv.org/abs/2208.03170} {arXiv:2208.03170 [gr-qc]}
  \BibitemShut {NoStop}%
\bibitem [{\citenamefont {Haddad}\ and\ \citenamefont
  {Helset}(2020)}]{Haddad:2020que}%
  \BibitemOpen
  \bibfield  {author} {\bibinfo {author} {\bibfnamefont {K.}~\bibnamefont
  {Haddad}}\ and\ \bibinfo {author} {\bibfnamefont {A.}~\bibnamefont
  {Helset}},\ }\href {\doibase 10.1007/JHEP12(2020)024} {\bibfield  {journal}
  {\bibinfo  {journal} {JHEP}\ }\textbf {\bibinfo {volume} {12}},\ \bibinfo
  {pages} {024} (\bibinfo {year} {2020})},\ \Eprint
  {http://arxiv.org/abs/2008.04920} {arXiv:2008.04920 [hep-th]} \BibitemShut
  {NoStop}%
\bibitem [{\citenamefont {Aoude}\ \emph {et~al.}(2021)\citenamefont {Aoude},
  \citenamefont {Haddad},\ and\ \citenamefont {Helset}}]{Aoude:2020ygw}%
  \BibitemOpen
  \bibfield  {author} {\bibinfo {author} {\bibfnamefont {R.}~\bibnamefont
  {Aoude}}, \bibinfo {author} {\bibfnamefont {K.}~\bibnamefont {Haddad}}, \
  and\ \bibinfo {author} {\bibfnamefont {A.}~\bibnamefont {Helset}},\ }\href
  {\doibase 10.1007/JHEP03(2021)097} {\bibfield  {journal} {\bibinfo  {journal}
  {JHEP}\ }\textbf {\bibinfo {volume} {03}},\ \bibinfo {pages} {097} (\bibinfo
  {year} {2021})},\ \Eprint {http://arxiv.org/abs/2012.05256} {arXiv:2012.05256
  [hep-th]} \BibitemShut {NoStop}%
\bibitem [{\citenamefont {Levi}\ and\ \citenamefont
  {Yin}(2022)}]{Levi:2022rrq}%
  \BibitemOpen
  \bibfield  {author} {\bibinfo {author} {\bibfnamefont {M.}~\bibnamefont
  {Levi}}\ and\ \bibinfo {author} {\bibfnamefont {Z.}~\bibnamefont {Yin}},\
  }\href@noop {} {\  (\bibinfo {year} {2022})},\ \Eprint
  {http://arxiv.org/abs/2211.14018} {arXiv:2211.14018 [hep-th]} \BibitemShut
  {NoStop}%
\end{thebibliography}%

\end{document}